\documentclass[12pt]{article}
\begin{document}
\newcommand{\beq}{\begin{equation}}
\newcommand{\eeq}{\end{equation}}
\newcommand{\beqa}{\begin{eqnarray}}
\newcommand{\eeqa}{\end{eqnarray}}
\newcommand{\beqar}{\begin{eqnarray*}}
\newcommand{\eeqar}{\end{eqnarray*}}
\newcommand{\al}{\alpha}
\newcommand{\be}{\beta}
\newcommand{\del}{\delta}
\newcommand{\D}{\Delta}
\newcommand{\eps}{\epsilon}
\newcommand{\ga}{\gamma}
\newcommand{\Ga}{\Gamma}
\newcommand{\ka}{\kappa}
\newcommand{\nn}{\nonumber}
\newcommand{\inn}{\!\cdot\!}
\newcommand{\h}{\eta}
\newcommand{\ii}{\iota}
\newcommand{\kk}{\varphi}
\newcommand\F{{}_3V_2}
\newcommand{\la}{\lambda}
\newcommand{\La}{\Lambda}
\newcommand{\na}{\prt}
\newcommand{\Om}{\Omega}
\newcommand{\om}{\omega}
\newcommand{\p}{\phi}
\newcommand{\sig}{\sigma}
\renewcommand{\t}{\theta}
\newcommand{\z}{\zeta}
\newcommand{\ssc}{\scriptscriptstyle}
\newcommand{\eg}{{\it e.g.,}\ }
\newcommand{\ie}{{\it i.e.,}\ }
\newcommand{\labell}[1]{\label{#1}} 
\newcommand{\reef}[1]{(\ref{#1})}
\newcommand{\labels}[1]{\vskip-2ex$_{#1}$\label{#1}} 
\newcommand\prt{\partial}
\newcommand\veps{\varepsilon}
\newcommand{\pol}{\varepsilon}
\newcommand\vp{\varphi}
\newcommand\ls{\ell_s}
\newcommand\cF{{\cal F}}
\newcommand\cA{{\cal A}}
\newcommand\cS{{\cal S}}
\newcommand\cT{{\cal T}}
\newcommand\cV{{\cal V}}
\newcommand\cL{{\cal L}}
\newcommand\cM{{\cal M}}
\newcommand\cN{{\cal N}}
\newcommand\cG{{\cal G}}
\newcommand\cH{{\cal H}}
\newcommand\cI{{\cal I}}
\newcommand\cJ{{\cal J}}
\newcommand\cl{{\iota}}
\newcommand\cK{{\cal K}}
\newcommand\cQ{{\cal Q}}
\newcommand\cg{{\it g}}
\newcommand\cR{{\cal R}}
\newcommand\cB{{\cal B}}
\newcommand\cO{{\cal O}}
\newcommand\tcO{{\tilde {{\cal O}}}}
\newcommand\bg{\bar{g}}
\newcommand\bb{\bar{b}}
\newcommand\bH{\bar{H}}
\newcommand\bX{\bar{X}}
\newcommand\bK{\bar{K}}
\newcommand\bR{\bar{R}}
\newcommand\bZ{\bar{Z}}
\newcommand\ba{\bar{a}}
\newcommand\bxi{\bar{\xi}}
\newcommand\bphi{\bar{\phi}}
\newcommand\bpsi{\bar{\psi}}
\newcommand\bprt{\bar{\prt}}
\newcommand\bet{\bar{\eta}}
\newcommand\btau{\bar{\tau}}
\newcommand\hF{\hat{F}}
\newcommand\hA{\hat{A}}
\newcommand\hT{\hat{T}}
\newcommand\htau{\hat{\tau}}
\newcommand\hD{\hat{D}}
\newcommand\hf{\hat{f}}
\newcommand\hg{\hat{g}}
\newcommand\hp{\hat{\phi}}
\newcommand\hi{\hat{i}}
\newcommand\ha{\hat{a}}
\newcommand\hb{\hat{b}}
\newcommand\hc{\hat{c}}
\newcommand\hQ{\hat{Q}}
\newcommand\hP{\hat{\Phi}}
\newcommand\hS{\hat{S}}
\newcommand\hX{\hat{X}}
\newcommand\tL{\tilde{\cal L}}
\newcommand\hL{\hat{\cal L}}
\newcommand\tG{{\widetilde G}}
\newcommand\tg{{\widetilde g}}
\newcommand\tphi{{\widetilde \phi}}
\newcommand\tPhi{{\widetilde \Phi}}
\newcommand\td{{\tilde d}}
\newcommand\te{{\tilde e}}
\newcommand\tk{{\tilde k}}
\newcommand\tf{{\tilde f}}
\newcommand\ta{{\tilde a}}
\newcommand\tb{{\tilde b}}
\newcommand\tR{{\tilde R}}
\newcommand\teta{{\tilde \eta}}
\newcommand\tF{{\widetilde F}}
\newcommand\tK{{\widetilde K}}
\newcommand\tE{{\widetilde E}}
\newcommand\tpsi{{\tilde \psi}}
\newcommand\tX{{\widetilde X}}
\newcommand\tD{{\widetilde D}}
\newcommand\tO{{\widetilde O}}
\newcommand\tS{{\tilde S}}
\newcommand\tB{{\widetilde B}}
\newcommand\tA{{\widetilde A}}
\newcommand\tT{{\widetilde T}}
\newcommand\tC{{\widetilde C}}
\newcommand\tV{{\widetilde V}}
\newcommand\thF{{\widetilde {\hat {F}}}}
\newcommand\Tr{{\rm Tr}}
\newcommand\tr{{\rm tr}}
\newcommand\STr{{\rm STr}}
\newcommand\hR{\hat{R}}
\newcommand\M[2]{M^{#1}{}_{#2}}

\newcommand\bS{\textbf{ S}}
\newcommand\bI{\textbf{ I}}
\newcommand\bJ{\textbf{ J}}
\newcommand\bTD{\textbf{ TD}}

\begin{titlepage}
\begin{center}

\vskip 2 cm
{\LARGE \bf   Surface terms in effective action of O-plane   \\  \vskip 0.25 cm  at order $\alpha'^2$
 }\\
\vskip 1.25 cm
 
Yasser Akou\footnote{akou@mail.um.ac.ir} and   Mohammad R. Garousi\footnote{garousi@um.ac.ir}

\vskip 1 cm
{{\it Department of Physics, Faculty of Science, Ferdowsi University of Mashhad\\}{\it P.O. Box 1436, Mashhad, Iran}\\}
\vskip .1 cm
 \end{center}

\begin{abstract}
The  effective action of   string theory   has both bulk and boundary terms if the spacetime is an open  manifold.  Recently, the known  classical effective action of string theory at the leading order of $\alpha'$ and its corresponding boundary action have been reproduced by constraining the effective actions to be invariant  under gauge transformations  and under string duality transformations. In this paper, we  use this idea to find the classical effective action of the O-plane and its corresponding boundary terms in type II superstring theories  at order $\alpha'^2$ and for  NS-NS couplings. We find that  these constraints fix the bulk action and its corresponding boundary terms up to one overall factor. They also produce three multiplets in the boundary action that their coefficients are independent of the bulk couplings under the string dualities. 
  
\end{abstract}
\end{titlepage}

\section{Introduction}

 Perturbative string theory is a quantum theory of gravity   with  a finite number of massless fields and a  tower of infinite number of  massive fields reflecting the stringy nature of the gravity at the weak coupling. String theory on the spacetime manifolds with boundary is conjectured to be dual to a  gauge theory on the boundary \cite{Maldacena:1997re}. The string theory and its non-perturbative objects are  usually explored by studying their low-energy effective actions  which include the massless fields and their  covariant derivatives. For the open spacetime manifolds, the effective actions have both bulk and boundary terms, \ie $\bS_{\rm eff}+\prt\!\! \bS_{\rm eff}$.
 They should  be produced by specific techniques in string theory. 
 
 There are various approaches  for calculating the   bulk   effective action  $\bS_{\rm eff}$, \eg  the S-matrix approach \cite{Scherk:1974mc,Yoneya:1974jg},  the  sigma-model approach \cite{Callan:1985ia,Fradkin:1984pq}, the Double Field Theory   approach  \cite{ Siegel:1993th,Hull:2009mi} and the duality  approach \cite{Ferrara:1989bc,Font:1990gx,Green:1997tv,Green:2016tfs,Garousi:2017fbe,Green:2019rhz}.
In the duality approach, the  consistency of the effective actions with gauge transformations and with T- and S-duality transformations are imposed to find the higher derivative couplings. The Double Field Theory and T-duality  approaches are based on the observation made by Sen in the context of closed string field theory \cite{Sen:1991zi} that the classical effective action of bosonic string theory should be invariant under T-duality to all orders in $\alpha'$. Similar observation has been made for the hetrotic string theory in \cite{Hohm:2014sxa}.
 
 In the T-duality  approach,  by removing total derivative terms, using field redefinitions and using Bianchi identities, one first find the  minimum number of  independent and gauge invariant couplings in  the string frame action $\!\!\bS_{\rm eff}$ at each order of $\alpha'$. Then  one reduces the spacetime on a circle, \ie $M^{(D)}=S^{(1)}\times M^{(D-1)}$. The  T-duality  \cite{Giveon:1994fu,Alvarez:1994dn} is imposed as a constraint on the reduction of the effective action on the  circle to find the coefficients of the independent couplings, \ie the  effective action satisfies the following constraint:
  \beqa
 S_{\rm eff}(\psi)-S_{\rm eff}(\psi')&=&{\rm TD} \labell{TS}
 \eeqa
where $ S_{\rm eff}$ is the reduction of the effective action on the circle, $\psi$ represents all  massless fields in the base space $M^{(D-1)}$ and $\psi'$ represents  their transformations under the T-duality  transformations which are the Buscher rules \cite{Buscher:1987sk,Buscher:1987qj} and their higher derivative corrections. They form a  $Z_2$-subgroup of $O(1,1;R)$.  On the right-hand side, TD represents some total derivative terms  in the base space which may not be invariant under the T-duality. They become zero for the closed spacetime manifolds using the Stokes's theorem. This approach has been used in \cite{Garousi:2019wgz,Garousi:2019mca} to find  effective action of the bosonic string theory at orders $ \alpha',\, \alpha'^2$.  This approach has been also used in \cite{Garousi:2020mqn,Garousi:2020gio} to construct NS-NS couplings in type II superstring  effective action at order $\alpha'^3$ 

The constraint \reef{TS} for the effective action of the non-perturbative D$_p$-brane/O$_p$-plane objects is such that  $S_{\rm eff}(\psi)$ represents the reduction of $(p-1)$-brane action along the circle transverse to the brane, \ie $M^{(D)}=M^{(p)}\times M^{(D-p)}$ where  $(p-1)$-brane is in the subspace $M^{(p)}$ and $M^{(D-p)}=S^{(1)}\times M^{(D-p-1)}$,  and  $S_{\rm eff}(\psi')$ represents T-duality transformation of the reduction of $p$-brane action along the circle tangent  to the brane, \ie $M^{(D)}=M^{(p+1)}\times M^{(D-p-1)}$ where  $p$-brane is in the subspace $M^{(p+1)}$ and $M^{(p+1)}=S^{(1)}\times M^{(p)}$.  This approach has been  used to construct the O$_p$-plane effective action at order $\alpha'^2$ in type II superstring theory for zero R-R field  in \cite{Robbins:2014ara,Garousi:2014oya}, and for linear R-R field  in \cite{Mashhadi:2020mzf}. The latter couplings include the well-known anomalous coupling $C\wedge \Tr(R\wedge R)$, as well as some non-anomalous  couplings involving the R-R field strengths. 

The type IIB superstring theory has S-duality, hence, its effective action should be invariant under the S-duality as well. To have an S-duality invariant effective action one should  include to the tree-level effective action  the non-perturbative and string loop effects \cite{Green:1997tv,Green:2016tfs}. They  are required to make the tree-level effective action to be invariant under the S-duality  group $SL(2,Z)$. Even the  tree-level effective action at a given order of $\alpha'$  should be also consistent with S-duality in the sense that up to an overall dilaton factor, the action should be invariant under the S-duality  group $SL(2,R)$. To study the S-duality, one first should change the string-frame metric to the Einstein-frame metric, \ie $G_{\mu\nu}=e^{\phi/2}G_{\mu\nu}^{(E)}$. Then up to some total derivative terms the effective action should be written as an S-duality invariant form, \ie
\beqa
\bS_{\rm eff}(G,\phi,B,C^{(0)},C^{(2)},C^{(4)})&=&\bS_{\rm eff}(G^E,\tau,\bar{\tau},\cH,C^{(4)})+\bTD\labell{TSS}
\eeqa 
where the Einstein-frame metric and R-R four-form are invariant under the S-duality, $\cH$ which includes the B-field and the R-R two-form, transforms as doublet and $\tau$ which includes the dilaton and the R-R scalar, transforms as modular transformation. On the right-hand side of above equation, $\!\!\bTD $ again represents some total derivative terms   which may not be invariant under the S-duality. They however become zero for the closed spacetime manifolds using the Stokes's theorem. Since the R-R four-form couples to the non-perturbative D$_3$-brane and O$_3$-plane objects, up to some total derivative terms,  the effective action of these objects should be also invariant under the S-duality \cite{Bachas:1999um}.

When the spacetime manifold has boundary $\prt M^{(D)}$, the total derivative terms on the right-hand sides of \reef{TS} and \reef{TSS} can not be ignored. If one ignores them then the effective action would not be invariant under the T-duality and S-duality. In fact, for the open spacetime manifold,  the total derivative terms in the original spacetime and in the base space  have physical effects and, hence,  should not be ignored. On the other hand,  there might be some couplings $\prt\!\! \bS_{\rm eff}$ at the boundary of the spacetime that one should take into account  to have a fully duality invariant effective action. At the leading order of $\alpha'$, requiring the effective action to be invariant under the gauge transformations and under the T- and S-duality transformations, one can fix the couplings up to an overall normalisation factor \cite{Garousi:2019xlf}. In fact, the total derivative terms on the right-hand sides of the duality constraints \reef{TS} and \reef{TSS} at the leading order of $\alpha'$ are cancelled if one includes  the Gibbons-Hawking-York  boundary term \cite{York:1972sj,Gibbons:1976ue} in the boundary action.  At the higher orders of $\alpha'$,   the gauge and duality constraints may also fix both the bulk  and  the  boundary actions. 

 Using the Stokes's theorem, the total derivative terms in the bulk  action $\!\!\bS_{\rm eff}$ can be transferred to the boundary action $\prt\!\! \bS_{\rm eff}$ to produce couplings that are proportional to the unit vector of the boundary. As a result, one can write the bulk action without total derivative terms even for the  spacetime manifolds which have boundary. Hence the extension of the T-duality constraint \reef{TS} to the spacetime with boundary has two parts. One part is exactly as in \reef{TS} in which the bulk action $\!\!\bS_{\rm eff}$  has no total derivative term, however, the  total derivative terms in the base space $M^{(D-1)}$ which appear on the right-hand side of \reef{TS},  are transferred to the boundary $\prt M^{(D-1)}$ in the base space using the Stokes's theorem. We  call them $\prt {\rm TD}$. In the second part one first write all independent gauge invariant couplings in the boundary action $\prt\!\! \bS_{\rm eff}$ including the couplings which are proportional to the unit vector. Then  one should add $\prt {\rm TD}$ to the T-duality constraint on the boundary action, \ie
 \beqa
\prt {\rm TD}+\prt S_{\rm eff}(\psi)-\prt S_{\rm eff}(\psi')&=&{\cal TD}\labell{TT}
 \eeqa
where ${\cal TD}$ represents some boundary total derivative terms. Since the boundary of boundary, \ie $\prt\prt M^{(D-1)}$,   is zero  ${\cal TD}$ becomes zero after using the Stokes's theorem. 
The sum of the bulk constraint \reef{TS} and the boundary constraint \reef{TT} means the total bulk and boundary actions are invariant under the T-duality, \ie 
 \beqa
 S_{\rm eff}(\psi)+\prt S_{\rm eff}(\psi)&=&S_{\rm eff}(\psi')+\prt S_{\rm eff}(\psi')\labell{TTS}
 \eeqa 
up to some total derivative terms in the boundary of base space $\prt M^{(D-1)}$ which are zero by the Stokes's theorem. The above T-duality constraint has been used in \cite{Garousi:2019xlf} to reproduce the  bulk and boundary couplings at the leading order of $\alpha'$. It reproduces the known bulk couplings and its corresponding  the Gibbons-Hawking-York  boundary term. However, it produces an extra T-dual multiplet in the boundary as well.

The  T-duality constraint \reef{TTS} for D$_p$-brane/O$_p$-plane is such that when spacetime has boundary $\prt M^{(D)}$, the $p$-branes  may end on the boundary, \ie $\prt M^{(D)}=\prt M^{(p+1)}\times M^{(D-p-1)}$. Hence their corresponding low-energy effective action should have boundary terms as well. In this case, $\prt S_{\rm eff}(\psi)$   represents the reduction  of the $(p-1)$-brane  boundary action along the circle transverse to the brane,   \ie $\prt M^{(D)}=\prt M^{(p)}\times M^{(D-p)}$ where  the $(p-1)$-brane boundary action is in the subspace $\prt M^{(p)}$ and $M^{(D-p)}=S^{(1)}\times M^{(D-p-1)}$, and $\prt S_{\rm eff}(\psi')$ represents T-duality transformation of the reduction  of the $p$-brane  boundary action along the circle tangent  to the brane, \ie $\prt M^{(D)}=\prt M^{(p+1)}\times M^{(D-p-1)}$ where  the $p$-brane boundary action is in the subspace $\prt M^{(p+1)}$ and $\prt M^{(p+1)}=S^{(1)}\times \prt M^{(p)}$.

Similarly, the extension of the S-duality constraint \reef{TSS} to the spacetime with boundary has two parts. One part is exactly as in \reef{TSS} in which the bulk action $\!\!\bS_{\rm eff}$ in the string frame  has no total derivative term, however, the  total derivative terms  $ \!\!{\bTD}$ resulting from transforming the string frame action to the Einstein frame  are transferred to the boundary  using the Stokes's theorem. We  call them  $\prt \!\!{\bTD}$. In the second part, one should combine them  with the  boundary action to be written in the S-duality invariant form, \ie
 \beqa
&&\prt {\rm \!\!\bTD}(n,G,\phi,B,C^{(0)},C^{(2)},C^{(4)})+\prt \!\!\bS_{\rm eff}(n,G,\phi,B,C^{(0)},C^{(2)},C^{(4)})\nn\\
&&=\prt \!\!\bS_{\rm eff}(G^E,\tau,\bar{\tau},\cH,C^{(4)})+\!\!\textbf{ td}\nn
 \eeqa
where $\!\!\textbf{ td}$ represents some  total derivative terms, however, since the boundary of boundary is zero, they are zero by using the Stokes's theorem. The sum of the bulk constraint \reef{TSS} and the above boundary constraint  means the total bulk and boundary actions are invariant under the S-duality, \ie 
\beqa
\bS_{\rm eff}+ \prt\!\! \bS_{\rm eff}= \!\!\bS_{\rm eff}(G^E,\tau,\bar{\tau},\cH,C^{(4)})+\prt \!\!\bS_{\rm eff}(G^E,\tau,\bar{\tau},\cH,C^{(4)})\labell{SS}
\eeqa
up to some total derivative terms in the boundary $\prt M^{(D)}$ which are zero by the Stokes's theorem. The above S-duality constraint has been used in \cite{Garousi:2019xlf} on the couplings that the T-duality produces at the leading order of  $\alpha'$. This constraint removes the extra  couplings in the boundary that the T-duality produces. 

The  S-duality constraint \reef{SS} for D$_3$-brane/O$_3$-plane is such that the combination of world-volume action and its boundary terms should be written in an S-duality invariant form up to some total derivative terms in the world-volume boundary  $\prt M^{(4)}$ which are zero by the Stokes's theorem..

In this paper, we are going to apply the T-duality constraint \reef{TTS} and the S-duality constraint \reef{SS} on the  effective actions of O-plane when spacetime has boundary. We are interested in NS-NS couplings of O-planes of  type II superstring theory. At the leading order of $\alpha'$ there is no boundary term and the  bulk action which is gien by DBI action is  invariant under T-duality and S-duality (see \eg \cite{Garousi:2017fbe}). The first corrections to the DBI action is at order $\alpha'^2$.   At this order, the T-duality transformations are given only by the Buscher rules because  the first corrections to the effective action of type II superstring theory are at order $\alpha'^3$. To  study the S-duality at order $\alpha'^2$, one needs to take into account R-R fields as well in which we are not interested in this paper. However, it has been observed in \cite{Garousi:2013qka} that it is impossible to combine couplings in the Einstein frame involving  odd number of dilatons and zero B-field with corresponding R-R couplings to be written in an S-duality invariant form. Hence the S-duality constraint on the NS-NS couplings is such that the O$_3$-plane couplings with zero B-field which involve  odd number of dilatons must be zero.  The T-duality constraint  as well as  this S-duality constraint may fix the NS-NS couplings in the bulk and boundary actions of O-planes. 

The outline of the paper is as follows:  In section 2.1, we first impose gauge symmetry to show that there are 48 independent bulk couplings. In section 2.2, we impose T-duality to fix the 48 couplings up to an overall factor, and up to some total derivative terms in the base space which are transferred to the boundary by using the Stokes's theorem.  In section 2.3, we show that the bulk couplings that are fixed by the gauge symmetry and the T-duality, are consistent with  S-duality up to some total derivative terms  which are transferred to the boundary by using the Stokes's theorem. In section 3.1, we first impose gauge symmetry to show that there are 78 independent boundary couplings. In section 3.2, we show that  the T-duality can not fix all parameters. In fact we find, apart from the boundary couplings that are needed to make the total derivative terms in the bulk to be invariant under the T-duality, there are 17 boundary  multiplets that are  T-duality invariant. In section 3.3,  we impose the S-duality constraint on the T-duality invariant  couplings. We find,  apart from the boundary couplings that are needed to make the total derivative terms in the bulk to be invariant under the T-duality and S-duality, there are also three other boundary multipletes that are invariant under the T-duality and S-duality. In section 4, we briefly discuss our results.
 
 \section{Bulk couplings}
 
 The NS-NS couplings in the  O$_p$-plane bulk action at order $\alpha'^2$ have been found in \cite{Robbins:2014ara,Garousi:2014oya} by the T-duality method. The total derivative terms in the base space, \ie the TD on the right-hand side of \reef{TS}, are needed for the calculations of the boundary action in \reef{TT}. So we reproduce the bulk couplings here again to find the corresponding total derivative terms in the base space. To this end, we need first to find minimum number of independent and gauge invariant terms at order $\alpha'^2$ and then reduce them on a circle to apply the T-duality constraint \reef{TS}. So let us  find how many independent  gauge invariant couplings are in the bulk.

\subsection{Minimal gauge invariant   couplings in the bulk}

In this subsection we would like to find all independent and  gauge invariant   couplings on the O$_p$-plane bulk action involving the   NS-NS fields at order $\alpha'^2$ in the string frame, \ie
\beqa
\bS_p&=&-\frac{T_p\pi^2\alpha'^2}{48}\int_{M^{(p+1)}} d^{p+1}\sigma\, e^{-\phi}\sqrt{-\tg}\,{\cal L}_p\labell{Gen}
\eeqa
where the 10-dimensional spacetime is written as $M^{(10)}=M^{(p+1)}\times M^{(9-p)}$ and the O$_p$-plane is along the subspace $M^{(p+1)}$. In above equation,  $\tg$ is determinant of the pull-back metric 
\beqa
\tg_{ab}&=&\frac{\prt X^\mu}{\prt\sigma^a}\frac{\prt X^\nu}{\prt\sigma^b} G_{\mu\nu}
\eeqa
The O$_p$-plane is specified in the spacetime by vectors $X^{\mu}(\sigma^a)$, $T_p$ is tension of O$_p$-plane and ${\cal L}_p$  is the Lagrangian we are after which includes all independent couplings. 

As it has been argued in \cite{Robbins:2014ara},  since we are interested in O$_p$-plane as a probe, it does not have back reaction on the spacetime. As a result, the massless closed string fields must satisfy the bulk equations of motion at order $\alpha'^0$, \ie
\beqa
&&0=R +4\nabla_\mu\nabla^\mu \phi -4\nabla_\mu\phi\nabla^\mu \phi-\frac{1}{12}H^{\mu\nu\rho}H_{\mu\nu\rho}+\cdots\nn\\
&&0=R_{\mu\nu} +2\nabla_\mu\nabla_\nu \phi-\frac{1}{4} H_{\mu}^{\rho\sigma}H_{\nu\rho\sigma}+\cdots\nn\\
&& 0= \nabla_\mu\nabla^\mu \phi -2\nabla_\mu \phi\nabla^\mu \phi +\frac{1}{12}H^{\mu\nu\rho}H_{\mu\nu\rho}+\cdots\nn\\
&&0=\nabla^\rho H_{\mu\nu\rho}- 2\nabla^\rho \phi H_{\mu\nu\rho}\labell{eom}
\eeqa
where dots  represent terms involving the R-R fields in which we are not interested in this paper. In the third line we use subtraction of the first equation and the  contraction  of second equation with  metric $ G^{\mu\nu} $. To impose these equations, we remove  $ R $,  $ R_{\mu\nu} $, $ \nabla_\mu\nabla^\mu \phi $ and $ \nabla^\mu H_{\mu \nu\rho}$ and their derivatives from the Lagrangian $\cL_p$.
As a result,  one can rewrite the terms in the world-volume theory which have  contraction of two transverse indices, \eg $\nabla_i\nabla^i\Phi$,  $R_{i\mu}{}^i{}_\nu$, or $\nabla^i H_{i\mu\nu}$ in terms of contraction of two world-volume indices, \eg  $\nabla_a\nabla^a\Phi$,  $R_{a\mu}{}^a{}_\nu$, or $\nabla^a H_{a\mu\nu}$. This indicates that the former couplings are not independent.  Moreover,  the O$_p$-plane effective action has no open string couplings, no couplings that have odd number of transverse indices on metric and dilaton and their corresponding derivatives, and no couplings that have even number of transverse indices on B-field and its corresponding derivatives \cite{Polchinski:1996na}. This orientifold projection makes the construction of the O-plane effective action to be much more easier than the construction of the D-brane action at a given order of $\alpha'$.

 The couplings involving the Riemann curvature and its derivatives and the couplings involving derivatives of $H=dB$  satisfy the following Bianchi identities
\beqa
R_{\mu[\nu\alpha\beta]}&=&0\nn\\
\nabla_{[\mu}R_{\nu\alpha]\beta\gamma}&=&0\nn\\
dH&=&0\labell{Bian}
\eeqa
Moreover, the couplings involving the commutator of two covariant derivatives of a tensor are  not independent of the couplings involving the contraction of this tensor with the  Riemann curvature, \ie  
\beqa
[\nabla, \nabla ]{\cal O}&=&R{\cal O}\labell{RQ}
\eeqa
This indicates that if one considers all couplings involving the Riemann curvature, then only one ordering of covariant derivatives is needed to be considered as independent coupling\footnote{We have used the package ''xAct" \cite{Nutma:2013zea} for performing the calculations in this paper.}.

To find all  independent and  gauge invariant couplings at order $\alpha'^2$,  we first consider all even-parity contractions of $\tG,\bot, H$, $ \nabla H$, $ \nabla\nabla H$, $\nabla\Phi$, $\nabla\nabla\Phi$, $\nabla\nabla\nabla\Phi$, $\nabla\nabla\nabla\nabla\Phi$, $R$, $\nabla R$, $\nabla\nabla R$ at four-derivative order, where the first fundamental form $\tG^{\mu\nu}$  and the tensor  $\bot^{\mu\nu}$   project the spacetime tensors along the O-plane and orthogonal to the O-plane, respectively.  We then remove the terms which are projected out by the orientifold  projection and by the equations of motion. We call the remaining terms, with coefficients $a'_1, a'_2, \cdots $,  the Lagrangian  $L_p$. Not all terms in this Lagrangian, however, are  independent. Some of them are related by total derivative terms and by Bianchi identities \reef{Bian} and \reef{RQ}. To remove the redundancy corresponding to the total derivative terms, we add to  $L_p$ all total derivative terms at order $\alpha'^2$ with arbitrary coefficients. To this end we first write   all even-parity contractions of     $\tG,\bot, H$, $ \nabla H$, $\nabla\Phi$, $\nabla\nabla\Phi$, $\nabla\nabla\nabla\Phi$,  $R$, $\nabla R$ at three-derivative order with one free world-volume index.  Then we remove the terms which are projected out by the orientifold projection and by the equations of motion as we have done for $L_p$. We call the remaining terms, with arbitrary coefficients,  the vector $I_a$. The total derivative terms are then
\beqa
\alpha'^2\int d^{p+1}\sigma\,\sqrt{-\tg}J&=&\alpha'^2\int d^{p+1}\sigma\,\sqrt{-\tg}\,\tg^{ab}\nabla_a ( e^{-\phi} I_b)\labell{J}
\eeqa
where $\tg^{ab}$ is inverse of the pull-back metric.

Adding the total derivative terms with arbitrary coefficients  to $L_p$, one finds the same Lagrangian     but with different parameters $a_1, a_2, \cdots$. We call the new Lagrangian  ${\cal L}_p$. Hence 
\beqa
\Delta-J&=&0\labell{DL}
\eeqa
where $\Delta={\cal L}_p-L_p$ is the same as $L_p$ but with coefficients $\delta a_1,\delta a_2,\cdots$ where $\delta a_i= a_i-a'_i$. Solving the above equation, one finds some linear  relations between  only $\delta a_1,\delta a_2,\cdots$ which indicate how the couplings are related among themselves by the total derivative terms. The above equation also gives some relation between the coefficients of the total derivative terms and $\delta a_1,\delta a_2,\cdots$ in which we are not interested.

However, to accurately solve the equation \reef{DL} one should write it in terms of independent couplings, \ie    one has to consider the terms in  $\Delta$ and in total derivatives which are not related to each other by the Bianchi identities \reef{Bian}. To impose the Bianchi identities in gauge invariant form, one may contract the  left-hand side of each Bianchi identity with field strengths of dilaton, B-field and metric to produce terms at order $\alpha'^2$. The coefficients of these terms are arbitrary. Adding these terms to the equation \reef{DL}, then one can solve the equation to find the linear relations between   only $\delta a_1,\delta a_2,\cdots$. Alternatively, to impose the  Bianchi identities in non-gauge invariant form, one may rewrite the terms in \reef{DL} in  the local frame in which the first derivative of metric is zero, and  rewrite the terms in \reef{DL} which have derivatives of $H$ in terms of B-field, \ie $H=dB$.  In this way,  the Bianchi identities satisfy automatically \cite{Garousi:2019cdn}. In fact, writing the couplings in terms of potential rather than field strength, there would be no Bianchi identity at all. We find that this latter approach is easier to impose the Bianchi identities by computer. Moreover, in this approach one does not need to introduce a large number of arbitrary parameters to include the Bianchi identities to the equation \reef{DL}.    

Using the above  steps, one can rewrite the different terms on the left-hand side of \reef{DL} in terms of independent but non-gauge invariant couplings. Some combinations of  the parameters appear as coefficients of the independent couplings.  The solution to the equation \reef{DL} which corresponds to setting all these coefficients to zero,  then has two parts. One part is 48 relations between only $\delta a_i$'s, and the other part is some relations between the coefficients of the total derivative terms and $\delta a_i$'s in which we are not interested. The number of relations in the first part gives the number of independent couplings in ${\cal L}_p$. In a particular scheme, one may set some of the coefficients in  $L_p$ to zero, however, after replacing the non-zero terms in \reef{DL}, the number of   relations between only $\delta a_i$'s should not be changed, \ie there must be always 48 relations.   We set the coefficients of the couplings in which each term has more than two derivatives, to zero.  After setting this coefficients to zero, there are still 48 relations between  $\delta a_i$'s.  This means we are  allowed  to  remove these terms.  We choose some other coefficients to zero such that the remaining coefficients satisfy the 48 relations $\delta a_i=0$.    In this way one can find the minimum number of gauge invariant couplings.  One particular choice for the 48  gauge invariant couplings is the following:
\beqa
{\cal L}_p\!\!\!\!&\!\!\!\!\!\!\!\!\!\!\!\!\!=\!\!\!\!\!\!\!\!\!\!\!\!\!\!\!&  a_{1}\  H_{a}{}^{cj} H^{abi} H_{b}{}^{d}{}_{j} H_{cdi} + a_{2}\  H_{a}{}^{c}{}_{i} H^{abi} H_{b}{}^{dj} H_{cdj} + a_{3}\  H_{ab}{}^{j} H^{abi} H_{cdj} H^{cd}{}_{i} \nonumber \\ 
&& + a_{4}\  H_{abi} H^{abi} H_{cdj} H^{cdj} + a_{5}\  H_{a}{}^{cj} H^{abi} H_{bc}{}^{k} H_{ijk} + a_{6}\  H_{ab}{}^{j} H^{abi} H_{i}{}^{kl} H_{jkl} \nonumber \\ 
&& + a_{7}\  H_{abi} H^{abi} H_{jkl} H^{jkl} + a_{8}\  H_{i}{}^{lm} H^{ijk} H_{jl}{}^{n} H_{kmn} + a_{9}\  H_{ij}{}^{l} H^{ijk} H_{k}{}^{mn} H_{lmn} \nonumber \\ 
&& + a_{10}\  H_{ijk} H^{ijk} H_{lmn} H^{lmn} + a_{11}\  H^{abi} H^{cd}{}_{i} R_{abcd} + a_{12}\  H^{abi} H_{i}{}^{jk} R_{abjk} \nonumber \\ 
&& + a_{13}\  H_{ijk} H^{ijk} R^{ab}{}_{ab} + a_{14}\  R_{abcd} R^{abcd} + a_{15}\  R_{abij} R^{abij} + a_{16}\  R_{aibj} R^{aibj} \nonumber \\ 
&& + a_{17}\  H_{ij}{}^{l} H^{ijk} R^{a}{}_{kal} + a_{18}\  H_{a}{}^{c}{}_{i} H^{abi} R_{b}{}^{d}{}_{cd} + a_{19}\  R^{ab}{}_{a}{}^{c} R_{b}{}^{d}{}_{cd}  + a_{20}\  H_{a}{}^{cj} H^{abi} R_{bicj} \nonumber \\ 
&&+ a_{21}\  R^{ai}{}_{a}{}^{j} R^{b}{}_{ibj} + a_{22}\  H_{abi} H^{abi} R^{cd}{}_{cd} + a_{23}\  R^{ab}{}_{ab} R^{cd}{}_{cd}  + a_{24}\  H_{ab}{}^{j} H^{abi} R^{c}{}_{icj}\nonumber \\ 
&& + a_{25}\  R_{ijkl} R^{ijkl} + a_{26}\  H_{i}{}^{lm} H^{ijk} R_{jklm}  + a_{27}\  \nabla_{a}H_{bci} \nabla^{a}H^{bci} + a_{28}\  \nabla_{a}H_{ijk} \nabla^{a}H^{ijk} \nonumber \\ 
&&+ a_{29}\  H_{bci} H^{bci} \nabla_{a}\phi \nabla^{a}\phi  + a_{30}\  H_{ijk} H^{ijk} \nabla_{a}\phi \nabla^{a}\phi + a_{31}\  H_{bci} H^{bci} \nabla^{a}\nabla_{a}\phi \nonumber \\ 
&&+ a_{32}\  H_{ijk} H^{ijk} \nabla^{a}\nabla_{a}\phi  + a_{33}\  R^{bc}{}_{bc} \nabla^{a}\nabla_{a}\phi + a_{34}\  H_{a}{}^{ci} H_{bci} \nabla^{a}\nabla^{b}\phi + a_{35}\  R_{a}{}^{c}{}_{bc} \nabla^{a}\nabla^{b}\phi \nonumber \\ 
&& + a_{36}\  \nabla^{a}\nabla^{b}\phi \nabla_{b}\nabla_{a}\phi + a_{37}\  H_{bci} \nabla^{a}\phi \nabla^{b}H_{a}{}^{ci} + a_{38}\  H_{a}{}^{ci} H_{bci} \nabla^{a}\phi \nabla^{b}\phi  + a_{39}\  R_{a}{}^{c}{}_{bc} \nabla^{a}\phi \nabla^{b}\phi \nonumber \\ 
&&+ a_{40}\  \nabla^{a}\phi \nabla_{b}\nabla_{a}\phi \nabla^{b}\phi  + a_{41}\  \nabla^{a}\nabla_{a}\phi \nabla^{b}\nabla_{b}\phi + a_{42}\  \nabla^{a}H_{a}{}^{bi} \nabla^{c}H_{bci} + a_{43}\  \nabla_{i}H_{abc} \nabla^{i}H^{abc} \nonumber \\ 
&& + a_{44}\  H_{abj} H^{ab}{}_{i} \nabla^{i}\nabla^{j}\phi + a_{45}\  H_{i}{}^{kl} H_{jkl} \nabla^{i}\nabla^{j}\phi + a_{46}\  R^{a}{}_{iaj} \nabla^{i}\nabla^{j}\phi \nonumber \\ 
&& + a_{47}\  \nabla^{i}H^{ajk} \nabla_{j}H_{aik} + a_{48}\  \nabla^{i}\nabla^{j}\phi \nabla_{j}\nabla_{i}\phi\labell{48a}
\eeqa
where $a_1,\cdots, a_{48}$ are 48 arbitrary $p$-independent coefficients that should be fixed by the duality constraint. In writing the above couplings we have used the fact that the first fundamental form $\tG^{\mu\nu}$ for O-plane has non-zero components only for world-volume indices, and tensor  $ \perp^{\mu\nu} $ has non-zero components only for transverse indices. For example, the last term above in terms of 10-dimensional indices is
\beqa
\nabla^{i}\nabla^{j}\phi \nabla_{j}\nabla_{i}\phi&=& \perp^{\mu\nu}\perp^{\rho\sigma}\nabla_\mu\nabla_\rho\phi\nabla_\sigma\nabla_\nu\phi
\eeqa
Similarly for all other terms in \reef{48a}.

Since the above string-frame couplings involve NS-NS fields which transform into each others under T-duality, the T-duality constraint should produce relations between all the 48  coefficients in \reef{48a}.  On the other hand, the S-duality relates the above couplings to the couplings involving R-R fields in which we are not interested in this paper. As we argued in the Introduction section, the S-duality  on the NS-NS couplings  constrains  the Einstein frame couplings involving zero B-field and  odd number of dilaton  to be zero. In the next subsection, we impose the T-duality constraint to the above couplings to find relations between the coefficients. In fact, as we will see, this constraint fixes all coefficients up to an overall factor. It also produces some total derivative terms in the base space which are needed for studying the T-duality of the boundary action.  In the subsequent subsection we impose the S-duality constraint on the resulting coefficients. Since all parameters are already fixed by the T-duality constraint, the S-duality satisfies automatically up to some  total derivative terms that should be included in the study of S-duality of the boundary action. 

\subsection{T-duality constraint in the bulk}

In this subsection we are going to impose the T-duality constraint \reef{TS} on the gauge invariant couplings \reef{48a} to  fix their parameters. To find the reduction $ S_{\rm eff}(\psi)$ we need to dimensionally reduce O$_{(p-1)}$-plane bulk action along the circle orthogonal to the O-plane (transverse reduction), and to find $ S_{\rm eff}(\psi')$ we need to dimensionally reduce O$_p$-plane action along the circle tangent to the O-plane (world-volume reduction) and then transform it under the T-duality. The reduction of the  spacetime fields $G_{\mu\nu}, B_{\mu\nu}, \phi$ and their derivatives which appear in \reef{48a}, are independent of orientation of the  O-plane. However, the reduction of the first fundamental form $\tG^{\mu\nu}$ and the tensor  $\bot^{\mu\nu}$  which also appear in the couplings \reef{48a}, do depend on the orientation of the  O-plane.

When one of the spatial dimensions is circle with coordinate $y$, \ie $M^{(10)}=S^{(1)}\times M^{(9)}$,  the reduction of  metric $ G_{\mu\nu} $ and $ B_{\mu\nu} $ are \cite{Maharana:1992my}
\beqa
G_{\mu\nu}=
\left(\matrix{
g_{\alpha \beta}+e^{\varphi}g_\alpha g_\beta   &  e^{\varphi}g_\alpha &\cr
e^{\varphi}g_\beta  &  e^{\varphi}&}\right)
\qquad;\qquad B_{\mu\nu}=\left(\matrix{
\bar{b}_{\alpha \beta}-\frac{1}{2} g_\alpha b_\beta+\frac{1}{2}g_\beta b_\alpha   &   b_\alpha&\cr
-b_\beta & 0&
}\right)\labell{GB}
\eeqa
Inverse of this metric is
\beqa
G^{\mu\nu}=
\left(\matrix{
g^{\alpha \beta} & -g^\alpha &\cr
-g^\beta & e^{-\varphi}+g_\sigma g^\sigma &
}\right)\labell{IG}
\eeqa
Using these reductions, it is straightforward to calculate the reduction of the spacetime tensors 
 $ R_{\mu\nu\rho\sigma} $, $ \nabla_\mu H_{\nu\rho\sigma} $,  $ H_{\mu\nu\rho} $,  $ \nabla_\mu \phi $, and $ \nabla_\mu\nabla_\nu \phi $ which appear in the couplings \reef{48a}. For example the reduction of $ \nabla_\mu\nabla_\nu \phi $ when both indices are in the 9-dimensional base space is
\beqa
\nabla_{\mu  }\nabla_{\nu  }\phi&=&\frac{1}{2} e^{\varphi } g^{\beta  \alpha  } g_{\nu  } \nabla_{\alpha  }\phi  \nabla_{\beta  }g_{\mu  } + \frac{1}{2} e^{\varphi } g^{\beta  \alpha  } g_{\mu  } \nabla_{\alpha  }\phi  \nabla_{\beta  }g_{\nu  } + \frac{1}{2} e^{\varphi } g^{\beta  \alpha  } g_{\mu  } g_{\nu  } \nabla_{\alpha  }\phi  \nabla_{\beta  }\varphi \nn\\
&& \qquad\qquad  - \frac{1}{2} e^{\varphi } g^{\alpha  \beta  } g_{\nu  } \nabla_{\alpha  }\phi  \nabla_{\mu  }g_{\beta  } - \frac{1}{2} e^{\varphi } g^{\alpha  \beta  } g_{\mu  } \nabla_{\alpha  }\phi  \nabla_{\nu  }g_{\beta  } + \nabla_{\nu  }\nabla_{\mu }\phi
\eeqa
One can find the expression for the  reduction of all other tensors  in \cite{Robbins:2014ara}.

When O$_p$-plane is along the $y$-direction, \ie $M^{(10)}=M^{(p+1)}\times M^{(9-p)}$ and $M^{(p+1)}=S^{(1)}\times M^{(p)}$,  the reduction of pull-back metric  $ \tilde{g}_{ab} $ and its inverse are
\beqa
\tilde{g}_{ab}=
\left(\matrix{
{g}_{\tilde{a} \tilde{b}}+e^{\varphi}{g}_{\tilde{a}}{g}_{\tilde{b}} & e^{\varphi}{g}_{\tilde{a}}&\cr
e^{\varphi}{g}_{\tilde{b}} & e^{\varphi}&
}\right)\qquad;\qquad \tilde{g}^{ab}=
\left(\matrix{
g^{\tilde{a} \tilde{b}} & -g^{\tilde{a}} &\cr
-g^{\tilde{b}} & e^{-\varphi}+g_{\tilde{c}} g^{\tilde{c} }&
}\right)
\eeqa
where the indices $\ta,\tb$ are world-volume indices that do not include the world-volume index $y$, \ie they are belong to $M^{(p)}$. In above equation we  have used the static gauge and assumed the O-plane is at the origin, \ie
\beqa
X^a=\sigma^a&;&  X^i=0 \labell{static}
\eeqa
where the world-volume index $a$ belong to $M^{(p+1)}$ and the transverse index $i$ belong to $M^{(9-p)}$. The reduction of the first fundamental form  $ \tilde{G}^{\mu\nu}=\frac{\partial X^{\mu}}{\partial \sigma^{a}}\frac{\partial X^{\nu}}{\partial \sigma^{b}}\tilde{g}^{ab} $ and $ \perp^{\mu\nu}=G^{\mu\nu}-\tilde{G}^{\mu\nu}  $ in this case have the following non-zero components:
\beqa
\tilde{G}^{ab}=
\left(\matrix{
{g}^{\tilde{a} \tilde{b}} & -{g}^{\tilde{a}}&\cr
-{g}^{\tilde{b}} & e^{-\varphi}+{g}_{\tilde{c}} {g}^{\tilde{c}}&}\right)
 \qquad; \qquad \perp^{ij}=
g^{i j}\labell{ww}
\eeqa 
where we have used the fact that $g^{\ta i}$ and vector $g_i$ are projected out by the orientifold projection.

When O$_{(p-1)}$-plane is orthogonal to  the $y$-direction, \ie  $M^{(10)}=M^{(p)}\times M^{(10-p)}$ and $M^{(10-p)}=S^{(1)}\times M^{(9-p)}$,  $g^{\ta i}$ and the vector $g_{\ta}$ are projected out by the orientifold projection. Then the reduction of the pull-back metric becomes  $ \tilde{g}_{ab}={g}_{\tilde{a} \tilde{b}} $, and the non-zero components of the first fundamental form and $\perp^{\mu\nu}$ are
\beqa
\tilde{G}^{\ta\tb}={g}^{\ta\tb}
\qquad ;\qquad \perp^{\tilde{i} \tilde{j}}=\left(\matrix{
g^{{i} {j}} & -g^{{i}}&\cr
-g^{{j}} & e^{-\varphi}+g_{{k}}g^{{k}}&}\right)\labell{tr}
\eeqa 
where the indices $\tilde{i},\tilde{j}$ are transverse indices that  include the transverse index $y$, \ie they are belong to $M^{(10-p)}$. Note that the determinate of the pull-back metric is gauge invariant in both cases, \ie when O$_p$-plane is along the $y$-direction it is $\sqrt{-\tg}=e^{\varphi/2}\sqrt{-g}$, and when  O$_{(p-1)}$-plane is orthogonal to  the $y$-direction it is  $\sqrt{-\tg}=\sqrt{-g}$.

Using the above reductions, one can calculate reduction of each gauge invariant coupling in \reef{48a} when O-plane is along or orthogonal to the circle. Since the 10-dimensional couplings are gauge invariant, one expects the dimensional reduction of the couplings to be  gauge invariant under various 9-dimensional gauge transformations. In particular they should be invariant under  the $U(1)\times U(1)$ gauge transformations corresponding to the two vectors $g_{\mu}, b_{\mu}$.   This observation has been used  in \cite{Garousi:2019mca} to simplify  greatly the complexity of the calculations at six derivatives order. Using this trick,  one should keep the terms in the reductions of various tensors which are invariant under the $U(1)\times U(1)$ gauge transformations.  The gauge invariant terms in the reduction of the spacetime tensors 
 $ R_{\mu\nu\rho\sigma} $, $ \nabla_\mu H_{\nu\rho\sigma} $,  $ H_{\mu\nu\rho} $,  $ \nabla_\mu \phi $, and $ \nabla_\mu\nabla_\nu \phi $ are
\beqa
\nabla_{\mu }\nabla_{\nu }\phi &=&\nabla_{\mu }\nabla_{\nu }\phi \nn\\
\nabla_{\mu }\nabla_{y }\phi& =& \frac{1}{2} e^{\varphi} V^{\alpha}{}_  \mu    \nabla_{\alpha }\phi\nn\\
 \nabla_{y}\nabla_{y }\phi &=&\frac{1}{2} e^{\varphi} \nabla^{\alpha }\phi \nabla_{\alpha }\varphi\nn\\
 \nabla_{\mu }\phi &=& \nabla_{\mu }\phi\nn\\
 \nabla_{y}\phi &=&0\nn\\
 \nabla_{y }H_{\nu  \rho  y}&=&- \frac{1}{2} e^{\varphi} V_{\rho }{}^{\alpha } W_{\nu  \alpha } + \frac{1}{2} e^{\varphi} V_{\nu }{}^{\alpha } W_{\rho  \alpha }  + \frac{1}{2} e^{\varphi} \bar{H}_{\nu  \rho  \alpha } \nabla^{\alpha }\varphi\nn\\
  \nabla_{y }H_{\nu  \rho  \sigma } & =&  \frac{1}{2} \Big(e^{\varphi} V_{\rho }{}^{\alpha } \bar{H}_{\nu  \sigma  \alpha }  -  e^{\varphi} V_{\sigma }{}^{\alpha } \bar{H}_{\nu  \rho  \alpha }-  e^{\varphi} V_{\nu }{}^{\alpha } \bar{H}_{\rho  \sigma  \alpha } -  W_{\rho  \sigma } \nabla_{\nu }\varphi +  W_{\nu  \sigma } \nabla_{\rho }\varphi -   W_{\nu  \rho } \nabla_{\sigma }\varphi\Big)\nn\\
  \nabla_{\mu }H_{\nu  \rho  y } & =& - \frac{1}{2} e^{\varphi} V_{\mu }{}^{\alpha } \bar{H}_{\nu  \rho  \alpha } + \nabla_{\mu }W_{\nu  \rho } -  \frac{1}{2} W_{\nu  \rho } \nabla_{\mu }\varphi\nn\\
  \nabla_{\mu }H_{\nu  \rho  \sigma }& =& \frac{1}{2} V_{\mu  \sigma } W_{\nu  \rho } -  \frac{1}{2} V_{\mu  \rho } W_{\nu  \sigma } + \frac{1}{2} V_{\mu  \nu } W_{\rho  \sigma }  + \nabla_{\mu }\bar{H}_{\nu  \rho  \sigma }\nn\\
 H_{\nu  \rho  \sigma } &= &\bar{H}_{\nu  \rho  \sigma }\nn\\
 H_{\mu  \nu  \mathbf{y} }& = &W_{\mu  \nu }\nn\\
 R_{\mu  \nu  \rho  \sigma }&=& \hat{R}_{\mu  \nu  \rho  \sigma }+ \frac{1}{4} e^{\varphi} V_{\mu  \sigma } V_{\nu  \rho } -  \frac{1}{4} e^{\varphi} V_{\mu  \rho } V_{\nu  \sigma } -  \frac{1}{2} e^{\varphi} V_{\mu  \nu } V_{\rho  \sigma } \nn \\
 R_{\mu  \nu  \rho  y }& = &\frac{1}{4} e^{\varphi} V_{\nu  \rho } \nabla_{\mu }\varphi -  \frac{1}{4} e^{\varphi} V_{\mu  \rho } \nabla_{\nu }\varphi -  \frac{1}{2} e^{\varphi} \nabla_{\rho }V_{\mu  \nu } -  \frac{1}{2} e^{\varphi} V_{\mu  \nu } \nabla_{\rho }\varphi\nn\\
  R_{\mu  y \nu  y}&=& \frac{1}{4} e^{2 \varphi} V_{\mu }{}^{\rho } V_{\nu  \rho }  -  \frac{1}{4} e^{\varphi} \nabla_{\mu }\varphi \nabla_{\nu }\varphi -  \frac{1}{2} e^{\varphi} \nabla_{\nu }\nabla_{\mu }\varphi\labell{red}
\eeqa
where $ \bar{H}_{\mu\nu\rho}\equiv 3\partial_{[\mu}\bar{b}_{\nu\rho]}-\frac{3}{2}g_{[\mu}W_{\nu\rho]}-\frac{3}{2}b_{[\mu}V_{\nu\rho]} $,  $ V_{\mu\nu}=\partial_{\mu}g_{\nu}-\partial_{\nu}g_{\mu} $ and $ W_{\mu\nu}=\partial_{\mu}b_{\nu}-\partial_{\nu}b_{\mu} $ are 9-dimensional field strengths. The base space field strengths $ \bar{H}$, $V$ and $W$  satisfy the following  Bianchi identities:
\beqa
d\bar{H}&=&-\frac{3}{2}V\wedge W\nn\\
dV&=&0\nn\\
dW&=&0\labell{Hff}
\eeqa
Note that the field strengths  $\bar{H}, W$ have odd parity.
Similarly,  the  gauge invariant components of the tensors in \reef{ww} are
\beqa
\tilde{G}^{ab}=
\left(\matrix{
{g}^{\tilde{a} \tilde{b}} & 0&\cr
0 & e^{-\varphi}&}\!\!\!\!\!\!\right)
 \qquad; \qquad \perp^{ij}=
g^{i j}\labell{GB1}
\eeqa 
and the  gauge invariant components of the tensors in \reef{tr} are
\beqa
\tilde{G}^{\ta\tb}={g}^{\ta\tb}
\qquad ;\qquad \perp^{\tilde{i} \tilde{j}}=\left(\matrix{
g^{ij} &0 &\cr
0 & e^{-\varphi}&}\!\!\!\!\!\!\right)\labell{GB2}
\eeqa 
 Note  that the $yy$ component of the metric needed for producing scalars from the tensors on the right-hand side of reductions \reef{red} is $e^{-\varphi}$. Then one can easily observe that in the couplings in the base space the tensors $V, W$ always appear as $e^{\varphi/2}V, e^{-\varphi/2}W$.

Using the above gauge invariant parts of the reductions, one can calculate reduction of various terms in \reef{48a} along or orthogonal to the circle. For example, the reduction of last term in \reef{48a} when O$_p$-plane is along the circle is 
\beqa
 \perp^{\mu\nu}\perp^{\rho\sigma}\nabla_\mu\nabla_\rho\phi\nabla_\sigma\nabla_\nu\phi&=&\nabla^i\nabla^j\phi\nabla_j\nabla_i\phi
 \eeqa
 The reduction of this term when $O_{(p-1)}$-plane is orthogonal to the circle is 
\beqa
 \perp^{\mu\nu}\perp^{\rho\sigma}\nabla_\mu\nabla_\rho\phi\nabla_\sigma\nabla_\nu\phi&=&\nabla^i\nabla^j\phi\nabla_j\nabla_i\phi+\frac{1}{4}(\nabla^\alpha\phi\nabla_\alpha\varphi)^2
 \eeqa 
Similarly one can calculate reduction of all other 10-dimensional covariant terms in \reef{48a}. It is important to note that if one keeps all gauge invariant and non-gauge invariant terms in the reduction of tensors, one  would find the same result for the reduction of 10-dimension covariant couplings. 

After reduction, one has to impose the orientifold projection which means the  O-plane couplings in the base space do not have  couplings with odd number of transverse indices on metric, $\nabla\phi$, $\nabla\varphi$,  and their corresponding derivatives, and do not have couplings with   even number of transverse indices on $\bar{H}$ and its derivatives. When O-plane is along (orthogonal) the  circle, the reduction of O-plane couplings do not have odd number of transverse indices on $V$ ($W$) and its derivatives, and do not have couplings with even number of transverse indices on $W$ ($V$) and its derivatives. 

After applying the above O-plane conditions, one has to also impose T-duality transformations which are the Buscher rules \cite{Buscher:1987sk,Buscher:1987qj}.   The base space metric and $\bar{H}$ are invariant under T-duality and the other fields transform as
\beqa
\phi\to \phi -\frac{1}{2}\varphi \ \ , \qquad  \varphi \to -\varphi\ \ ,   \qquad V_{\mu\nu}\longleftrightarrow W_{\mu\nu}.\labell{Buscher}
\eeqa
Using the above transformations, then one can calculate the left-hand side of T-duality constraint \reef{TS}. Note that the  overall factor of $e^{-\phi}\sqrt{-\tg}$ in  O$_p$-plane action \reef{Gen} transforms under the  T-duality in the world-volume reduction, to $e^{-\phi}\sqrt{-g}$ which is the same as the transverse reduction of the corresponding term in  O$_{(p-1)}$-plane. So the T-duality constraint \reef{TS} is only on the couplings in the Lagrangian \reef{48a}.

To construct the total derivative terms in the constraint \reef{TS}, we use the observation made in \cite{Garousi:2019mca} that the T-duality constraint \reef{TS} for flat base space and for curved base space produces identical constraint on the coefficients of the gauge invariant couplings in the bosonic effective action at orders $\alpha',\alpha'^2$.  This is as expected because the effective action should be independent of the details of the geometry of the base space. Hence, we continue  our calculations on the T-duality constraint \reef{TS} for flat base space. To construct the most general total derivative terms for the right-hand side of \reef{TS}, we consider all even-parity  contractions of the base space tensors $\tG,\bot,\prt\phi$, $\prt\prt\phi$, $\prt\prt\prt\phi$, $\prt\varphi$, $\prt\prt\varphi$, $\prt\prt\prt\varphi$, $\bar{H}$, $\prt\bar{H}$, $e^{\varphi/2}V$, $e^{\varphi/2}\prt V$, $e^{-\varphi/2}W$ and $e^{-\varphi/2}\prt W$ at three derivatives with one free world-volume index.  Then we remove the terms which are projected out by the orientifold projection. We call the remaining terms, with arbitrary coefficients,  the vector ${\cal I}_\ta$.  The most general gauge invariant even-parity total derivative terms in the base space is  then
\beqa
TD&=&-\frac{T_{p-1}\pi^2\alpha'^2}{48}\int d^{p}\sigma\,\sqrt{-g}\,g^{\ta\tb}\prt_\ta ( e^{-\phi} {\cal I}_\tb)\labell{j}
\eeqa
where $ {\cal I}_\ta$ is 
\beqa
{\cal I}_\td&=&u_{4}\  V^{\ta i} \bar{H}_{\td\tb i} W_{\ta}{}^{\tb } + u_{1}\  V_{\td}{}^{i} \bar{H}_{\ta\tb i} W^{\ta\tb} + u_{2}\  V^{\ta i} \bar{H}_{\ta\tb i} W_{\td}{}^{\tb } + u_{3}\  V^{\ta i} \bar{H}_{daj} W_{i}{}^{j} + u_{5}\  V_{\td}{}^{i} \bar{H}_{ijk} W^{jk}\nn\\
&& + u_{6}\  e^{\varphi} V_{\td}{}^{i} \partial_{\ta}V^{\ta}{}_{i}  + u_{7}\  e^{\varphi} V^{\ta i} \partial_{\ta}V_{di} + u_{8}\  \partial_{\ta}\partial^{\ta}\partial_{\td}\varphi + u_{9}\  \partial_{\ta}\partial^{\ta}\partial_{\td}\phi + u_{10}\  e^{\varphi} V_{\ta i} V_{\td}{}^{i} \partial^{\ta}\varphi \nonumber \\ 
&& + u_{12}\  \bar{H}_{\ta\tb i} \bar{H}_{\td}{}^{bi} \partial^{\ta}\varphi  + u_{11}\  e^{-\varphi} W_{\ta\tb} W_{\td}{}^{\tb } \partial^{\ta}\varphi + u_{13}\  e^{\varphi} V_{\ta i} V_{\td}{}^{i} \partial^{\ta}\phi + u_{15}\  \bar{H}_{\ta\tb i} \bar{H}_{\td}{}^{bi} \partial^{\ta}\phi \nonumber \\ 
&& + u_{14}\  e^{-\varphi} W_{\ta\tb} W_{\td}{}^{\tb } \partial^{\ta}\phi + u_{18}\  \bar{H}_{\td}{}^{\ta i} \partial_{\tb }\bar{H}_{\ta}{}^{\tb }{}_{i} + u_{19}\  \bar{H}^{\ta\tb i} \partial_{\tb }\bar{H}_{dai} + u_{16}\  e^{-\varphi} W_{\td}{}^{\ta} \partial_{\tb }W_{\ta}{}^{\tb }\nonumber \\ 
&&  + u_{17}\  e^{-\varphi} W^{\ta\tb} \partial_{\tb }W_{\td\ta} + u_{20}\  e^{\varphi} V^{\ta i} \partial_{\td}V_{\ta i}+ u_{23}\  \bar{H}^{\ta\tb i} \partial_{\td}\bar{H}_{\ta\tb i} + u_{24}\  \bar{H}^{ijk} \partial_{\td}\bar{H}_{ijk}\nonumber \\ 
&&  + u_{21}\  e^{-\varphi} W^{\ta\tb} \partial_{\td}W_{\ta\tb} + u_{22}\  e^{-\varphi} W^{ij} \partial_{\td}W_{ij} + u_{25}\  e^{\varphi} V_{\ta i} V^{\ta i} \partial_{\td}\varphi  + u_{28}\  \bar{H}_{\ta\tb i} \bar{H}^{\ta\tb i} \partial_{\td}\varphi \nonumber \\ 
&& + u_{29}\  \bar{H}_{ijk} \bar{H}^{ijk} \partial_{\td}\varphi + u_{26}\  e^{-\varphi} W_{\ta\tb} W^{\ta\tb} \partial_{\td}\varphi + u_{27}\  e^{-\varphi} W_{ij} W^{ij} \partial_{\td}\varphi + u_{30}\  \partial_{\ta}\partial^{\ta}\varphi \partial_{\td}\varphi \nonumber \\ 
&& + u_{31}\  \partial_{\ta}\partial^{\ta}\phi \partial_{\td}\varphi + u_{32}\  \partial_{\ta}\varphi \partial^{\ta}\varphi \partial_{\td}\varphi + u_{33}\  \partial_{\ta}\phi \partial^{\ta}\varphi \partial_{\td}\varphi + u_{34}\  \partial_{\ta}\phi \partial^{\ta}\phi \partial_{\td}\varphi \nonumber \\ 
&& + u_{35}\  e^{\varphi} V_{\ta i} V^{\ta i} \partial_{\td}\phi + u_{38}\  \bar{H}_{\ta\tb i} \bar{H}^{\ta\tb i} \partial_{\td}\phi + u_{39}\  \bar{H}_{ijk} \bar{H}^{ijk} \partial_{\td}\phi + u_{36}\  e^{-\varphi} W_{\ta\tb} W^{\ta\tb} \partial_{\td}\phi \nonumber \\ 
&& + u_{37}\  e^{-\varphi} W_{ij} W^{ij} \partial_{\td}\phi + u_{40}\  \partial_{\ta}\partial^{\ta}\varphi \partial_{\td}\phi + u_{41}\  \partial_{\ta}\partial^{\ta}\phi \partial_{\td}\phi + u_{42}\  \partial_{\ta}\varphi \partial^{\ta}\varphi \partial_{\td}\phi + u_{43}\  \partial_{\ta}\phi \partial^{\ta}\varphi \partial_{\td}\phi \nonumber \\ 
&& + u_{44}\  \partial_{\ta}\phi \partial^{\ta}\phi \partial_{\td}\phi + u_{45}\  \partial^{\ta}\varphi \partial_{\td}\partial_{\ta}\varphi + u_{46}\  \partial^{\ta}\phi \partial_{\td}\partial_{\ta}\varphi + u_{47}\  \partial^{\ta}\varphi \partial_{\td}\partial_{\ta}\phi + u_{48}\  \partial^{\ta}\phi \partial_{\td}\partial_{\ta}\phi \nonumber \\ 
&& + u_{49}\  \partial_{\td}\partial_{\ta}\partial^{\ta}\varphi + u_{50}\  \partial_{\td}\partial_{\ta}\partial^{\ta}\phi + u_{51}\  e^{\varphi} V^{\ta i} \partial_{i}V_{\td\ta} + u_{53}\  \bar{H}^{\ta\tb i} \partial_{i}\bar{H}_{\td\ta\tb} + u_{52}\  e^{-\varphi} W_{\td}{}^{\ta} \partial_{i}W_{\ta}{}^{i} \nonumber \\ 
&& + u_{54}\  \partial_{\td}\varphi \partial_{i}\partial^{i}\varphi + u_{55}\  \partial_{\td}\phi \partial_{i}\partial^{i}\varphi + u_{56}\  \partial_{\td}\varphi \partial_{i}\partial^{i}\phi + u_{57}\  \partial_{\td}\phi \partial_{i}\partial^{i}\phi + u_{58}\  \partial_{i}\partial^{i}\partial_{\td}\varphi \nonumber \\ 
&& + u_{59}\  \partial_{i}\partial^{i}\partial_{\td}\phi + u_{60}\  e^{\varphi} V_{\td}{}^{i} \partial_{j}V_{i}{}^{j} + u_{62}\  \bar{H}_{\td}{}^{\ta i} \partial_{j}\bar{H}_{\ta i}{}^{j} + u_{61}\  e^{-\varphi} W^{ij} \partial_{j}W_{\td i} + u_{63}\  \bar{H}^{ijk} \partial_{k}\bar{H}_{\td ij} \nn
\eeqa
The parameters  $u_1,\cdots, u_{63}$ are yet some  arbitrary coefficients. Note that the base space is $M^{(9)}=M^{(p)}\times M^{(9-p)}$,  the indices $\ta,\tb$ belong to $M^{(p)}$ and the indices $i,j$ belong to $M^{(9-p)}$. Replacing the above total derivative terms to the right-hand  side of the T-duality constraint \reef{TS}, one should then write the couplings in the form of independent structures by imposing the Bianchi identities \reef{Hff} in the base space. Here again we write the field strengths $\bar{H}, V,W$ in terms of potentials $\bar{b}_{\mu\nu},g_\mu,b_\mu$ to satisfy the Bianchi identities automatically.

Writing the couplings in the T-duality constraint  \reef{TS} in terms of independent  and non-gauge invariant structures, then one makes the coefficients of the independent structures which include the parameters of the gauge invariant Lagrangian \reef{48a} and the above total derivative terms, to be zero. These linear equations produce the following 47 relations between  the 48 parameters of the Lagrangian \reef{48a}:
\beqa
&&a_{23}\ \to 0, a_{22}\ \to 0, a_{19}\ \to 12 a_{28}, a_{18}\ \to -6 a_{28},  a_{14}\ \to -6 a_{28}, a_{11}\ \to 6 a_{28},\nn\\
&& a_{44}\ \to 9 a_{28}, a_{4}\ \to 0,  a_{3}\ \to -\frac{3}{2} a_{28}, a_{2}\ \to 0, a_{1}\ \to -\frac{3}{4} a_{28},  a_{24}\ \to 9 a_{28}, a_{20}\ \to 0,\nn\\
&& a_{5}\ \to a_{28}, a_{21}\ \to -12 a_{28}, a_{16}\ \to 0,  a_{12}\ \to -6 a_{28}, a_{15}\ \to 6 a_{28}, a_{13}\ \to 0, a_{7}\ \to 0, \nn\\
&& a_{6}\ \to \frac{3}{2} a_{28}, a_{17}\ \to -3 a_{28}, a_{26}\ \to 0, a_{25}\ \to 0,  a_{10}\ \to 0, a_{9}\ \to 0, a_{8}\ \to -\frac{1}{4} a_{28},\nn\\
&& a_{42}\ \to 0,  a_{27}\ \to -3 a_{28}, a_{37}\ \to 0, a_{47}\ \to 0, a_{46}\ \to -24 a_{28}, a_{40}\ \to 0,  a_{41}\ \to 0,\nn\\
&& a_{36}\ \to 12 a_{28}, a_{35}\ \to 24 a_{28}, a_{34}\ \to -6 a_{28},  a_{43}\ \to 2 a_{28}, a_{48}\ \to -12 a_{28}, a_{45}\ \to -3 a_{28}, \nn\\
&&a_{29}\ \to 0,  a_{39}\ \to 0, a_{38}\ \to 0, a_{30}\ \to 0, a_{33}\ \to 0, a_{31}\ \to 0, a_{32}\ \to 0\labell{aaa}
\eeqa
Which fix the bulk effective action up to one overall factor $a_{28}$, \ie
\beqa
{\cal L}_p&\!\!\!\!\!\!\!\!\!\!\!\!\!=\!\!\!\!\!\!\!\!\!\!\!\!\!\!\!&a_{28}\Big[- \frac{3}{4} H_{a}{}^{cj} H^{abi} H_{b}{}^{d}{}_{j} H_{cdi} -  \frac{3}{2} H_{ab}{}^{j} H^{abi} H_{cdj} H^{cd}{}_{i} + H_{a}{}^{cj} H^{abi} H_{bc}{}^{k} H_{ijk} \nonumber \\ 
&& + \frac{3}{2} H_{ab}{}^{j} H^{abi} H_{i}{}^{kl} H_{jkl} -  \frac{1}{4} H_{i}{}^{lm} H^{ijk} H_{jl}{}^{n} H_{kmn}  + 6 H^{abi} H^{cd}{}_{i} R_{abcd} \nonumber \\ 
&&- 6 H^{abi} H_{i}{}^{jk} R_{abjk} - 6 R_{abcd} R^{abcd} + 6 R_{abij} R^{abij} - 6 H_{a}{}^{ci} H_{bci} \mathcal{R}^{ab} \nonumber \\ 
&& + 12 \mathcal{R}_{ab} \mathcal{R}^{ab} + 9 H_{abj} H^{ab}{}_{i} \mathcal{R}^{ij} - 3 H_{i}{}^{kl} H_{jkl} \mathcal{R}^{ij} - 12 \mathcal{R}_{ij} \mathcal{R}^{ij} \nonumber \\ 
&& + \nabla_{a}H_{ijk} \nabla^{a}H^{ijk} - 3 \nabla_{c}H_{abi} \nabla^{c}H^{abi} + 2 \nabla_{i}H_{abc} \nabla^{i}H^{abc} \Big]\labell{bulkf}
\eeqa
where $ \mathcal{R}_{\mu\nu}=\tG^{\rho\sigma}{R}_{\rho\mu\sigma\nu}+\nabla_{\mu}\nabla_{\nu} \phi $. This is exactly the action that has been found in \cite{Garousi:2014oya}. For $a_{28}=-\frac{1}{6}$, it is consistent with S-matrix element of two vertex operators \cite{Bachas:1999um}.  In finding the above action we have imposed the equations of motion in finding the independent gauge invariant couplings in \reef{48a}. If one does not impose the equations of motion to find the independent couplings, then the T-duality would produce the above T-duality invariant multiple  with coefficient $a_{28}$ and 10 other T-duality invariant multiples which includes terms like  $\nabla_i\nabla^i\Phi$, or $R_{i\mu}{}^i{}_\nu$.

The linear equations also fix the following relations between the parameters of the total derivative terms and $a_{28}$:
\beqa
&& u_{10}\ \to 6 a_{28},\ u_{13}\ \to 0,\ u_{14}\ \to -u_{11} ,\ u_{16}\ \to  6 a_{28},\ u_{17}\ \to 0,\ u_{18}\ \to -u_{12},\ u_{2}\ \to 0,\ \nn\\
&&u_{20}\ \to  6 a_{28},\ u_{21}\ \to 0,\ u_{22}\ \to 18 a_{28} + 2 u_{1},\ u_{24}\ \to  6 a_{28},\ u_{27}\ \to  9 a_{28} + u_{1},\ u_{29}\ \to \frac{1}{2}u_{23},\ \nn\\
&&u_{3}\ \to -12 a_{28} - 2 u_{1},\ u_{32}\ \to  3 a_{28} + \frac{1}{2}u_{25},\ u_{34}\ \to  6 a_{28},\ u_{35}\ \to 0,\ u_{36}\ \to 0,\ u_{37}\ \to 0,\ \nn\\
&&u_{38}\ \to 0,\ u_{39}\ \to 0,\ u_{4} \to\  0,\ u_{40}\ \to -\frac{15}{2} a_{28},\ u_{41}\ \to \frac{3}{2} a_{28},\ u_{42}\ \to -6 a_{28},\ u_{44}\ \to -3 a_{28},\ \nn\\
&& u_{45}\ \to  6 a_{28},\ u_{46}\ \to -u_{43},\ u_{47}\ \to 0,\ u_{48}\ \to 0,\ u_{49}\ \to 0,\ u_{5}\ \to  3 a_{28} +   3 u_{28},\ u_{50}\ \to 0,\ \nn\\
&&u_{51}\ \to 0,\ u_{52}\ \to 0,\ u_{53}\ \to 0,\ u_{54}\ \to 0,\ u_{55}\ \to -u_{43},\ u_{56}\ \to 0,\ u_{57}\ \to 0,\ u_{58}\ \to u_{43},\ \nn\\
&&u_{59}\ \to 0,\ u_{6}\ \to -6 a_{28},\ u_{60}\ \to  6 a_{28},\ u_{61}\ \to u_{43},\ u_{62}\ \to -u_{43},\ u_{63}\ \to 0,\ u_{64}\ \to -u_{11},\ \nn\\
&&u_{65}\ \to -u_{12},\ u_{66}\ \to u_{15},\ u_{67}\ \to u_{19},\ u_{68}\ \to u_{26},\ u_{69}\ \to -9 a_{28} - u_{1},\ u_{7}\ \to  6 a_{28} - u_{26},\ \nn\\
&&u_{70}\ \to u_{30},\ u_{71}\ \to 0,\ u_{72}\ \to 0,\ u_{73}\ \to 0,\ u_{74}\ \to 0,\ u_{75}\ \to 0,\ u_{76}\ \to 0,\ u_{77}\ \to 0,\ \nn\\
&&u_{78}\ \to 2 u_{31},\ u_{79}\ \to 2 u_{33},\ u_{8}\ \to -u_{30},\ u_{80}\ \to -u_{15},\ u_{81}\ \to -u_{19},\ u_{82}\ \to -3 u_{28},\ \nn\\
&&u_{9}\ \to -12 a_{28}
\eeqa
The parameters  which are not fix in terms of $a_{28}$ are cancelled when one replaces the  above relations on the right-hand side of \reef{j} and imposed the Bianchi identities. So the unfixed parameters represent the redundancy of the couplings in \reef{j}. Hence one can set them to zero. The fixed parameters then produce the following vector 

\beqa
{\cal I}_\td&=&a_{28}\Big[ 2 V^{\tilde{a}i} W_{\tilde{d}}{}^{\tilde{b}} \bar{H}_{\tilde{a}\tilde{b}i} -  \frac{1}{2} V_{\tilde{d}}{}^{i} W^{jk} \bar{H}_{ijk} + e^{\varphi} V_{\tilde{d}}{}^{i} \partial_{\tilde{a}}V^{\tilde{a}}{}_{i} -  e^{\varphi} V^{\tilde{a}i} \partial_{\tilde{a}}V_{\tilde{d}i} + 2 e^{\varphi} V_{\tilde{a}i} V_{\tilde{d}}{}^{i} \partial^{\tilde{a}}\varphi \nonumber \\ 
&& -  e^{-\varphi} W_{\tilde{a}\tilde{b}} W_{\tilde{d}}{}^{\tilde{b}} \partial^{\tilde{a}}\varphi -  e^{\varphi} V_{\tilde{a}i} V_{\tilde{d}}{}^{i} \partial^{\tilde{a}}\phi -  e^{-\varphi} W_{\tilde{a}\tilde{b}} W_{\tilde{d}}{}^{\tilde{b}} \partial^{\tilde{a}}\phi -  e^{-\varphi} W_{\tilde{d}}{}^{\tilde{a}} \partial_{\tilde{b}}W_{\tilde{a}}{}^{\tilde{b}} - 3 \bar{H}^{\tilde{a}\tilde{b}i} \partial_{\tilde{b}}\bar{H}_{\tilde{d}\tilde{a}i}\nonumber \\ 
&&  -  \frac{1}{2} e^{-\varphi} W^{\tilde{a}\tilde{b}} \partial_{\tilde{d}}W_{\tilde{a}\tilde{b}} -  \frac{3}{2} \bar{H}^{\tilde{a}\tilde{b}i} \partial_{\tilde{d}}\bar{H}_{\tilde{a}\tilde{b}i} -  e^{\varphi} V_{\tilde{a}i} V^{\tilde{a}i} \partial_{\tilde{d}}\varphi  + \frac{5}{4} e^{-\varphi} W_{\tilde{a}\tilde{b}} W^{\tilde{a}\tilde{b}} \partial_{\tilde{d}}\varphi -  \frac{1}{4} e^{-\varphi} W_{ij} W^{ij} \partial_{\tilde{d}}\varphi \nonumber \\ 
&&+ \partial_{\tilde{a}}\partial^{\tilde{a}}\varphi \partial_{\tilde{d}}\varphi + \frac{1}{2} \partial_{\tilde{a}}\varphi \partial^{\tilde{a}}\varphi \partial_{\tilde{d}}\varphi -  \partial_{\tilde{a}}\phi \partial^{\tilde{a}}\varphi \partial_{\tilde{d}}\varphi -  \partial^{\tilde{a}}\varphi \partial_{\tilde{d}}\partial_{\tilde{a}}\varphi + \frac{3}{2} \bar{H}^{\tilde{a}\tilde{b}i} \partial_{i}\bar{H}_{\tilde{d}\tilde{a}\tilde{b}}\Big]
\eeqa
Note that only for simplicity we have assumed the base space is flat. If it is not flat, then the partial derivatives in  above equation  would be covariant derivative. In fact we have performed the calculations for curved base space and find the same Lagrangian \reef{bulkf} and the same total derivative as above in which the partial derivatives are replaced by covariant derivatives.

Now we assume the subspace $M^{(p)}$ in the base space $M^{(9)}=M^{(p)}\times M^{(9-p)}$ has boundary $\prt M^{(p)}$, \ie $\prt M^{(10)}=S^{(1)}\times \prt M^{(9)}$ and $\prt M^{(9)}=\prt M^{(p)}\times M^{(9-p)}$. The Stokes's theorem in this subspace is (see Appendix)
 \beqa
\int_{M^{(p)}} d^{p}\sigma\sqrt{-g}g^{\ta\tb}\prt_\ta ( e^{-\phi}{\cal I}_\tb)&=&\int_{\prt {M^{(p)}}}d^{p-1}\tau\,e^{-\phi}\sqrt{|\bg|}\,g^{\ta\tb}n_{\ta}  {\cal I}_{\tb}\labell{Stokes}
\eeqa
where   $n^\ta$ is the normal vector to the boundary $\prt M^{(p)}$ which  is outward-pointing (inward-pointing) if the boundary is spacelike (timelike),  and the boundary in the static gauge is specified by the  functions  $\sigma^\ta=\sigma^\ta(\tau^{\ba})$.  In the square root on the right-hand side $\bg$ is determinant of    the induced metric, \ie
\beqa
\bg_{\ba\bb}&=&\frac{\prt \sigma^{\ta}}{\prt\tau^{\ba}}\frac{\prt \sigma^{\tb}}{ \prt\tau^{\bb}}  g_{\ta\tb}\labell{bg}
\eeqa
 The coordinates of the boundary $\prt M^{(p)}$ are $\tau^0,\tau^1,\cdots, \tau^{p-2}$.
Using the above  Stokes's theorem, one finds that the contribution of the total derivative terms in the boundary  is
\beqa
\prt TD&=&-\frac{T_{p-1}\pi^2\alpha'^2 a_{28}}{48}\int_{\prt M^{(p)}}d^{p-1}\tau\,e^{-\phi}\sqrt{|\bg|}\,n^{\td}\Big[  2 V^{\tilde{a}i} W_{\tilde{d}}{}^{\tilde{b}} \bar{H}_{\tilde{a}\tilde{b}i} -  \frac{1}{2} V_{\tilde{d}}{}^{i} W^{jk} \bar{H}_{ijk} + e^{\varphi} V_{\tilde{d}}{}^{i} \partial_{\tilde{a}}V^{\tilde{a}}{}_{i}\nonumber \\ 
&& -  e^{\varphi} V^{\tilde{a}i} \partial_{\tilde{a}}V_{\tilde{d}i} + 2 e^{\varphi} V_{\tilde{a}i} V_{\tilde{d}}{}^{i} \partial^{\tilde{a}}\varphi -  e^{-\varphi} W_{\tilde{a}\tilde{b}} W_{\tilde{d}}{}^{\tilde{b}} \partial^{\tilde{a}}\varphi -  e^{\varphi} V_{\tilde{a}i} V_{\tilde{d}}{}^{i} \partial^{\tilde{a}}\phi -  e^{-\varphi} W_{\tilde{a}\tilde{b}} W_{\tilde{d}}{}^{\tilde{b}} \partial^{\tilde{a}}\phi \nonumber \\ 
&&-  e^{-\varphi} W_{\tilde{d}}{}^{\tilde{a}} \partial_{\tilde{b}}W_{\tilde{a}}{}^{\tilde{b}} - 3 \bar{H}^{\tilde{a}\tilde{b}i} \partial_{\tilde{b}}\bar{H}_{\tilde{d}\tilde{a}i}  -  \frac{1}{2} e^{-\varphi} W^{\tilde{a}\tilde{b}} \partial_{\tilde{d}}W_{\tilde{a}\tilde{b}} -  \frac{3}{2} \bar{H}^{\tilde{a}\tilde{b}i} \partial_{\tilde{d}}\bar{H}_{\tilde{a}\tilde{b}i} -  e^{\varphi} V_{\tilde{a}i} V^{\tilde{a}i} \partial_{\tilde{d}}\varphi  \nonumber \\ 
&&+ \frac{5}{4} e^{-\varphi} W_{\tilde{a}\tilde{b}} W^{\tilde{a}\tilde{b}} \partial_{\tilde{d}}\varphi -  \frac{1}{4} e^{-\varphi} W_{ij} W^{ij} \partial_{\tilde{d}}\varphi + \partial_{\tilde{a}}\partial^{\tilde{a}}\varphi \partial_{\tilde{d}}\varphi + \frac{1}{2} \partial_{\tilde{a}}\varphi \partial^{\tilde{a}}\varphi \partial_{\tilde{d}}\varphi -  \partial_{\tilde{a}}\phi \partial^{\tilde{a}}\varphi \partial_{\tilde{d}}\varphi \nonumber \\ 
&&-  \partial^{\tilde{a}}\varphi \partial_{\tilde{d}}\partial_{\tilde{a}}\varphi + \frac{3}{2} \bar{H}^{\tilde{a}\tilde{b}i} \partial_{i}\bar{H}_{\tilde{d}\tilde{a}\tilde{b}} \Big]\labell{PTD}
\eeqa
In each term the tensors $n_{\alpha}, V_{\alpha\beta}, W_{\alpha\beta}, \cdots$ in the base space $M^{(9)}$ are contracted with projections $\tilde{G}^{\alpha\beta}=\frac{\prt X^{\alpha}}{\prt \sigma^{\ta}}\frac{\prt X^{\beta}}{\prt \sigma^{\tb}}g^{\ta\tb}$ and $\bot^{\alpha\beta}=G^{\alpha\beta}-\tilde{G}^{\alpha\beta}$ at the boundary. Note that in the static gauge $X^{\ta}=\sigma^{\ta}$ and $X^i=0$. The above boundary terms  are zero if the subspace $M^{(p)}$  has no boundary. However, if it has boundary $\prt M^{(p)}$, then the above terms should be included in the T-duality constraint of the boundary action to have full T-duality in the bulk and boundary. We will consider the above terms in the T-duality of the boundary action in section 3.

\subsection{S-duality constraint in the bulk}

We have seen that the coefficients of the gauge invariant couplings in \reef{48a} are all fixed up to an overall factor by imposing the T-duality constraint. Hence the resulting couplings should be consistent with S-duality for the case of O$_3$-plane up to some total derivative terms. To fully have an S-duality invariant action for O$_3$-plane one should include appropriate R-R couplings in which we are not interested in this paper.  However, the S-duality has also constraint on the couplings involving only metric and dilaton. In the Einstein frame, \ie $G_{\mu\nu}=e^{\phi/2}G_{\mu\nu}^{(E)}$, there must be no such couplings involving odd number of dilatons because they can not be combined with appropriate R-R scalar couplings to make S-duality invariant \cite{Garousi:2013qka}. Note that the couplings involving B-field and odd number of dilaton can be combined with appropriate R-R  couplings to be written in S-duality invariant form. Hence, we are going to check that,  up to some  total derivative terms,  in the Einstein frame there should be no couplings involving metric and odd number of dilaton. The total derivative terms should be transferred to the boundary using the Stokes's theorem.

The overall factor $e^{-\phi}\sqrt{-\tg}$ in the string frame action \reef{Gen} transforms to the following factor in the Einstein frame:
\beqa
e^{\frac{p-3}{4}\phi}\sqrt{-\tg^E}
\eeqa
which is invariant under the S-duality for $p=3$. Hence the Lagrangian \reef{bulkf} should be consistent with the S-duality separately.  The string-frame  Lagrangian \reef{bulkf}  transforms to the following Lagrangian in the Einstein frame:
\beqa
{\cal L}_p^E&\!\!\!\!\!\!\!\!\!\!\!\!\!=\!\!\!\!\!\!\!\!\!\!\!\!\!\!\!& a_{28}e^{-\phi}\Bigg[-6 R_{abcd} R^{abcd} + 6 R_{abij} R^{abij} + 12 R^{ab}{}_{a}{}^{c} R_{b}{}^{d}{}_{cd} - 12 R^{ai}{}_{a}{}^{j} R^{b}{}_{ibj} - 6 R^{bc}{}_{bc} \nabla_{a}\nabla^{a}\phi\nn\\&& + 6 R^{bi}{}_{bi} \nabla_{a}\nabla^{a}\phi + (9 -  \frac{3}{2} p) R^{bc}{}_{bc} \nabla_{a}\phi \nabla^{a}\phi + \frac{3}{2} (-4 + p) R^{bi}{}_{bi} \nabla_{a}\phi \nabla^{a}\phi \nn\\&&+ 3 (-5 + p) \nabla_{a}\nabla^{a}\phi \nabla_{b}\nabla^{b}\phi + \frac{3}{8} (67 - 30 p + 3 p^2) \nabla_{a}\phi \nabla^{a}\phi \nabla_{b}\nabla^{b}\phi \nn\\&&+ \frac{3}{2} (-11 + p) R_{a}{}^{c}{}_{bc} \nabla^{a}\phi \nabla^{b}\phi + \frac{3}{32} (-64 + 57 p - 14 p^2 + p^3) \nabla_{a}\phi \nabla^{a}\phi \nabla_{b}\phi \nabla^{b}\phi \nn\\&&-  \frac{3}{8} (47 - 16 p + p^2) \nabla^{a}\phi \nabla_{b}\nabla_{a}\phi \nabla^{b}\phi - 6 (p-7) R_{a}{}^{c}{}_{bc} \nabla^{b}\nabla^{a}\phi \nn\\&&+ \frac{3}{4} (27 - 12 p + p^2) \nabla_{b}\nabla_{a}\phi \nabla^{b}\nabla^{a}\phi -  \frac{3}{2} (p-3) \nabla_{a}\nabla^{a}\phi \nabla_{i}\nabla^{i}\phi + 6 (p-3) R^{a}{}_{iaj} \nabla^{j}\nabla^{i}\phi \nn\\&&-  \frac{3}{8} (12 - 7 p + p^2) \nabla_{a}\phi \nabla^{a}\phi \nabla_{i}\nabla^{i}\phi -  \frac{3}{4} (p-3)^2 \nabla_{j}\nabla_{i}\phi \nabla^{j}\nabla^{i}\phi + \cdots\Bigg]\labell{lp}
\eeqa
where we have imposed  the O-plane condition that  $ \nabla_{i}\phi =0 $. In above Lagrangian dots represent the couplings including $ H $ and its derivatives. For $p=3$, it becomes
\beqa
{\cal L}_3^E&\!\!\!\!\!\!\!\!\!\!\!\!\!=\!\!\!\!\!\!\!\!\!\!\!\!\!\!\!& a_{28}e^{-\phi}\Bigg[-6 R_{abcd} R^{abcd} + 6 R_{abij} R^{abij} + 12 R^{ab}{}_{a}{}^{c} R_{b}{}^{d}{}_{cd} - 12 R^{ai}{}_{a}{}^{j} R^{b}{}_{ibj} - 6 R^{bc}{}_{bc} \nabla_{a}\nabla^{a}\phi\nn\\&& + 6 R^{bi}{}_{bi} \nabla_{a}\nabla^{a}\phi + \frac{9}{2} R^{bc}{}_{bc} \nabla_{a}\phi \nabla^{a}\phi -  \frac{3}{2} R^{bi}{}_{bi} \nabla_{a}\phi \nabla^{a}\phi - 6 \nabla_{a}\nabla^{a}\phi \nabla_{b}\nabla^{b}\phi \nn\\&&+ \frac{3}{2} \nabla_{a}\phi \nabla^{a}\phi \nabla_{b}\nabla^{b}\phi - 12 R_{a}{}^{c}{}_{bc} \nabla^{a}\phi \nabla^{b}\phi + \frac{3}{4} \nabla_{a}\phi \nabla^{a}\phi \nabla_{b}\phi \nabla^{b}\phi\nn\\&& - 3 \nabla^{a}\phi \nabla_{b}\nabla_{a}\phi \nabla^{b}\phi + 24 R_{a}{}^{c}{}_{bc} \nabla^{b}\nabla^{a}\phi+\cdots\Bigg]\labell{l3}
\eeqa
 To study the S-duality of these terms, one should include the R-R couplings in which we are not interested in this paper. To make the overall factor $e^{-\phi}$ to be invariant under the S-duality, one should include loop and non-perturbative effects \cite{Bachas:1999um}. The terms in the bracket  which have odd number of dilaton must be zero up to  some total derivative terms. Since we have already imposed the equations of motion in the string frame, we have to impose the equations of motion in the Einstein frame as well. The equations of motion are
\beqa
R^{\mu\nu}-\frac{1}{2}\partial^{\mu}\phi \partial^{\nu}\phi+\cdots = 0  \ \ , \qquad {\rm and} \qquad \nabla_{\mu}\partial^{\mu}\phi =0\labell{eomE}
\eeqa
where dots represent terms that involve $H$. Using the above Einstein frame equations of motion, one can rewrite the Lagrangian \reef{l3}  as 
\beqa
{\cal L}_3^E&\!\!\!\!\!\!\!\!\!\!\!\!\!=\!\!\!\!\!\!\!\!\!\!\!\!\!\!\!&a_{28}e^{-\phi}\Bigg[-6 R_{abcd} R^{abcd} + 6 R_{abij} R^{abij} + 12 R^{ab}{}_{a}{}^{c} R_{b}{}^{d}{}_{cd} - 12 R^{ai}{}_{a}{}^{j} R^{b}{}_{ibj} + 6 R^{bc}{}_{bc} \nabla_{a}\phi \nabla^{a}\phi \nn\\
&&\qquad\qquad - 6 \nabla_{a}\nabla^{a}\phi \nabla_{b}\nabla^{b}\phi - 12 R_{a}{}^{c}{}_{bc} \nabla^{a}\phi \nabla^{b}\phi -12 R^{bd}{}_{bd} \nabla_{a}\phi \nabla^{a}\phi -  \alpha \nabla_{a}\nabla_{b}\nabla^{b}\phi \nabla^{a}\phi \nn\\
&&\qquad\qquad- \alpha \nabla_{a}\nabla^{a}\phi \nabla_{b}\nabla^{b}\phi - 2 (6 +\alpha) \nabla^{a}\phi \nabla_{b}\nabla^{b}\nabla_{a}\phi + 24 R_{a}{}^{d}{}_{bd} \nabla^{a}\phi \nabla^{b}\phi \nn\\
&&\qquad\qquad+ ( \frac{9}{2} +  \alpha) \nabla_{a}\phi \nabla^{a}\phi \nabla_{b}\phi \nabla^{b}\phi - 2 (6 + \alpha) \nabla_{b}\nabla_{a}\phi \nabla^{b}\nabla^{a}\phi + \ldots \Big] \nn\\
&& +a_{28}\nabla^{d}\Big[e^{-\phi}\Big(24 R_{d}{}^{b}{}_{ab} \nabla^{a}\phi - 12 R^{ab}{}_{ab} \nabla_{d}\phi + \alpha \nabla_{a}\nabla^{a}\phi \nabla_{d}\phi \nn\\
&&\qquad\qquad  + (\frac{9}{2} + \alpha) \nabla_{a}\phi \nabla^{a}\phi \nabla_{d}\phi + 2 (6 + \alpha) \nabla^{a}\phi \nabla_{d}\nabla_{a}\phi\Big)\Big]\labell{LE}
\eeqa
where $\alpha$ is an arbitrary parameter. The terms in the first bracket are consistent with the S-duality for any value for the parameter $\alpha$. The terms in the second bracket are not consistent with the S-duality because they have couplings with odd number of dilaton. However, using the Stokes's theorem they becomes zero if spacetime has no boundary and they are transferred to the boundary if the spacetime has boundary.  

Now we assume the subspace $M^{(4)}$ in the spacetime $M^{(10)}=M^{(4)}\times M^{(6)}$ has boundary $\prt M^{(4)}$, \ie $\prt M^{(10)}=\prt M^{(4)}\times M^{(6)}$. The Stokes's theorem in the world-volume of O$_3$-plane in the Einstein frame is (see Appendix)
 \beqa
\int_{M^{(4)}} d^{4}\sigma\sqrt{-\tg^E}\,\nabla_a  V^a&=&\int_{\prt M^{(4)}}d^{3}\tau\,\sqrt{|\hg^E|}\,n_a^EV^a\labell{Stokes2}
\eeqa
where   $n_a^E$ is the normal vector to the boundary $\prt M^{(4)}$ in the Einstein frame  and the boundary is specified by the  functions  $\sigma^a=\sigma^a(\tau^{\ha})$.   The coordinates of the boundary  are $\tau^0,\tau^1,\tau^2, \tau^3$. In the square root on the right-hand side $\hg^E_{\ha\hb}$ is    the induced metric in the coordinates $\tau^{\ha}$, \ie 
$\hg^E_{\ha\hb}=\frac{\prt \sigma^{a}}{\prt\tau^{\ha}}\frac{\prt \sigma^{b}}{ \prt\tau^{\hb}}  \tg^E_{ab}$.
Then the  total derivative terms in the last line of \reef{LE} produce the following  boundary  terms:
\beqa
\prt \!\!\bTD&=&-\frac{T_3\pi^2\alpha'^2a_{28}}{48}\int_{\prt M^{(4)}}d^{3}\tau\,e^{-\phi}\sqrt{|\hg^E|}\,n^E_c\Big[  2 (6 + \alpha) \nabla^{a}\phi \nabla^{c}\nabla_{a}\phi + \alpha \nabla_{a}\nabla^{a}\phi \nabla^{c}\phi \nn\\
&&\quad+24 R^{c}{}^{b}{}_{ab} \nabla^{a}\phi - 12 R^{ab}{}_{ab} \nabla^{c}\phi + (\frac{9}{2} + \alpha) \nabla_{a}\phi \nabla^{a}\phi \nabla^{c}\phi \Big]\labell{PTD2}
\eeqa
They  involve the projection tensor  $ \tilde{G}^{\mu\nu}$ evaluated at the boundary of O$_3$-plane. 
Here again the overall dilaton factor can be extended to an S-duality invariant form by including the loop and non-perturbative effects \cite{Bachas:1999um}. While the terms in the first line can be extended to an S-duality invariant form by including the R-R couplings, the terms in the last line have odd number of  dilaton which can not be extended to the S-duality invariant form. They should be cancelled  with the  appropriate terms in the boundary action to be consistent with the S-duality. The consistency of the boundary action with the S-duality may then fix the parameter $\alpha$. We are going to consider the boundary action in the next section. 

\section{Boundary couplings}

When spacetime has boundary, the O$_p$-planes in this manifold may end on the boundary. For example, if one writes the spacetime as $M^{(10)}=M^{(p+1)}\times M^{(9-p)}$  where the  O$_p$-plane is along the subspace $M^{(p+1)}$ and this subspace has boundary $\prt M^{(p+1)}$, then  the effective action of O$_p$-plane at specific order of $\alpha'$ has world-volume couplings on the bulk of the O$_p$-plane, \ie in  $M^{(p+1)}$ , as well as boundary couplings on the boundary of the O$_p$-plane, \ie in $\prt M^{(p+1)}$. We have seen in the previous section that the invariance under gauge transformations and under T-duality transformations  constructs  the bulk action at order $\alpha'^2$. The T-duality constraint however is not fully satisfied. It produces some total derivative terms in the 9-dimensional base space $M^{(9)}= M^{(p)}\times M^{(9-p)}$ which is not zero. When base space has boundary, \ie $\prt M^{(9)}=\prt M^{(p)}\times M^{(9-p)}$, they produce some couplings in the boundary $\prt M^{(p)}$  which are proportional to the  unit vector $n^{\ta}$ orthogonal to the boundary, \ie \reef{PTD}. They should be included in the T-duality of the boundary action. Similarly, writing  the spacetime as $M^{(10)}=M^{(4)}\times M^{(6)}$  with boundary $\prt M^{(10)}=\prt M^{(4)}\times M^{(6)}$, we have seen that the bulk O$_3$-plane couplings that the T-duality produces are consistent with the S-duality provided that the  boundary terms \reef{PTD2} are included in the S-duality of the boundary action. In this section we are going to study string duality of the boundary action.

To impose the T-duality constraint \reef{TT} on the boundary action, we need first to find minimum number of independent and gauge invariant couplings at three derivative order on the boundary and then reduce them on the circle to apply the T-duality constraint \reef{TS}. So let us  find how many independent  gauge invariant couplings are in the boundary.

\subsection{Minimal gauge invariant   couplings in the boundary}

In this subsection we would like to find all independent and  gauge invariant   couplings on the boundary of O$_p$-plane  involving   NS-NS fields at order $\alpha'^2$ in the string frame. Inspired by the boundary couplings \reef{PTD2} in the Einstein frame for $p=3$ case, one realizes that the effective boundary action in the string frame should be as  
\beqa
\prt\!\! \bS_p&=&-\frac{T_p\pi^2\alpha'^2}{48}\int_{\prt M^{(p+1)}} d^{p}\tau\, e^{-\phi}\sqrt{|\hg|}\,\prt{\cal L}_p\labell{Genb}
\eeqa
where $\hg$ is the determinant of the induced  metric on the boundary of O$_p$-plane, \ie 
\beqa
\hg_{\ha\hb}&=&\frac{\prt \sigma^{a}}{\prt \tau^{\ha}}\frac{\prt \sigma^{b}}{\prt \tau^{\hb}}\tg_{ab}\labell{hg}
\eeqa
The boundary of O$_p$-plane is specified by  the vectors     $\sigma^{a}(\tau^{\ha})$ where $\tau^0,\tau^1,\cdots \tau^{p-1}$ are coordinates of the boundary,  and $\prt {\cal L}_p$ in \reef{Genb} is the boundary Lagrangian at three-derivative order which includes all  couplings  involving the projection tensors  $ \tilde{G}^{\mu\nu}$ and $ \bot^{\mu\nu}$ evaluated at the boundary of O$_p$-plane. 

Since the boundary of spacetime has a unite normal vector $n^{\mu}$, the boundary Lagrangian  $\prt {\cal L}_p$  should include this vector as well as the  tensors $K_{\mu\nu}$, $H_{\mu\nu\rho}$, $R_{\mu\nu\rho\sigma}$, $\nabla_{\mu}\phi$ and their derivatives. They should be contracted with the projection tensors  $ \tilde{G}^{\mu\nu}$ and $ \bot^{\mu\nu}$.   The extrinsic curvature  of boundary, \ie  $K_{\mu\nu}$, is defined as $ K_{\mu\nu}=P^{\alpha}_{\ \mu}P^{\beta}_{\ \nu}\nabla_{(\alpha}n_{\beta)} $ where the projection tensor  $ P^{\mu\nu}$  is 
$P^{\mu\nu}=G^{\mu\nu}-n^\mu n^\nu$. Using the fact that $n^\mu$ is unit vector orthogonal to the boundary, \ie 
\beqa
n^{\mu}=(\nabla_\alpha f\nabla^\alpha f)^{-1/2}\nabla^\mu f\labell{nf}
\eeqa
 where boundary is specified by the function $f$ to be a constant $f^*$,  one can rewrite $K_{\mu\nu}$ as 
\beqa
K_{\mu\nu}=\nabla_{\mu}n_{\nu}-n_{\mu}a_\mu
\eeqa
where $a_\nu= n^{\rho}\nabla_{\rho}n_{\nu}$ is acceleration. It satisfies the relation $n^\mu a_\mu=0$. Note that the extrinsic curvature is symmetric and satisfies $n^\mu K_{\mu\nu}=0$ which can easily be seen by writing it in terms of function $f$. Using this symmetry and $n^\mu n_\mu=1$, one finds the most general couplings have the structures  
 $KH^2, $ $ \ n H \nabla H,$ $ \ K R,$ $ \  n \nabla R,$ $ \  n(\nabla \phi)^3, $ $ \  K(\nabla \phi)^2,  $ $ \ n \nabla\nabla\nabla\phi,$ $ \ K\nabla\nabla\phi,$ $ \ n H^2 \nabla\phi, $ $  \ n \nabla\phi \nabla\nabla\phi, $ $ \ \nabla K \nabla\phi,$ $ \ \nabla\nabla K , $ $\  K^3 , $ $ K\nabla K n ,$ $ n^2 \nabla K \nabla\phi ,$ $ n^2 \nabla\nabla K ,$ $ n K^2 \nabla\phi ,$ $ n^2 K H^2 ,$ $ n^2 K R ,$ $ n^2 K (\nabla\phi)^2 ,$ $ n^2 K \nabla\nabla\phi ,$ $ n^3 \nabla R ,$ $ n^3 H \nabla H ,$ $ n^3 R \nabla\phi ,$ $ n^3 H^2 \nabla\phi ,$ $ n^3 \nabla\nabla\nabla\phi ,$ $ n^3 \nabla\nabla\phi \nabla\phi ,$ $ n^3 (\nabla\phi)^3 ,$ $ n^4 kH^2, $ $ \ n^5 H \nabla H,$ $ \ n^4 K R,$ $ \  n^5 \nabla R,$ $ \  n^5(\nabla \phi)^3, $ $ \ n^4 K(\nabla \phi)^2,  $ $ \ n^5 \nabla\nabla\nabla\phi,$ $ \ n^4 K\nabla\nabla\phi,$ $ \ n^5 H^2 \nabla\phi, $ $  \ n^5 \nabla\phi \nabla\nabla\phi, $ $ \ n^4 \nabla K \nabla\phi,$ $ \ n^4 \nabla\nabla K , $ $\  n^2 K^3 , $ $ K\nabla K n^3 ,$ $ n^3 K^2 \nabla\phi$. 
One should impose the  equations of motion \reef{eom} and the orientifold  projections for the bulk fields as in the bulk action. The orientifold projection for the boundary fields requires to remove the following boundary terms:  
\beqa
&& K_{bi} =0, \ \nabla_{b} K_{ai} =0, \ \nabla_{i} K_{ab}=0,  \ \nabla_{j} K_{ik}=0 ,  \ \  \nabla_{a}\nabla_{b} K_{ci}=0 \nn\\
&& \nabla_{a}\nabla_{j} K_{ik}=0  ,  \ \nabla_{j}\nabla_{l} K_{ai}=0 ,\  n_i=0\labell{pro}
\eeqa
After imposing the equations of motion and the orientifold projection, one finds that the corresponding Lagrangian has 108 couplings. We call this Lagranian, with coefficients $b'_1,b'_2, \cdots, b'_{108}$, the boundary Lagrangian $\prt L_p$. Not all terms in this Lagrangian, however,  are  independent. Some of them are related by total derivative terms and by the Bianchi identities \reef{Bian} and \reef{RQ}. The unit vector also satisfies the relation 
 \beqa
n_{[\mu}\nabla_{\nu}n_{\rho]}&=&0\labell{nn}
\eeqa
which can easily be seen by writing it in terms of function $f$ using \reef{nf}. 

To remove the redundancy corresponding to the total derivative terms, we add to  $\prt L_p$ all total derivative terms at order $\alpha'^2$ with arbitrary coefficients.  
In this case, however,  the total derivative terms in the boundary have different structure  than the total derivative terms in the bulk. According to the Stokes's theorem, the total derivative terms in the boundary which have the following structure are zero (see Appendix):
\beqa
\alpha'^2\int_{\prt M^{(p+1)}}d^{p}\tau \sqrt{|\hat{g} |}\cJ=\alpha'^2\int_{\prt M^{(p+1)}}d^{p}\tau \sqrt{|\hat{g} |} n_{a}\nabla_{b}(e^{-\phi}\mathcal{F}^{ab}) =0
\labell{bstokes}
\eeqa
where  $ \mathcal{F}^{ab} $ is an arbitrary antisymmetric tensor constructed from $n,K,\nabla K,H^2, \nabla\phi,\nabla\nabla\phi$ at two-derivative order, \ie 
\beqa
\mathcal{F}_{de}=&&z_{1}\ \Big(H_{abi} H_{e}{}^{bi} n^{a} n_{d} -  H_{abi} H_{d}{}^{bi} n^{a} n_{e}\Big)\ + z_{2}\ \Big(K^{b}{}_{b} K_{ea} n^{a} n_{d} -  K^{b}{}_{b} K_{da} n^{a} n_{e}\Big)\ \nonumber \\ 
&& + z_{3}\ \Big(K_{ab} K_{e}{}^{b} n^{a} n_{d} -  K_{ab} K_{d}{}^{b} n^{a} n_{e}\Big)\ + z_{4}\ \Big(K_{ea} K^{i}{}_{i} n^{a} n_{d} -  K_{da} K^{i}{}_{i} n^{a} n_{e}\Big)\ \nonumber \\ 
&& + z_{5}\ \Big(K_{bc} K_{ea} n^{a} n^{b} n^{c} n_{d} -  K_{bc} K_{da} n^{a} n^{b} n^{c} n_{e}\Big)\ + z_{6}\ \Big(n^{a} n_{e} R_{d}{}^{b}{}_{ab} -  n^{a} n_{d} R_{e}{}^{b}{}_{ab}\Big)\ \nonumber \\ 
&& + z_{7}\ \Big(n_{e} \nabla_{a}K_{d}{}^{a} -  n_{d} \nabla_{a}K_{e}{}^{a}\Big)\ + z_{8}\ \Big(K_{eb} n^{a} n^{b} n_{d} \nabla_{a}\phi -  K_{db} n^{a} n^{b} n_{e} \nabla_{a}\phi\Big)\ \nonumber \\ 
&& + z_{9}\ \Big(K_{ea} n_{d} \nabla^{a}\phi -  K_{da} n_{e} \nabla^{a}\phi\Big)\ + z_{10}\ \Big(n^{a} n^{b} n_{e} \nabla_{b}K_{da} -  n^{a} n^{b} n_{d} \nabla_{b}K_{ea}\Big)\ \nonumber \\ 
&& + z_{11}\ \Big(n^{a} n^{b} n_{e} \nabla_{d}K_{ab} -  n^{a} n^{b} n_{d} \nabla_{e}K_{ab}\Big)\ + z_{12}\ \Big(n_{e} \nabla_{d}K^{a}{}_{a} -  n_{d} \nabla_{e}K^{a}{}_{a}\Big)\ \nonumber \\ 
&& + z_{13}\ \Big(n^{a} \nabla_{d}K_{ea} -  n^{a} \nabla_{e}K_{da}\Big)\ + z_{14}\ \Big(n_{e} \nabla_{d}K^{i}{}_{i} -  n_{d} \nabla_{e}K^{i}{}_{i}\Big)\ \nonumber \\ 
&& + z_{15}\ \Big(K_{ea} n^{a} \nabla_{d}\phi -  K_{da} n^{a} \nabla_{e}\phi\Big)\ + z_{16}\ \Big(K^{a}{}_{a} n_{e} \nabla_{d}\phi -  K^{a}{}_{a} n_{d} \nabla_{e}\phi\Big)\ \nonumber \\ 
&& + z_{17}\ \Big(K^{i}{}_{i} n_{e} \nabla_{d}\phi -  K^{i}{}_{i} n_{d} \nabla_{e}\phi\Big)\ + z_{18}\ \Big(K_{ab} n^{a} n^{b} n_{e} \nabla_{d}\phi -  K_{ab} n^{a} n^{b} n_{d} \nabla_{e}\phi\Big)\ \nonumber \\ 
&& + z_{19}\ \Big(n^{a} n_{e} \nabla_{a}\phi \nabla_{d}\phi -  n^{a} n_{d} \nabla_{a}\phi \nabla_{e}\phi\Big)\ + z_{20}\ \Big(n^{a} n_{e} \nabla_{d}\nabla_{a}\phi -  n^{a} n_{d} \nabla_{e}\nabla_{a}\phi\Big)\ \nonumber \\ 
&& + z_{21}\ \Big(n_{e} \nabla_{i}K_{d}{}^{i} -  n_{d} \nabla_{i}K_{e}{}^{i}\Big)
\eeqa
where $z_1,\cdots, z_{21}$ are arbitrary parameters. It is important to note that these total derivative terms connect the boundary terms which have different number of the unit vector.
 For example, consider  $ \mathcal{F}_{{d}{e}}= n_{{e}} \nabla_{i}K_{{d}}{}^{i} -  n_{{d}} \nabla_{i}K_{{e}}{}^{i} $. Then the   boundary integral of the following    term is zero
\beqa  
n^{{e}}\nabla^{{d}}\Big(e^{-\phi}\Big(n_{{e}} \nabla_{i}K_{{d}}{}^{i} -  n_{{d}} \nabla_{i}K_{{e}}{}^{i}\Big) =&&\!\!\!\!\!\!\!\! e^{-\phi}\Big[- n^{{d}} n^{{e}} \nabla_{{e}}\nabla_{i}K_{{d}}{}^{i} + \nabla_{{e}}\nabla_{i}K^{{e}i} -  n^{{d}} \nabla_{{e}}n^{{e}} \nabla_{i}K_{{d}}{}^{i} \nn\\
&&+ n^{{d}} n^{{e}} \nabla_{{d}}\phi \nabla_{i}K_{{e}}{}^{i} 
 -  \nabla^{{e}}\phi \nabla_{i}K_{{e}}{}^{i} \Big]
\eeqa
where we have used the fact that $n^\mu$ is unite vector orthogonal to the boundary and $n^\mu K_{\mu\nu}=0$. The right-hand side then gives a relation between one $n$ and three $n$'s.

Adding the total derivative terms with arbitrary coefficients  to $\prt L_p$, one finds the same Lagrangian     but with different parameters $b_1, b_2, \cdots$. We call the new Lagrangian  $\prt {\cal L}_p$. Hence 
\beqa
{\bf \Delta}-\cJ&=&0\labell{DLb}
\eeqa
where ${\bf\Delta}=\prt{\cal L}_p-\prt L_p$ is the same as $\prt L_p$ but with coefficients $\delta b_1,\delta b_2,\cdots$ where $\delta b_i= b_i-b'_i$. Solving the above equation, one finds some linear  relations between  only $\delta b_1,\delta b_2,\cdots$ which indicate how the couplings are related among themselves by the total derivative terms. The above equation also gives some relation between the coefficients of the total derivative terms and $\delta b_1,\delta b_2,\cdots$ in which we are not interested. 

To impose in \reef{DLb} the Bianchi identities \reef{Bian}, \reef{RQ} we go to the local frame, and to impose the identities corresponding to the unit vector we write $n^\mu$ in terms of function $f$ using \reef{nf}. Then one finds 78 independent couplings.  In this case, there is no scheme in which the couplings involve only terms with one and two derivatives, \eg there is no scheme which has no $\nabla R$. One particular choice for the 78 gauge invariant boundary couplings is the following:
\beqa
\prt\cL_p&\!\!\!\!\!=\!\!\!\!\!& b_{1}\  H_{bci} H^{bci} K^{a}{}_{a} + b_{2}\  H_{ijk} H^{ijk} K^{a}{}_{a} + b_{3}\  H_{a}{}^{ci} H_{bci} K^{ab} + b_{4}\  K_{a}{}^{c} K^{ab} K_{bc} + b_{5}\  K^{a}{}_{a} K_{bc} K^{bc} \nonumber \\ 
&& + b_{6}\  K^{a}{}_{a} K^{b}{}_{b} K^{c}{}_{c} + b_{7}\  H_{abj} H^{abj} K^{i}{}_{i} + b_{8}\  H_{jkl} H^{jkl} K^{i}{}_{i} + b_{9}\  K^{a}{}_{a} K^{b}{}_{b} K^{i}{}_{i} \nonumber \\ 
&& + b_{10}\  H_{abj} H^{ab}{}_{i} K^{ij} + b_{11}\  H_{i}{}^{kl} H_{jkl} K^{ij} + b_{12}\  K_{i}{}^{k} K^{ij} K_{jk} + b_{13}\  K^{a}{}_{a} K^{i}{}_{i} K^{j}{}_{j} \nonumber \\ 
&& + b_{15}\  K^{i}{}_{i} K^{j}{}_{j} K^{k}{}_{k} + b_{16}\  H_{a}{}^{di} H_{bdi} K^{c}{}_{c} n^{a} n^{b} + b_{17}\  H_{ac}{}^{i} H_{bdi} K^{cd} n^{a} n^{b} + b_{18}\  H_{a}{}^{cj} H_{bcj} K^{i}{}_{i} n^{a} n^{b} \nonumber \\ 
&& + b_{19}\  H_{a}{}^{c}{}_{i} H_{bcj} K^{ij} n^{a} n^{b} + b_{20}\  K^{cd} n^{a} n^{b} R_{acbd} + b_{21}\  K^{ab} R_{a}{}^{c}{}_{bc} + b_{22}\  K^{i}{}_{i} n^{a} n^{b} R_{a}{}^{c}{}_{bc} \nonumber \\ 
&& + b_{23}\  K^{c}{}_{c} n^{a} n^{b} R_{a}{}^{d}{}_{bd} + b_{24}\  K^{ij} n^{a} n^{b} R_{aibj} + b_{25}\  K^{i}{}_{i} R^{ab}{}_{ab} + b_{26}\  K^{ij} R^{a}{}_{iaj} + b_{27}\  K^{a}{}_{a} R^{bc}{}_{bc} \nonumber \\ 
&& + b_{29}\  \nabla_{a}\nabla^{a}K^{i}{}_{i} + b_{28}\  n^{a} n^{b} \nabla_{b}\nabla_{a}K^{i}{}_{i} + b_{30}\  H^{bci} n^{a} \nabla_{a}H_{bci} + b_{31}\  K^{i}{}_{i} n^{a} \nabla_{a}K^{b}{}_{b} \nonumber \\ 
&& + b_{34}\  K^{ij} n^{a} \nabla_{a}K_{ij} + b_{32}\  K^{b}{}_{b} n^{a} \nabla_{a}K^{i}{}_{i} + b_{33}\  K^{i}{}_{i} n^{a} \nabla_{a}K^{j}{}_{j} + b_{35}\  n^{a} \nabla_{a}R^{bc}{}_{bc} \nonumber \\ 
&& + b_{37}\  H_{bci} H^{bci} n^{a} \nabla_{a}\phi + b_{38}\  H_{ijk} H^{ijk} n^{a} \nabla_{a}\phi + b_{39}\  K^{b}{}_{b} K^{c}{}_{c} n^{a} \nabla_{a}\phi + b_{40}\  K^{b}{}_{b} K^{i}{}_{i} n^{a} \nabla_{a}\phi \nonumber \\ 
&& + b_{41}\  K^{i}{}_{i} K^{j}{}_{j} n^{a} \nabla_{a}\phi + b_{42}\  n^{a} R^{bc}{}_{bc} \nabla_{a}\phi + b_{43}\  n^{a} \nabla_{a}\nabla_{b}\nabla^{b}\phi + b_{71}\  \nabla_{a}\nabla_{i}K^{ai} \nonumber \\ 
&& + b_{45}\  \nabla_{a}K^{b}{}_{b} \nabla^{a}\phi + b_{46}\  \nabla_{a}K^{i}{}_{i} \nabla^{a}\phi + b_{47}\  K^{b}{}_{b} \nabla_{a}\phi \nabla^{a}\phi + b_{48}\  K^{i}{}_{i} \nabla_{a}\phi \nabla^{a}\phi \nonumber \\ 
&& + b_{51}\  K^{i}{}_{i} n^{a} \nabla_{b}K_{a}{}^{b} + b_{52}\  K^{c}{}_{c} n^{a} n^{b} \nabla_{a}\phi \nabla_{b}\phi + b_{53}\  K^{i}{}_{i} n^{a} n^{b} \nabla_{a}\phi \nabla_{b}\phi \nonumber \\ 
&& + b_{54}\  \nabla_{b}\nabla_{a}K^{ab} + b_{55}\  n^{a} n^{b} \nabla_{b}\nabla_{a}K^{c}{}_{c} + b_{56}\  K^{ab} \nabla_{b}\nabla_{a}\phi + b_{57}\  K^{c}{}_{c} n^{a} n^{b} \nabla_{b}\nabla_{a}\phi \nonumber \\ 
&& + b_{58}\  K^{i}{}_{i} n^{a} n^{b} \nabla_{b}\nabla_{a}\phi + b_{59}\  \nabla_{b}\nabla^{b}K^{a}{}_{a} + b_{60}\  n^{a} n^{b} \nabla_{b}\nabla_{c}K_{a}{}^{c} + b_{44}\  n^{a} n^{b} \nabla_{b}\nabla_{i}K_{a}{}^{i} \nonumber \\ 
&& + b_{61}\  H_{a}{}^{ci} H_{bci} n^{a} \nabla^{b}\phi + b_{62}\  n^{a} R_{a}{}^{c}{}_{bc} \nabla^{b}\phi + b_{63}\  n^{a} \nabla_{a}\phi \nabla_{b}\phi \nabla^{b}\phi + b_{64}\  n^{a} \nabla_{b}\nabla_{a}\phi \nabla^{b}\phi \nonumber \\ 
&& -  b_{49}\  H^{bci} n^{a} \nabla_{c}H_{abi} + b_{68}\  H_{a}{}^{bi} n^{a} \nabla_{c}H_{b}{}^{c}{}_{i} + b_{50}\  H_{a}{}^{di} n^{a} n^{b} n^{c} \nabla_{c}H_{bdi} + b_{65}\  n^{a} n^{b} \nabla_{a}\phi \nabla_{c}K_{b}{}^{c} \nonumber \\ 
&& + b_{36}\  n^{a} n^{b} n^{c} \nabla_{c}R_{a}{}^{d}{}_{bd} + b_{66}\  n^{a} n^{b} n^{c} \nabla_{a}\phi \nabla_{b}\phi \nabla_{c}\phi + b_{67}\  n^{a} n^{b} n^{c} \nabla_{a}\phi \nabla_{c}\nabla_{b}\phi \nonumber \\ 
&& + b_{69}\  n^{a} n^{b} \nabla_{b}K_{ac} \nabla^{c}\phi + b_{70}\  n^{a} n^{b} n^{c} n^{d} \nabla_{d}\nabla_{c}K_{ab} + b_{73}\  K^{b}{}_{b} n^{a} \nabla_{i}K_{a}{}^{i} \nonumber \\ 
&& + b_{74}\  n^{a} n^{b} \nabla_{a}\phi \nabla_{i}K_{b}{}^{i} + b_{78}\  \nabla_{i}\nabla^{i}K^{a}{}_{a} + b_{75}\  \nabla_{j}\nabla_{i}K^{ij} + b_{76}\  K^{ij} \nabla_{j}\nabla_{i}\phi \nonumber \\ 
&& + b_{77}\  \nabla_{j}\nabla^{j}K^{i}{}_{i} + b_{72}\  H^{ijk} n^{a} \nabla_{k}H_{aij}+ b_{14}\  K^{i}{}_{i} K_{jk} K^{jk} \labell{b73}
\eeqa
where $b_1,\cdots, b_{78}$ are 78 arbitrary $p$-independent coefficients that may be fixed by the duality constraints.

\subsection{T-Duality  constraint in the boundary}
In this subsection we are going to impose the T-duality constraint \reef{TT} on the gauge invariant couplings \reef{b73} to  fix their parameters.  To find $ \prt S_{\rm eff}(\psi)$ we need to dimensionally reduce O$_{(p-1)}$-plane boundary action along the circle orthogonal to the O-plane (transverse reduction), \ie $\prt M^{(10)}=\prt M^{(p)}\times M^{(10-p)}$ and $M^{(10-p)}=S^{(1)}\times M^{(9-p)}$, and to find $ \prt S_{\rm eff}(\psi')$ we need to dimensionally reduce O$_p$-plane boundary action along the circle tangent to the O-plane (world-volume reduction), \ie  $\prt M^{(10)}=\prt M^{(p+1)}\times M^{(9-p)}$ and $\prt M^{(p+1)}=S^{(1)}\times \prt M^{(p)}$. 

To find the T-duality  of the pull-back metric \reef{hg} in the boundary, we assume the killing direction in the boundary space to be  $y$. It is implicitly assumed in the T-duality prescription that everything should  be independent of the killing coordinate $y$, hence, the boundary should be  specified as $\sigma^a(\tau^{\ha})=(y,\sigma^{\ta}(\tau^{\ba}))$. Then one can show that  when O$_p$-plane is along the $y$-direction the reduction of  $e^{-\phi}\sqrt{|\hg|}=e^{-\phi+\varphi/2}\sqrt{|\bg|}$ where $\bg$ is determinant of the induced metric  \reef{bg}, and when  O$_{(p-1)}$-plane is orthogonal to  the $y$-direction the reduction of  $e^{-\phi}\sqrt{|\hg|}=e^{-\phi}\sqrt{|\bg|}$. The former transforms under the T-duality transformation \reef{Buscher} to  the latter. Hence, to find the T-duality constraints on the boundary action \reef{Genb}, one should consider only the T-duality constraint on  $\prt\cL_p$ in \reef{b73}.

The reductions of projection tensors $\tG,\, \bot$ and the spacetime tensors $\nabla\phi,\, H,\, R$ and their covariant derivative are exactly as we have found in the subsection 2.2, so we need to find  reduction of the boundary tensor $K_{\mu\nu}$ and its covariant derivatives, and the reduction of the unite vector $n^\mu$ which appear in \reef{b73}. Using the fact that everything should be independent of the killing coordinate $y$, one finds $n^y\sim \prt^y f=0$ and $n^\mu$ when $\mu$ is not the $y$-index, is the unite vector orthogonal to the boundary in the base space.  The reduction of the  extrinsic curvature and its derivatives again have both $U(1)\times U(1)$ gauge invarinat part and non-gauge invariant part. The non-gauge invariant part will be cancelled in any covariant couplings. Hence, we need to keep only the gauge invariant part of the reduction of the extrinsic curvature and its covariant derivatives. Writing each tensor in terms of metric and $n^\mu$, and using the reductions \reef{GB} and \reef{IG}, one finds the gauge invariant part of the reductions when the base space is flat, are  
\beqa
K_{\mu\nu} = &&\!\!\!\!\!\!\! \hat{K}_{\mu\nu} \labell{rK}\\
K_{\mu y} = &&\!\!\!\!\!\!\! \frac{1}{2} e^{\varphi} n^{\alpha}V_{\alpha \mu} \equiv \frac{1}{2} e^{\varphi} \mathcal{V}_{\mu}\nn\\
K_{y y} = &&\!\!\!\!\!\!\! \frac{1}{2} e^{\varphi} n^{\alpha}\partial_{\alpha} \varphi \equiv\frac{1}{2} e^{\varphi} \Theta \nn\\
\nabla_{\rho} K_{\mu \nu} = &&\!\!\!\!\!\!\! e^{\varphi}\Big(- \frac{1}{4} V_{\nu  \rho } \mathcal{V}_{\mu } -  \frac{1}{4} V_{\mu  \rho } \mathcal{V}_{\nu } +e^{-\varphi}\partial_{\rho }\hat{K}_{\mu  \nu } \Big)\nn\\
\hat{\nabla}_{\rho} K_{\nu y} = &&\!\!\!\!\!\!\! e^{\varphi}\Big(- \frac {1}{2} V_{\rho  \alpha } \hat{K}_{\nu }^{\ \alpha } -  \frac {1}{4} V_{\nu  \rho } \Theta +\frac {1}{2} \partial_{\rho }\mathcal{V}_{\nu } + \frac {1}{4} \mathcal{V}_{\nu } \partial_{\rho }\varphi \Big)\nn\\
\nabla_{y} K_{\mu \nu} = &&\!\!\!\!\!\!\! e^{\varphi}\Big(- \frac{1}{2} V_{\nu  \alpha } \hat{K}^{\alpha }{}_{\mu } -  \frac{1}{2} V_{\mu  \alpha } \hat{K}^{\alpha }{}_{\nu } -  \frac{1}{4} \mathcal{V}_{\nu } \partial_{\mu }\varphi -  \frac{1}{4} \mathcal{V}_{\mu } \partial_{\nu }\varphi\Big)\nn\\
\nabla_{y} K_{\mu y} = &&\!\!\!\!\!\!\! \frac{1}{2} e^{\varphi}\Big(- \frac{1}{2} e^{\varphi} V_{\mu  \alpha } \mathcal{V}^{\alpha } +  \hat{K}_{\mu }^{\ \alpha } \partial_{\alpha }\varphi -  \frac{1}{2} \Theta \partial_{\mu }\varphi \Big)\nn\\
\nabla_{\rho} K_{y y} = &&\!\!\!\!\!\!\! \frac{1}{2} e^{\varphi}\Big(- e^{\varphi} V_{\rho  \beta } \mathcal{V}^{\beta } + \partial_{\rho }\Theta \Big)\nn\\
\nabla_{y} K_{y y} = &&\!\!\!\!\!\!\! \frac{1}{2} e^{\varphi}\Big(e^{\varphi} \mathcal{V}^{\beta } \partial_{\beta }\varphi \Big)\nn\\
\nabla_{\sigma}\nabla_{\rho} K_{\mu\nu} = &&\!\!\!\!\!\!\! e^{\varphi}\Big(\frac{1}{4} V_{\mu  \sigma } V_{\rho  \alpha } \hat{K}^{\alpha }{}_{\nu } + \frac{1}{4} V_{\mu  \alpha } V_{\rho  \sigma } \hat{K}^{\alpha }{}_{\nu } + \frac{1}{4} V_{\nu  \sigma } V_{\rho  \alpha } \hat{K}_{\mu }{}^{\alpha } + \frac{1}{4} V_{\nu  \alpha } V_{\rho  \sigma } \hat{K}_{\mu }{}^{\alpha } \nonumber \\ 
&& + \frac{1}{8} V_{\mu  \sigma } V_{\nu  \rho } \Theta + \frac{1}{8} V_{\mu  \rho } V_{\nu  \sigma } \Theta + \frac{1}{8} V_{\rho  \sigma } \mathcal{V}_{\nu } \partial_{\mu }\varphi + \frac{1}{8} V_{\rho  \sigma } \mathcal{V}_{\mu } \partial_{\nu }\varphi -  \frac{1}{4} V_{\nu  \sigma } \partial_{\rho }\mathcal{V}_{\mu } \nonumber \\ 
&& -  \frac{1}{4} V_{\mu  \sigma } \partial_{\rho }\mathcal{V}_{\nu } -  \frac{1}{8} V_{\nu  \sigma } \mathcal{V}_{\mu } \partial_{\rho }\varphi -  \frac{1}{8} V_{\mu  \sigma } \mathcal{V}_{\nu } \partial_{\rho }\varphi -  \frac{1}{4} \mathcal{V}_{\nu } \partial_{\sigma }V_{\mu  \rho } -  \frac{1}{4} \mathcal{V}_{\mu } \partial_{\sigma }V_{\nu  \rho } \nonumber \\ 
&& -  \frac{1}{4} V_{\nu  \rho } \partial_{\sigma }\mathcal{V}_{\mu } -  \frac{1}{4} V_{\mu  \rho } \partial_{\sigma }\mathcal{V}_{\nu } -  \frac{1}{4} V_{\nu  \rho } \mathcal{V}_{\mu } \partial_{\sigma }\varphi -  \frac{1}{4} V_{\mu  \rho } \mathcal{V}_{\nu } \partial_{\sigma }\varphi \nonumber \\ 
&& + e^{-\varphi}\partial_{\sigma }\partial_{\rho }\hat{K}_{\mu  \nu }\Big)\nn\\
\nabla_{y}\nabla_{\rho} K_{\mu\nu} = &&\!\!\!\!\!\!\! e^{\varphi}\Big(\frac{1}{8} e^{\varphi} V_{\mu  \rho } V_{\nu }{}^{\alpha } \mathcal{V}_{\alpha } + \frac{1}{8} e^{\varphi} V_{\mu }{}^{\alpha } V_{\nu  \rho } \mathcal{V}_{\alpha } -  \frac{1}{2} V_{\rho }{}^{\alpha } \partial_{\alpha }\hat{K}_{\mu  \nu } + \frac{1}{4} V_{\rho  \alpha } \hat{K}^{\alpha }{}_{\nu } \partial_{\mu }\varphi \nonumber \\ 
&& + \frac{1}{8} V_{\nu  \rho } \Theta \partial_{\mu }\varphi + \frac{1}{4} V_{\rho  \alpha } \hat{K}_{\mu }{}^{\alpha } \partial_{\nu }\varphi + \frac{1}{8} V_{\mu  \rho } \Theta \partial_{\nu }\varphi -  \frac{1}{2} V_{\mu }{}^{\alpha } \partial_{\rho }\hat{K}_{\alpha  \nu } \nonumber \\ 
&& -  \frac{1}{2} V_{\nu }{}^{\alpha } \partial_{\rho }\hat{K}_{\mu  \alpha } -  \frac{1}{4} \partial_{\nu }\varphi \partial_{\rho }\mathcal{V}_{\mu } -  \frac{1}{4} \partial_{\mu }\varphi \partial_{\rho }\mathcal{V}_{\nu } + \frac{1}{4} V_{\mu  \alpha } \hat{K}^{\alpha }{}_{\nu } \partial_{\rho }\varphi \nonumber \\ 
&& + \frac{1}{4} V_{\nu  \alpha } \hat{K}_{\mu }{}^{\alpha } \partial_{\rho }\varphi \Big)\nn\\
\nabla_{\sigma}\nabla_{y} K_{\mu\nu} = &&\!\!\!\!\!\!\! e^{\varphi}\Big(\frac{1}{8} e^{\varphi} V_{\mu  \sigma } V_{\nu }{}^{\alpha } \mathcal{V}_{\alpha } + \frac{1}{8} e^{\varphi} V_{\mu }{}^{\alpha } V_{\nu  \sigma } \mathcal{V}_{\alpha } + \frac{1}{8} e^{\varphi} V_{\nu }{}^{\alpha } V_{\sigma  \alpha } \mathcal{V}_{\mu } + \frac{1}{8} e^{\varphi} V_{\mu }{}^{\alpha } V_{\sigma  \alpha } \mathcal{V}_{\nu } \nonumber \\ 
&& -  \frac{1}{2} V_{\sigma }{}^{\alpha } \partial_{\alpha }\hat{K}_{\mu  \nu } -  \frac{1}{4} V_{\mu  \sigma } \hat{K}_{\alpha  \nu } \partial^{\alpha }\varphi -  \frac{1}{4} V_{\nu  \sigma } \hat{K}_{\mu  \alpha } \partial^{\alpha }\varphi + \frac{1}{8} V_{\nu  \sigma } \Theta \partial_{\mu }\varphi \nonumber \\ 
&& + \frac{1}{8} V_{\mu  \sigma } \Theta \partial_{\nu }\varphi -  \frac{1}{2} \hat{K}^{\alpha }{}_{\nu } \partial_{\sigma }V_{\mu  \alpha } -  \frac{1}{2} \hat{K}_{\mu }{}^{\alpha } \partial_{\sigma }V_{\nu  \alpha } -  \frac{1}{2} V_{\mu }{}^{\alpha } \partial_{\sigma }\hat{K}_{\alpha  \nu } \nonumber \\ 
&& -  \frac{1}{2} V_{\nu }{}^{\alpha } \partial_{\sigma }\hat{K}_{\mu  \alpha } -  \frac{1}{4} \partial_{\nu }\varphi \partial_{\sigma }\mathcal{V}_{\mu } -  \frac{1}{4} \partial_{\mu }\varphi \partial_{\sigma }\mathcal{V}_{\nu } -  \frac{1}{4} V_{\mu  \alpha } \hat{K}^{\alpha }{}_{\nu } \partial_{\sigma }\varphi \nonumber \\ 
&& -  \frac{1}{4} V_{\nu  \alpha } \hat{K}_{\mu }{}^{\alpha } \partial_{\sigma }\varphi -  \frac{1}{8} \mathcal{V}_{\nu } \partial_{\mu }\varphi \partial_{\sigma }\varphi -  \frac{1}{8} \mathcal{V}_{\mu } \partial_{\nu }\varphi \partial_{\sigma }\varphi -  \frac{1}{4} \mathcal{V}_{\nu } \partial_{\sigma }\partial_{\mu }\varphi \nonumber \\ 
&& -  \frac{1}{4} \mathcal{V}_{\mu } \partial_{\sigma }\partial_{\nu }\varphi \Big)\nn\\
\nabla_{\sigma}\nabla_{\rho} K_{\mu y} = &&\!\!\!\!\!\!\! e^{\varphi}\Big(\frac{1}{4} e^{\varphi} V_{\mu  \sigma } V_{\rho }{}^{\alpha } \mathcal{V}_{\alpha } + \frac{1}{8} e^{\varphi} V_{\mu }{}^{\alpha } V_{\rho  \sigma } \mathcal{V}_{\alpha } + \frac{1}{8} e^{\varphi} V_{\mu  \rho } V_{\sigma }{}^{\alpha } \mathcal{V}_{\alpha } \nonumber \\ 
&& -  \frac{1}{8} e^{\varphi} V_{\rho }{}^{\alpha } V_{\sigma  \alpha } \mathcal{V}_{\mu } -  \frac{1}{4} V_{\rho  \sigma } \hat{K}_{\mu  \alpha } \partial^{\alpha }\varphi + \frac{1}{8} V_{\rho  \sigma } \theta \partial_{\mu }\varphi -  \frac{1}{2} V_{\sigma }{}^{\alpha } \partial_{\rho }\hat{K}_{\mu  \alpha } \nonumber \\ 
&& -  \frac{1}{4} V_{\mu  \sigma } \partial_{\rho }\theta + \frac{1}{2} \partial_{\rho }\partial_{\sigma }\mathcal{V}_{\mu } -  \frac{1}{4} \theta \partial_{\sigma }V_{\mu  \rho } -  \frac{1}{2} \hat{K}_{\mu }{}^{\alpha } \partial_{\sigma }V_{\rho  \alpha } -  \frac{1}{2} V_{\rho }{}^{\alpha } \partial_{\sigma }\hat{K}_{\mu  \alpha } \nonumber \\ 
&& -  \frac{1}{4} V_{\mu  \rho } \partial_{\sigma }\theta + \frac{1}{4} \partial_{\rho }\varphi \partial_{\sigma }\mathcal{V}_{\mu } -  \frac{1}{4} V_{\rho  \alpha } \hat{K}_{\mu }{}^{\alpha } \partial_{\sigma }\varphi -  \frac{1}{8} V_{\mu  \rho } \theta \partial_{\sigma }\varphi \nonumber \\ 
&& + \frac{1}{4} \partial_{\rho }\mathcal{V}_{\mu } \partial_{\sigma }\varphi + \frac{1}{8} \mathcal{V}_{\mu } \partial_{\rho }\varphi \partial_{\sigma }\varphi + \frac{1}{4} \mathcal{V}_{\mu } \partial_{\sigma }\partial_{\rho }\varphi \Big) \nn\\
\nabla_{\mu}\nabla_{\nu} K_{y y} = &&\!\!\!\!\!\!\! e^{\varphi}\Big( \frac{1}{4} e^{\varphi} V_{\mu  \beta } V_{\nu  \alpha } \hat{K}^{\alpha  \beta } + \frac{1}{4} e^{\varphi} V_{\mu  \alpha } V_{\nu  \beta } \hat{K}^{\alpha  \beta } -  \frac{1}{4} e^{\varphi} V_{\mu }{}^{\alpha } V_{\nu  \alpha } \Theta + \frac{1}{4} e^{\varphi} V_{\mu  \nu } \mathcal{V}_{\alpha } \partial^{\alpha }\varphi \nonumber \\ 
&& -  \frac{1}{2} e^{\varphi} \mathcal{V}^{\beta } \partial_{\mu }V_{\nu  \beta } -  \frac{1}{2} e^{\varphi} V_{\nu }{}^{\beta } \partial_{\mu }\mathcal{V}_{\beta } -  \frac{1}{2} e^{\varphi} V_{\nu }{}^{\alpha } \mathcal{V}_{\alpha } \partial_{\mu }\varphi -  \frac{1}{2} e^{\varphi} V_{\mu }{}^{\beta } \partial_{\nu }\mathcal{V}_{\beta } \nonumber \\ 
&& -  \frac{1}{4} e^{\varphi} V_{\mu }{}^{\alpha } \mathcal{V}_{\alpha } \partial_{\nu }\varphi + \frac{1}{2} \partial_{\nu }\partial_{\mu }\Theta \Big) \nn\\
\nabla_{\mu}\nabla_{y} K_{\nu y} = &&\!\!\!\!\!\!\! e^{\varphi}\Big( \frac{1}{4} e^{\varphi} V_{\mu  \beta } V_{\nu  \alpha } \hat{K}^{\alpha  \beta } -  \frac{1}{2} e^{\varphi} V_{\alpha  \beta } V_{\mu }{}^{\beta } \hat{K}_{\nu }{}^{\alpha } + \frac{1}{8} e^{\varphi} V_{\mu }{}^{\alpha } V_{\nu  \alpha } \Theta \nonumber \\ 
&& + \frac{1}{4} e^{\varphi} V_{\mu  \nu } \mathcal{V}_{\alpha } \partial^{\alpha }\varphi -  \frac{1}{4} e^{\varphi} V_{\mu }{}^{\beta } \partial_{\beta }\mathcal{V}_{\nu } -  \frac{1}{4} e^{\varphi} \mathcal{V}^{\beta } \partial_{\mu }V_{\nu  \beta } + \frac{1}{2} \partial^{\alpha }\varphi \partial_{\mu }\hat{K}_{\nu  \alpha } \nonumber \\ 
&& -  \frac{1}{4} e^{\varphi} V_{\nu }{}^{\beta } \partial_{\mu }\mathcal{V}_{\beta } -  \frac{1}{4} e^{\varphi} V_{\nu }{}^{\alpha } \mathcal{V}_{\alpha } \partial_{\mu }\varphi + \frac{1}{2} \hat{K}_{\nu }{}^{\alpha } \partial_{\mu }\partial_{\alpha }\varphi \nonumber \\ 
&& + \frac{1}{8} e^{\varphi} V_{\mu }{}^{\alpha } \mathcal{V}_{\alpha } \partial_{\nu }\varphi -  \frac{1}{4} \partial_{\mu }\Theta \partial_{\nu }\varphi -  \frac{1}{4} \Theta \partial_{\nu }\partial_{\mu }\varphi  \Big) \nn\\
\nabla_{y}\nabla_{\mu} K_{\nu y} = &&\!\!\!\!\!\!\! e^{\varphi}\Big( \frac{1}{4} e^{\varphi} V_{\mu  \beta } V_{\nu  \alpha } \hat{K}^{\alpha  \beta } -  \frac{1}{4} e^{\varphi} V_{\alpha  \beta } V_{\mu }{}^{\beta } \hat{K}_{\nu }{}^{\alpha } + \frac{1}{8} e^{\varphi} V_{\mu  \nu } \mathcal{V}_{\alpha } \partial^{\alpha }\varphi \nonumber \\ 
&& -  \frac{1}{4} e^{\varphi} V_{\mu }{}^{\beta } \partial_{\beta }\mathcal{V}_{\nu } + \frac{1}{2} \partial^{\alpha }\varphi \partial_{\mu }\hat{K}_{\nu  \alpha } -  \frac{1}{4} e^{\varphi} V_{\nu }{}^{\beta } \partial_{\mu }\mathcal{V}_{\beta } -  \frac{1}{4} \hat{K}_{\nu  \alpha } \partial^{\alpha }\varphi \partial_{\mu }\varphi \nonumber \\ 
&& + \frac{1}{4} e^{\varphi} V_{\mu }{}^{\alpha } \mathcal{V}_{\alpha } \partial_{\nu }\varphi -  \frac{1}{4} \partial_{\mu }\Theta \partial_{\nu }\varphi + \frac{1}{8} \Theta \partial_{\mu }\varphi \partial_{\nu }\varphi \Big)\nn\\
\nabla_{y}\nabla_{y} K_{\mu \nu} = &&\!\!\!\!\!\!\! e^{\varphi}\Big(\frac{1}{2} e^{\varphi} V_{\mu  \alpha } V_{\nu  \beta } \hat{K}^{\alpha  \beta } -  \frac{1}{4} e^{\varphi} V_{\alpha  \beta } V_{\mu }{}^{\beta } \hat{K}^{\alpha }{}_{\nu } -  \frac{1}{4} e^{\varphi} V_{\alpha  \beta } V_{\nu }{}^{\beta } \hat{K}_{\mu }{}^{\alpha } \nonumber \\ 
&& + \frac{1}{2} \partial_{\alpha }\hat{K}_{\mu  \nu } \partial^{\alpha }\varphi + \frac{1}{4} e^{\varphi} V_{\nu }{}^{\alpha } \mathcal{V}_{\alpha } \partial_{\mu }\varphi -  \frac{1}{4} \hat{K}_{\alpha  \nu } \partial^{\alpha }\varphi \partial_{\mu }\varphi \nonumber \\ 
&& + \frac{1}{4} e^{\varphi} V_{\mu }{}^{\alpha } \mathcal{V}_{\alpha } \partial_{\nu }\varphi -  \frac{1}{4} \hat{K}_{\mu  \alpha } \partial^{\alpha }\varphi \partial_{\nu }\varphi + \frac{1}{4} \Theta \partial_{\mu }\varphi \partial_{\nu }\varphi \Big) \nn\\
\nabla_{\mu}\nabla_{y} K_{y y} = &&\!\!\!\!\!\!\! e^{2\varphi}\Big(\frac{1}{2} e^{\varphi} V_{\alpha  \beta } V_{\mu }{}^{\alpha } \mathcal{V}^{\beta } -  \frac{1}{4} V_{\mu  \beta } \hat{K}_{\alpha }{}^{\beta } \partial^{\alpha }\varphi -  \frac{1}{4} V_{\mu  \beta } \hat{K}^{\beta }{}_{\alpha } \partial^{\alpha }\varphi + \frac{1}{4} V_{\mu  \alpha } \Theta \partial^{\alpha }\varphi \nonumber \\ 
&& -  \frac{1}{4} V_{\mu }{}^{\beta } \partial_{\beta }\Theta + \frac{1}{2} \partial^{\beta }\varphi \partial_{\mu }\mathcal{V}_{\beta } + \frac{1}{4} \mathcal{V}_{\alpha } \partial^{\alpha }\varphi \partial_{\mu }\varphi + \frac{1}{2} \mathcal{V}^{\alpha } \partial_{\mu }\partial_{\alpha }\varphi \Big)\nn\\
\nabla_{y}\nabla_{\mu} K_{y y} = &&\!\!\!\!\!\!\! e^{2\varphi}\Big(\frac{1}{4} e^{\varphi} V_{\alpha  \beta } V_{\mu }{}^{\alpha } \mathcal{V}^{\beta } -  \frac{1}{4} V_{\mu  \beta } \hat{K}_{\alpha }{}^{\beta } \partial^{\alpha }\varphi -  \frac{1}{4} V_{\mu  \beta } \hat{K}^{\beta }{}_{\alpha } \partial^{\alpha }\varphi + \frac{1}{4} V_{\mu  \alpha } \Theta \partial^{\alpha }\varphi \nonumber \\ 
&& -  \frac{1}{4} V_{\mu }{}^{\beta } \partial_{\beta }\Theta + \frac{1}{2} \partial^{\beta }\varphi \partial_{\mu }\mathcal{V}_{\beta } \Big)\nn\\
\nabla_{y}\nabla_{y} K_{\mu y} = &&\!\!\!\!\!\!\! e^{2\varphi}\Big(\frac{1}{8} e^{\varphi} V_{\alpha  \beta } V_{\mu }{}^{\alpha } \mathcal{V}^{\beta } -  \frac{1}{2} V_{\mu  \beta } \hat{K}_{\alpha }{}^{\beta } \partial^{\alpha }\varphi -  \frac{1}{2} V_{\alpha  \beta } \hat{K}^{\beta }{}_{\mu } \partial^{\alpha }\varphi + \frac{1}{4} \partial_{\beta }\mathcal{V}_{\mu } \partial^{\beta }\varphi \nonumber \\ 
&& -  \frac{3}{8} \mathcal{V}_{\alpha } \partial^{\alpha }\varphi \partial_{\mu }\varphi \Big)\nn\\
\nabla_{y}\nabla_{y} K_{y y} = &&\!\!\!\!\!\!\! e^{2\varphi}\Big(- \frac{1}{2} e^{\varphi} V_{\alpha  \beta } \mathcal{V}^{\beta } \partial^{\alpha }\varphi -  \frac{1}{4} \Theta \partial_{\alpha }\varphi \partial^{\alpha }\varphi + \frac{1}{2} \hat{K}_{\alpha  \beta } \partial^{\alpha }\varphi \partial^{\beta }\varphi  + \frac{1}{4} \partial_{\beta }\Theta \partial^{\beta }\varphi \Big)\nn
\eeqa
where $ \hat{K}_{\mu\nu}= \partial_{\mu} n_{\nu}-n_{\mu}n^{\alpha}\partial_{\alpha} n_{\nu} $ is the 9-dimentional extrinsic curvature of the boundary in the base space. The indices on the right-hand side are contracted with base space metric $g^{\alpha\beta}$.  Using the above gauge invariant parts of the reductions and reductions \reef{red}, \reef{GB1}, \reef{GB2}, one can calculate reduction of various terms in \reef{b73}  along or orthogonal to the circle. For example, the reduction of $K^a{}_a$ and $K^i{}_i$  when O$_p$-plane is along the circle are 
\beqa
\tG^{\mu\nu}K_{\mu\nu}&=&\hat{K}^{\ta}{}_{\ta}+\frac{1}{2}  n^{\alpha } \nabla_{\alpha }\varphi \,\,\,;\,\,\,\bot^{\mu\nu }K_{\mu\nu}\,=\,\hat{K}^{i}{}_{i}
 \eeqa
 The reduction of these terms when $O_{(p-1)}$-plane is orthogonal to the circle are 
\beqa
\tG^{\mu\nu}K_{\mu\nu}&=&\hat{K}^{\ta}{}_{\ta}\,\,\,;\,\,\,\bot^{\mu\nu }K_{\mu\nu}\,=\,\hat{K}^{i}{}_{i}+\frac{1}{2}  n^{\alpha } \nabla_{\alpha }\varphi 
 \eeqa 
 Similarly one can calculate reduction of all  10-dimensional covariant terms in \reef{b73}. It is important to note that if one keeps all gauge invariant and non-gauge invariant terms in the reduction of tensors \reef{rK}, one  would find the same result for the reduction of 10-dimension covariant couplings.

Using the above reductions, then one can calculate $\prt S_{\rm eff}(\psi)$ which represents the reduction of the O$_{(p-1)}$-plane boundary action along the circle transverse to the O-plane, and $\prt S_{\rm eff}(\psi')$ which represents the T-duality transformation of the reduction of the O$_p$-plane boundary action along the circle tangent to the O-plane. The boundary term $\prt {\rm TD}$ in \reef{TT} is also given in \reef{PTD}.

We are free to add to the right-hand side of the constraint \reef{TT} the following total derivative terms in the boundary in the base space which are zero according to the Stokes's theorem (see Appendix):
\beqa
\int_{\prt M^{(p)}}d^{p-1}\tau \sqrt{|{\bg} |} n_{\ta}\prt_{\tb}(e^{-\phi}\mathcal{F}^{\ta\tb}) &=&0\labell{tot}
\eeqa
where  $ \mathcal{F}^{\ta\tb} $ is an arbitrary antisymmetric tensor constructed from the base space fields $n,\prt n,\prt\prt n$, $W^2$, $V^2$, $R$, $\prt\phi$, $\prt\prt\phi$ and $\bH^2$ at two-derivative order\footnote{Note that the antisymmetric tensor $\cF_{\ta\tb}$ should be constructed from the base space fields $n$, $\cV, \prt\cV,\Theta, \prt \Theta,\hat{K}, \prt\hat{K}$, $\cdots$. However, since $\prt\cV$ and $\prt\Theta$ include $\prt n$, one can consider only $n$, $\prt n$, $\prt\prt n$, $\cdots$.}, \ie
\beqa
\mathcal{F}_{\td\te}&=& y_{5}\ \Big(e^{\varphi} V_{\ta i} V_{\te}{}^{i} n^{\ta} n_{\td} -  e^{\varphi} V_{\ta i} V_{\td}{}^{i} n^{\ta} n_{\te}\Big)\ + y_{16}\ \Big(e^{- \varphi} n^{\ta} n_{\te} W_{\ta\tb} W_{\td}{}^{\tb} -  e^{- \varphi} n^{\ta} n_{\td} W_{\ta\tb} W_{\te}{}^{\tb}\Big)\ \nonumber \\ 
&& + y_{15}\ \Big(n^{\ta} n_{\te} \bH_{abi} \bH_{\td}{}^{bi} -  n^{\ta} n_{\td} \bH_{\ta\tb i} \bH_{\te}{}^{\tb i}\Big)\ + y_{3}\ \Big(n_{\te} \partial_{\ta}\partial^{\ta}n_{\td} -  n_{\td} \partial_{\ta}\partial^{\ta}n_{\te}\Big)\ \nonumber \\ 
&& + y_{2}\ \Big(n_{\te} \partial_{\ta}\partial_{\td}n^{\ta} -  n_{\td} \partial_{\ta}\partial_{\te}n^{\ta}\Big)\ + y_{1}\ \Big(n^{\ta} \partial_{\ta}\partial_{\td}n_{\te} -  n^{\ta} \partial_{\ta}\partial_{\te}n_{\td}\Big)\ \nonumber \\ 
&& + y_{45}\ \Big(n_{\te} \partial_{\ta}\varphi \partial^{\ta}n_{\td} -  n_{\td} \partial_{\ta}\varphi \partial^{\ta}n_{\te}\Big)\ + y_{44}\ \Big(n_{\te} \partial_{\ta}\phi \partial^{\ta}n_{\td} -  n_{\td} \partial_{\ta}\phi \partial^{\ta}n_{\te}\Big)\ \nonumber \\ 
&& + y_{14}\ \Big(n^{\ta} n_{\te} \partial_{\ta}n_{\td} \partial_{\tb}n^{\tb} -  n^{\ta} n_{\td} \partial_{\ta}n_{\te} \partial_{\tb}n^{\tb}\Big)\ + y_{13}\ \Big(n^{\ta} n^{\tb} n_{\te} \partial_{\ta}n_{\td} \partial_{\tb}\varphi -  n^{\ta} n^{\tb} n_{\td} \partial_{\ta}n_{\te} \partial_{\tb}\varphi\Big)\ \nonumber \\ 
&& + y_{12}\ \Big(n^{\ta} n^{\tb} n_{\te} \partial_{\ta}n_{\td} \partial_{\tb}\phi -  n^{\ta} n^{\tb} n_{\td} \partial_{\ta}n_{\te} \partial_{\tb}\phi\Big)\ + y_{11}\ \Big(n^{\ta} n^{\tb} n_{\te} \partial_{\tb}\partial_{\ta}n_{\td} -  n^{\ta} n^{\tb} n_{\td} \partial_{\tb}\partial_{\ta}n_{\te}\Big)\ \nonumber \\ 
&& + y_{10}\ \Big(n^{\ta} n^{\tb} n_{\te} \partial_{\tb}\partial_{\td}n_{\ta} -  n^{\ta} n^{\tb} n_{\td} \partial_{\tb}\partial_{\te}n_{\ta}\Big)\ + y_{9}\ \Big(n^{\ta} n_{\te} \partial_{\ta}n_{\tb} \partial^{\tb}n_{\td} -  n^{\ta} n_{\td} \partial_{\ta}n_{\tb} \partial^{\tb}n_{\te}\Big)\ \nonumber \\ 
&& + y_{42}\ \Big(\partial^{\ta}n_{\te} \partial_{\td}n_{\ta} -  \partial^{\ta}n_{\td} \partial_{\te}n_{\ta}\Big)\ + y_{41}\ \Big(n_{\te} \partial_{\ta}\varphi \partial_{\td}n^{\ta} -  n_{\td} \partial_{\ta}\varphi \partial_{\te}n^{\ta}\Big)\ \nonumber \\ 
&& + y_{40}\ \Big(n_{\te} \partial_{\ta}\phi \partial_{\td}n^{\ta} -  n_{\td} \partial_{\ta}\phi \partial_{\te}n^{\ta}\Big)\ + y_{8}\ \Big(n^{\ta} n_{\te} \partial_{\ta}n_{\tb} \partial_{\td}n^{\tb} -  n^{\ta} n_{\td} \partial_{\ta}n_{\tb} \partial_{\te}n^{\tb}\Big)\ \nonumber \\ 
&& + y_{39}\ \Big(\partial_{\ta}n^{\ta} \partial_{\td}n_{\te} -  \partial_{\ta}n^{\ta} \partial_{\te}n_{\td}\Big)\ + y_{38}\ \Big(n^{\ta} \partial_{\ta}\varphi \partial_{\td}n_{\te} -  n^{\ta} \partial_{\ta}\varphi \partial_{\te}n_{\td}\Big)\ \nonumber \\ 
&& + y_{37}\ \Big(n^{\ta} \partial_{\ta}\phi \partial_{\td}n_{\te} -  n^{\ta} \partial_{\ta}\phi \partial_{\te}n_{\td}\Big)\ + y_{36}\ \Big(n_{\te} \partial_{\ta}n^{\ta} \partial_{\td}\varphi -  n_{\td} \partial_{\ta}n^{\ta} \partial_{\te}\varphi\Big)\ \nonumber \\ 
&& + y_{35}\ \Big(n^{\ta} \partial_{\ta}n_{\te} \partial_{\td}\varphi -  n^{\ta} \partial_{\ta}n_{\td} \partial_{\te}\varphi\Big)\ + y_{34}\ \Big(n^{\ta} n_{\te} \partial_{\ta}\varphi \partial_{\td}\varphi -  n^{\ta} n_{\td} \partial_{\ta}\varphi \partial_{\te}\varphi\Big)\ \nonumber \\ 
&& + y_{33}\ \Big(n^{\ta} n_{\te} \partial_{\ta}\phi \partial_{\td}\varphi -  n^{\ta} n_{\td} \partial_{\ta}\phi \partial_{\te}\varphi\Big)\ + y_{31}\ \Big(n_{\te} \partial_{\ta}n^{\ta} \partial_{\td}\phi -  n_{\td} \partial_{\ta}n^{\ta} \partial_{\te}\phi\Big)\ \nonumber \\ 
&& + y_{30}\ \Big(n^{\ta} \partial_{\ta}n_{\te} \partial_{\td}\phi -  n^{\ta} \partial_{\ta}n_{\td} \partial_{\te}\phi\Big)\ + y_{29}\ \Big(n^{\ta} n_{\te} \partial_{\ta}\varphi \partial_{\td}\phi -  n^{\ta} n_{\td} \partial_{\ta}\varphi \partial_{\te}\phi\Big)\ \nonumber \\ 
&& + y_{28}\ \Big(n^{\ta} n_{\te} \partial_{\ta}\phi \partial_{\td}\phi -  n^{\ta} n_{\td} \partial_{\ta}\phi \partial_{\te}\phi\Big)\ + y_{32}\ \Big(\partial_{\td}\phi \partial_{\te}\varphi -  \partial_{\td}\varphi \partial_{\te}\phi\Big)\ \nonumber \\ 
&& + y_{26}\ \Big(n_{\te} \partial_{\td}\partial_{\ta}n^{\ta} -  n_{\td} \partial_{\te}\partial_{\ta}n^{\ta}\Big)\ + y_{25}\ \Big(n^{\ta} \partial_{\td}\partial_{\ta}n_{\te} -  n^{\ta} \partial_{\te}\partial_{\ta}n_{\td}\Big)\ \nonumber \\ 
&& + y_{24}\ \Big(n^{\ta} n_{\te} \partial_{\td}\partial_{\ta}\varphi -  n^{\ta} n_{\td} \partial_{\te}\partial_{\ta}\varphi\Big)\ + y_{23}\ \Big(n^{\ta} n_{\te} \partial_{\td}\partial_{\ta}\phi -  n^{\ta} n_{\td} \partial_{\te}\partial_{\ta}\phi\Big)\ \nonumber \\ 
&& + y_{6}\ \Big(n^{\ta} n^{\tb} n_{\te} \partial_{\td}\partial_{\tb}n_{\ta} -  n^{\ta} n^{\tb} n_{\td} \partial_{\te}\partial_{\tb}n_{\ta}\Big)\ + y_{22}\ \Big(n^{\ta} \partial_{\td}\partial_{\te}n_{\ta} -  n^{\ta} \partial_{\te}\partial_{\td}n_{\ta}\Big)\ \nonumber \\ 
&& + y_{27}\ \Big(n_{\te} \partial_{\td}\partial_{i}n^{i} -  n_{\td} \partial_{\te}\partial_{j}n^{j}\Big)\ + y_{43}\ \Big(n^{\ta} n_{\te} \partial_{\ta}n_{\td} \partial_{i}n^{i} -  n^{\ta} n_{\td} \partial_{\ta}n_{\te} \partial_{i}n^{i}\Big)\ \nonumber \\ 
&& + y_{19}\ \Big(\partial_{\td}n_{\te} \partial_{i}n^{i} -  \partial_{\te}n_{\td} \partial_{i}n^{i}\Big)\ + y_{18}\ \Big(n_{\te} \partial_{\td}\varphi \partial_{i}n^{i} -  n_{\td} \partial_{\te}\varphi \partial_{i}n^{i}\Big)\ \nonumber \\ 
&& + y_{7}\ \Big(n_{\te} \partial_{\td}\phi \partial_{i}n^{i} -  n_{\td} \partial_{\te}\phi \partial_{i}n^{i}\Big)\ + y_{21}\ \Big(n_{\te} \partial_{i}\partial^{i}n_{\td} -  n_{\td} \partial_{i}\partial^{i}n_{\te}\Big)\ \nonumber \\ 
&& + y_{20}\ \Big(n_{\te} \partial_{i}\partial_{\td}n^{i} -  n_{\td} \partial_{j}\partial_{\te}n^{j}\Big)
\eeqa
where $y_1,y_2,\cdots$ are arbitrary parameters.
After imposing the orientifold projection on \reef{tot}, we add it to the right-hand side of \reef{TT}.  Finally, one  should  write the couplings in the form of independent structures by imposing the Bianchi identities \reef{Hff} in the base space. Here again we write the field strengths $\bar{H}, V,W$ in terms of potentials $\bar{b}_{\mu\nu},g_\mu,b_\mu$ to satisfy the Bianchi identities automatically. Using the relation \reef{nf}, we also write the base space unit vector $n^\mu$ in terms of function $f$ to impose its  corresponding identities.

Writing all terms in the T-duality constraint  \reef{TT} in terms of independent  and non-gauge invariant structures, then one makes the coefficients of the independent structures which include the parameters of the gauge invariant Lagrangian \reef{b73}, $a_{28}$  and the arbitrary parameters in total derivative terms  \reef{tot}, to be zero. Unlike the bulk case,  not all parameters are fixed in terms of an overall factor. The linear equations in this case produce the following boundary Lagrangian which has $a_{28}$ and 17 other parameters:
\beqa
\prt \cL_p&\!\!\!\!\!=\!\!\!\!\!\!&   b_{50}\Big[H_{a}{}^{di} n^{a} n^{b} n^{c} \nabla_{c}H_{bdi}-2 K_{a}{}^{c} K^{ab} K_{bc} - 2 K_{i}{}^{k} K^{ij} K_{jk} - 2 K^{cd} n^{a} n^{b} R_{acbd} - 2 K^{ij} n^{a} n^{b} R_{aibj} \Big] \nonumber \\ 
&&+ b_{6} \Big[K^{a}{}_{a} K^{b}{}_{b} K^{c}{}_{c} - 3 K^{a}{}_{a} K^{b}{}_{b} K^{i}{}_{i} + 3 K^{a}{}_{a} K^{i}{}_{i} K^{j}{}_{j} -  K^{i}{}_{i} K^{j}{}_{j} K^{k}{}_{k}\Big] \nonumber \\ 
&& + b_{19} \Big[- \frac{4}{3} K_{a}{}^{c} K^{ab} K_{bc} + \frac{4}{3} K_{i}{}^{k} K^{ij} K_{jk} -  H_{ac}{}^{i} H_{bdi} K^{cd} n^{a} n^{b} + H_{a}{}^{c}{}_{i} H_{bcj} K^{ij} n^{a} n^{b}\Big] \nonumber \\ 
&& + b_{39} \Big[- K^{a}{}_{a} K^{b}{}_{b} K^{i}{}_{i} + 2 K^{a}{}_{a} K^{i}{}_{i} K^{j}{}_{j} -  K^{i}{}_{i} K^{j}{}_{j} K^{k}{}_{k}\nn\\
&&\qquad + K^{b}{}_{b} K^{c}{}_{c} n^{a} \nabla_{a}\phi - 2 K^{b}{}_{b} K^{i}{}_{i} n^{a} \nabla_{a}\phi  + K^{i}{}_{i} K^{j}{}_{j} n^{a} \nabla_{a}\phi\Big] \nonumber \\ 
&& + b_{23} \Big[K^{a}{}_{a} K_{bc} K^{bc} + K^{c}{}_{c} n^{a} n^{b} \mathcal{R}_{ab} -  K^{i}{}_{i} n^{a} n^{b} \mathcal{R}_{ab} + K^{i}{}_{i} n^{a} \nabla_{b}K_{a}{}^{b}\Big] \nonumber \\ 
&& + b_{55} \Big[n^{a} n^{b} \nabla_{b}\nabla_{a}K^{c}{}_{c} -  n^{a} n^{b} \nabla_{b}\nabla_{a}K^{i}{}_{i}\Big] \nn\\
&& + b_{46}\Big[- \nabla_{a}\nabla^{a}K^{i}{}_{i} -  \nabla_{a}K^{b}{}_{b} \nabla^{a}\phi + \nabla_{a}K^{i}{}_{i} \nabla^{a}\phi + \nabla_{b}\nabla^{b}K^{a}{}_{a}\Big] \nonumber \\ 
&& + b_{52}\Big[K^{a}{}_{a} K^{i}{}_{i} K^{j}{}_{j} -  \frac{2}{3} K^{i}{}_{i} K^{j}{}_{j} K^{k}{}_{k} - 2 K^{b}{}_{b} K^{i}{}_{i} n^{a} \nabla_{a}\phi\nn\\
&&\qquad  + K^{i}{}_{i} K^{j}{}_{j} n^{a} \nabla_{a}\phi + K^{c}{}_{c} n^{a} n^{b} \nabla_{a}\phi \nabla_{b}\phi -  \frac{1}{3} n^{a} n^{b} n^{c} \nabla_{a}\phi \nabla_{b}\phi \nabla_{c}\phi\Big] \nonumber \\ 
&& + b_{53} \Big[\frac{1}{3} K^{i}{}_{i} K^{j}{}_{j} K^{k}{}_{k} -  K^{i}{}_{i} K^{j}{}_{j} n^{a} \nabla_{a}\phi + K^{i}{}_{i} n^{a} n^{b} \nabla_{a}\phi \nabla_{b}\phi -  \frac{1}{3} n^{a} n^{b} n^{c} \nabla_{a}\phi \nabla_{b}\phi \nabla_{c}\phi\Big] \nonumber \\ 
&& + b_{34} \Big[- K_{i}{}^{k} K^{ij} K_{jk} -  K^{ij} n^{a} n^{b} R_{aibj} + K^{ij} n^{a} \nabla_{a}K_{ij} + n^{a} n^{b} \nabla_{b}\nabla_{i}K_{a}{}^{i} -  n^{a} n^{b} \nabla_{b}K_{ac} \nabla^{c}\phi\Big] \nonumber \\ 
&& + b_{32} \Big[- K^{ab} \mathcal{R}_{ab} + K^{i}{}_{i} n^{a} n^{b} \mathcal{R}_{ab} + K^{i}{}_{i} n^{a} \nabla_{a}K^{b}{}_{b} + K^{b}{}_{b} n^{a} \nabla_{a}K^{i}{}_{i} -  K^{i}{}_{i} n^{a} \nabla_{a}K^{j}{}_{j}\nn\\
&&\qquad  -  \frac{1}{2} n^{a} \nabla_{a}\mathcal{R}^{b}{}_{b} + \frac{1}{2} n^{a} \nabla_{a}\nabla_{b}\nabla^{b}\phi -  K^{i}{}_{i} n^{a} \nabla_{b}K_{a}{}^{b} + \nabla_{b}\nabla_{a}K^{ab} + n^{a} n^{b} \nabla_{b}\nabla_{a}K^{i}{}_{i}\nn\\
&&\qquad  + K^{ab} \nabla_{b}\nabla_{a}\phi -  K^{c}{}_{c} n^{a} n^{b} \nabla_{b}\nabla_{a}\phi -  \nabla_{b}\nabla^{b}K^{a}{}_{a} -  n^{a} n^{b} \nabla_{b}\nabla_{c}K_{a}{}^{c} \nn\\
&&\qquad + n^{a} n^{b} n^{c} \nabla_{c}\mathcal{R}_{ab} -  n^{a} n^{b} n^{c} \nabla_{c}\nabla_{b}\nabla_{a}\phi -  n^{a} n^{b} \nabla_{b}K_{ac} \nabla^{c}\phi\Big] \nonumber \\ 
&& + b_{70}\Big[ n^{a} n^{b} n^{c} n^{d} \nabla_{d}\nabla_{c}K_{ab}\Big]\labell{Lp1}\\
&&  + b_{16} \Big[2 K^{a}{}_{a} K_{bc} K^{bc} - 2 K^{i}{}_{i} K_{jk} K^{jk} + H_{a}{}^{di} H_{bdi} K^{c}{}_{c} n^{a} n^{b}\nn\\
&&\qquad  -  H_{a}{}^{cj} H_{bcj} K^{i}{}_{i} n^{a} n^{b} + 2 K^{i}{}_{i} n^{a} \nabla_{b}K_{a}{}^{b} - 2 K^{b}{}_{b} n^{a} \nabla_{i}K_{a}{}^{i}\Big]\nonumber \\ 
&& + b_{61} \Big[  2 K^{ab} \mathcal{R}_{ab} + 2 K^{ij} \mathcal{R}_{ij} + \frac{1}{2} H^{bci} n^{a} \nabla_{a}H_{bci} + \frac{1}{2} \nabla_{a}\nabla^{a}K^{i}{}_{i} -  \frac{1}{2} \nabla_{b}\nabla^{b}K^{a}{}_{a} \nn\\
&&\qquad + H_{a}{}^{ci} H_{bci} n^{a} \nabla^{b}\phi + H^{bci} n^{a} \nabla_{c}H_{abi} + H_{a}{}^{bi} n^{a} \nabla_{c}H_{b}{}^{c}{}_{i} + \frac{1}{2} \nabla_{i}\nabla^{i}K^{a}{}_{a} -  \frac{1}{2} \nabla_{j}\nabla^{j}K^{i}{}_{i} \nn\\
&&\qquad - H_{a}{}^{ci} H_{bci} K^{ab}-  \frac{1}{2} H^{ijk} n^{a} \nabla_{k}H_{aij}\Big] \nonumber \\ 
&& + b_{64} \Big[-2 K_{i}{}^{k} K^{ij} K_{jk} - 2 K^{ij} n^{a} n^{b} R_{aibj} + K^{ij} \mathcal{R}_{ij} + \frac{1}{4} H^{bci} n^{a} \nabla_{a}H_{bci} -  \frac{1}{2} n^{a} \nabla_{a}\mathcal{R}^{b}{}_{b}\nn\\
&&\qquad  + \frac{1}{4} \nabla_{a}\nabla^{a}K^{i}{}_{i} -  \frac{1}{2} n^{a} \nabla_{a}\nabla_{b}\nabla^{b}\phi -  \frac{1}{4} \nabla_{b}\nabla^{b}K^{a}{}_{a} + n^{a} \nabla_{b}\nabla_{a}\phi \nabla^{b}\phi -  K^{b}{}_{b} n^{a} \nabla_{i}K_{a}{}^{i} \nn\\
&&\qquad + n^{a} n^{b} \nabla_{a}\phi \nabla_{i}K_{b}{}^{i} + \frac{1}{4} \nabla_{i}\nabla^{i}K^{a}{}_{a} -  \frac{1}{4} \nabla_{j}\nabla^{j}K^{i}{}_{i} -  \frac{1}{4} H^{ijk} n^{a} \nabla_{k}H_{aij}\Big] \nonumber \\ 
&& + b_{65} \Big[  K^{a}{}_{a} K_{bc} K^{bc} -2 K_{a}{}^{c} K^{ab} K_{bc}- 2 K_{i}{}^{k} K^{ij} K_{jk} - 2 K^{cd} n^{a} n^{b} R_{acbd} - 2 K^{ij} n^{a} n^{b} R_{aibj} \nn\\
&&\qquad + K^{ab} \mathcal{R}_{ab} + K^{ij} \mathcal{R}_{ij} + \frac{1}{4} H^{bci} n^{a} \nabla_{a}H_{bci} + \frac{1}{4} \nabla_{a}\nabla^{a}(K^{i}{}_{i} - K^{b}{}_{b}) + n^{a} n^{b} \nabla_{a}\phi \nabla_{c}K_{b}{}^{c}\nn\\
&&\qquad  -  K^{b}{}_{b} n^{a} \nabla_{i}K_{a}{}^{i} + n^{a} n^{b} \nabla_{a}\phi \nabla_{i}K_{b}{}^{i} + \frac{1}{4} \nabla_{i}\nabla^{i}K^{a}{}_{a} -  \frac{1}{4} \nabla_{j}\nabla^{j}K^{i}{}_{i} -  \frac{1}{4} H^{ijk} n^{a} \nabla_{k}H_{aij}\Big] \nonumber \\ 
&& + b_{67} \Big[-2 K_{a}{}^{c} K^{ab} K_{bc} - 2 K_{i}{}^{k} K^{ij} K_{jk} - 2 K^{cd} n^{a} n^{b} R_{acbd} - 2 K^{ij} n^{a} n^{b} R_{aibj} + K^{ij} \mathcal{R}_{ij} \nn\\
&&\qquad + \frac{1}{4} H^{bci} n^{a} \nabla_{a}H_{bci}-  \frac{1}{2} n^{a} \nabla_{a}\mathcal{R}^{b}{}_{b} + \frac{1}{4} \nabla_{a}\nabla^{a}K^{i}{}_{i} + \frac{1}{2} n^{a} \nabla_{a}\nabla_{b}\nabla^{b}\phi + K^{ab} \nabla_{b}\nabla_{a}\phi\nn\\
&&\qquad  -  K^{c}{}_{c} n^{a} n^{b} \nabla_{b}\nabla_{a}\phi -  \frac{1}{4} \nabla_{b}\nabla^{b}K^{a}{}_{a} + n^{a} \mathcal{R}_{ab} \nabla^{b}\phi -  n^{a} \nabla_{b}\nabla_{a}\phi \nabla^{b}\phi + n^{a} n^{b} n^{c} \nabla_{c}\mathcal{R}_{ab} \nn\\
&&\qquad + n^{a} n^{b} n^{c} \nabla_{a}\phi \nabla_{c}\nabla_{b}\phi -  n^{a} n^{b} n^{c} \nabla_{c}\nabla_{b}\nabla_{a}\phi -  K^{b}{}_{b} n^{a} \nabla_{i}K_{a}{}^{i} + n^{a} n^{b} \nabla_{a}\phi \nabla_{i}K_{b}{}^{i} \nn\\
&&\qquad+ \frac{1}{4} \nabla_{i}\nabla^{i}K^{a}{}_{a}  -  \frac{1}{4} \nabla_{j}\nabla^{j}K^{i}{}_{i} -  \frac{1}{4} H^{ijk} n^{a} \nabla_{k}H_{aij}\Big] \nonumber \\ 
&& + a_{28} \Big[  9 H_{abj} H^{ab}{}_{i} K^{ij} -6 H_{a}{}^{ci} H_{bci} K^{ab}- 3 H_{i}{}^{kl} H_{jkl} K^{ij} + 48 K_{i}{}^{k} K^{ij} K_{jk} + 48 K^{ij} n^{a} n^{b} R_{aibj}\nn\\
&&\qquad  + 12 K^{ab} \mathcal{R}_{ab} - 36 K^{ij} \mathcal{R}_{ij} - 3 H^{bci} n^{a} \nabla_{a}H_{bci} - 3 \nabla_{a}\nabla^{a}K^{i}{}_{i} + 3 \nabla_{b}\nabla^{b}K^{a}{}_{a}  \nn\\
&&\qquad + 24 K^{b}{}_{b} n^{a} \nabla_{i}K_{a}{}^{i} - 24 n^{a} n^{b} \nabla_{a}\phi \nabla_{i}K_{b}{}^{i} - 3 \nabla_{i}\nabla^{i}K^{a}{}_{a} + 3 \nabla_{j}\nabla^{j}K^{i}{}_{i} + 3 H^{ijk} n^{a} \nabla_{k}H_{aij}\Big]\nn
\eeqa
where $ \mathcal{R}_{\mu\nu}=\tG^{\rho\sigma}{R}_{\rho\mu\sigma\nu}+\nabla_{\mu}\nabla_{\nu} \phi $. 
The above boundary Lagrangian is invariant under the T-duality for the above 18 parameters, \ie it satisfies  the T-duality constraint \reef{TT} for the following  base space boundary total derivative terms:
\beqa
\mathcal{F}_{\td\te}&=&b_{65}\ \Big(\frac{1}{2} e^{\varphi} V_{\ta i} V_{\te}{}^{i} n^{\ta} n_{\td} -  \frac{1}{2} e^{\varphi} V_{\ta i} V_{\td}{}^{i} n^{\ta} n_{\te} + \frac{1}{2} e^{-\varphi} n^{\ta} n_{\te} W_{\ta\tb} W_{\td}{}^{\tb} -  \frac{1}{2} e^{-\varphi} n^{\ta} n_{\td} W_{\ta\tb} W_{\te}{}^{\tb}\Big)\ \nonumber \\ 
&& + a_{28}\ \Big(-12 e^{\varphi} V_{\ta i} V_{\te}{}^{i} n^{\ta} n_{\td} + 12 e^{\varphi} V_{\ta i} V_{\td}{}^{i} n^{\ta} n_{\te} + 6 n^{\ta} n_{\te} \partial_{\ta}\varphi \partial_{\td}\varphi - 6 n^{\ta} n_{\td} \partial_{\ta}\varphi \partial_{\te}\varphi \Big)\ \nonumber \\ 
&& + b_{64}\ \Big(\frac{1}{2} e^{\varphi} V_{\ta i} V_{\te}{}^{i} n^{\ta} n_{\td} -  \frac{1}{2} e^{\varphi} V_{\ta i} V_{\td}{}^{i} n^{\ta} n_{\te} -  \frac{1}{4} n^{\ta} n_{\te} \partial_{\ta}\varphi \partial_{\td}\varphi + \frac{1}{4} n^{\ta} n_{\td} \partial_{\ta}\varphi \partial_{\te}\varphi \Big)\ \nonumber \\ 
&& + b_{67}\ \Big(\frac{1}{2} e^{\varphi} V_{\ta i} V_{\te}{}^{i} n^{\ta} n_{\td} -  \frac{1}{2} e^{\varphi} V_{\ta i} V_{\td}{}^{i} n^{\ta} n_{\te} -  \frac{1}{4} n^{\ta} n_{\te} \partial_{\ta}\varphi \partial_{\td}\varphi + \frac{1}{2} n^{\ta} n_{\te} \partial_{\td}\partial_{\ta}\varphi \nn\\
&&\qquad + \frac{1}{4} n^{\ta} n_{\td} \partial_{\ta}\varphi \partial_{\te}\varphi -  \frac{1}{2} n^{\ta} n_{\td} \partial_{\te}\partial_{\ta}\varphi \Big)
\eeqa
 The other multipletes in \reef{Lp1} are invariant under the T-duality without the total derivative terms in the boundary of base space.

The  Lagrangian \reef{Lp1} is not consistent with the  S-duality for all 18 parameters.  To have boundary couplings that their combinations with the bulk couplings \reef{bulkf} satisfy both the T-duality constraint \reef{TTS} and the S-duality constraint \reef{SS}, one has to consider a particular relations for the parameters.  In the next subsection we are going to find these relations. 

\subsection{S-duality constraint in the boundary}

As we mentioned before, the S-duality has a non-trivial constraint on the NS-NS couplings. In the Einstein frame, apart from the overall dilaton factor, there must be only even number of dilaton when B-field is zero. This constraint reduces the number of parameters in the T-duality invariant boundary Lagrangian \reef{Lp1}.

The overall factor $e^{-\phi}\sqrt{|\hg|}$ in the string frame action \reef{Genb} transforms to the following factor in the Einstein frame  $G_{\mu\nu}=e^{\phi/2}G_{\mu\nu}^E$ for the $O_3$-plane:
\beqa
e^{-\frac{1}{4}\phi}\sqrt{|\hg^E|}
\eeqa
On the other hand,   transformation of  the string frame Lagrangian $\cL_p$ to the Einstein frame Lagrangian has an overall dilaton factor $e^{-\frac{3}{4}\phi}$ for the gravity and dilaton couplings, \ie  $\cL_p(G,\phi)=e^{-\frac{3}{4}\phi}\cL_p^{E}(G^E,\phi)$. The couplings involving B-field, however, has another dilaton factor which is needed for making them to be invariant under the S-duality. Hence, the string frame boundary action \reef{Genb} transforms to the following Einstein frame Lagrangian for $O_3$-plane:
\beqa
\prt\!\! \bS_3&=&-\frac{T_3\pi^2\alpha'^2}{48}\int_{\prt M^{(4)}} d^{3}\tau\, e^{-\phi}\sqrt{|\hg^E|}\,\prt{\cal L}^E_3\labell{GenbE}
\eeqa
 To make the overall dilaton factor $e^{-\phi}$ to be invariant under the S-duality, one should include loop and non-perturbative effects \cite{Bachas:1999um}. It is straightforward to find the Lagrangian $\cL_3^E$ in the Einstein frame. One write each string frame term in \reef{Lp1} in terms of metric and its derivatives,  and then replaces  $G_{\mu\nu}=e^{\phi/2}G_{\mu\nu}^E$.  We then add to \reef{GenbE} the residual boundary terms in the bulk action, \ie the couplings \reef{PTD2}.  We are also free  to add   the following total derivative terms in the Einstein frame (see Appendix):
\beqa
\int_{\prt M^{(4)}}d^{3}\tau \sqrt{|\hg^E |} n^E_{a}\prt_{b}[e^{-\phi}(\mathcal{F}^E)^{ab}] &=&0\labell{tot1}
\eeqa
where  $ \mathcal{F}^{ab} $ is an arbitrary antisymmetric tensor constructed from the  boundary fields $n^E$, $\nabla n^E$, $\nabla\nabla n^E$, $R, \nabla R$, $\nabla\phi, \nabla\nabla\phi$ at two-derivative order, \ie
\beqa
\mathcal{F}^E_{de}=&&
  x_{2}\ \Big(n^{a} n_{e} R_{d}{}^{b}{}_{ab} -  n^{a} n_{d} R_{e}{}^{b}{}_{ab}\Big)\ \nonumber \\ 
&& + x_{3}\ \Big(n_{e} \nabla_{a}\nabla^{a}n_{d} -  n_{d} \nabla_{a}\nabla^{a}n_{e}\Big)\ + x_{4}\ \Big(n_{e} \nabla_{a}\phi \nabla^{a}n_{d} -  n_{d} \nabla_{a}\phi \nabla^{a}n_{e}\Big)\ \nonumber \\ 
&& + x_{5}\ \Big(n^{a} n_{e} \nabla_{a}n_{d} \nabla_{b}n^{b} -  n^{a} n_{d} \nabla_{a}n_{e} \nabla_{b}n^{b}\Big)\ \nonumber \\ 
&& + x_{6}\ \Big(n^{a} n^{b} n_{e} \nabla_{a}n_{d} \nabla_{b}\phi -  n^{a} n^{b} n_{d} \nabla_{a}n_{e} \nabla_{b}\phi\Big)\ \nonumber \\ 
&& + x_{7}\ \Big(n^{a} n^{b} n_{e} \nabla_{b}\nabla_{a}n_{d} -  n^{a} n^{b} n_{d} \nabla_{b}\nabla_{a}n_{e}\Big)\ \nonumber \\ 
&& + x_{8}\ \Big(n^{a} n_{e} \nabla_{a}n_{b} \nabla^{b}n_{d} -  n^{a} n_{d} \nabla_{a}n_{b} \nabla^{b}n_{e}\Big)\ \nonumber \\ 
&& + x_{9}\ \Big(n^{a} n_{e} \nabla_{b}n_{a} \nabla^{b}n_{d} -  n^{a} n_{d} \nabla_{b}n_{a} \nabla^{b}n_{e}\Big)\ \nonumber \\ 
&& + x_{10}\ \Big(n^{a} n^{b} n^{c} n_{e} \nabla_{a}n_{d} \nabla_{c}n_{b} -  n^{a} n^{b} n^{c} n_{d} \nabla_{a}n_{e} \nabla_{c}n_{b}\Big)\ \nonumber \\ 
&& + x_{11}\ \Big(\nabla^{a}n_{e} \nabla_{d}n_{a} -  \nabla^{a}n_{d} \nabla_{e}n_{a}\Big)\ + x_{12}\ \Big(n^{a} n_{e} \nabla_{b}n^{b} \nabla_{d}n_{a} -  n^{a} n_{d} \nabla_{b}n^{b} \nabla_{e}n_{a}\Big)\ \nonumber \\ 
&& + x_{13}\ \Big(n^{a} n^{b} n_{e} \nabla_{b}\phi \nabla_{d}n_{a} -  n^{a} n^{b} n_{d} \nabla_{b}\phi \nabla_{e}n_{a}\Big)\ \nonumber \\ 
&& + x_{14}\ \Big(n^{a} n^{b} n^{c} n_{e} \nabla_{c}n_{b} \nabla_{d}n_{a} -  n^{a} n^{b} n^{c} n_{d} \nabla_{c}n_{b} \nabla_{e}n_{a}\Big)\ \nonumber \\ 
&& + x_{15}\ \Big(n_{e} \nabla_{a}\phi \nabla_{d}n^{a} -  n_{d} \nabla_{a}\phi \nabla_{e}n^{a}\Big)\ \nonumber \\ 
&& + x_{16}\ \Big(n^{a} n^{b} \nabla_{a}n_{e} \nabla_{d}n_{b} -  n^{a} n^{b} \nabla_{a}n_{d} \nabla_{e}n_{b}\Big)\ \nonumber \\ 
&& + x_{17}\ \Big(n^{a} n_{e} \nabla_{a}n_{b} \nabla_{d}n^{b} -  n^{a} n_{d} \nabla_{a}n_{b} \nabla_{e}n^{b}\Big)\ \nonumber \\ 
&& + x_{18}\ \Big(n^{a} n_{e} \nabla_{b}n_{a} \nabla_{d}n^{b} -  n^{a} n_{d} \nabla_{b}n_{a} \nabla_{e}n^{b}\Big)\ \nonumber \\ 
&& + x_{19}\ \Big(\nabla_{a}n^{a} \nabla_{d}n_{e} -  \nabla_{a}n^{a} \nabla_{e}n_{d}\Big)\ + x_{20}\ \Big(n^{a} \nabla_{a}\phi \nabla_{d}n_{e} -  n^{a} \nabla_{a}\phi \nabla_{e}n_{d}\Big)\ \nonumber \\ 
&& + x_{21}\ \Big(n^{a} n^{b} \nabla_{b}n_{a} \nabla_{d}n_{e} -  n^{a} n^{b} \nabla_{b}n_{a} \nabla_{e}n_{d}\Big)\ \nonumber \\ 
&& + x_{22}\ \Big(n_{e} \nabla_{a}n^{a} \nabla_{d}\phi -  n_{d} \nabla_{a}n^{a} \nabla_{e}\phi\Big)\ + x_{23}\ \Big(n^{a} \nabla_{a}n_{e} \nabla_{d}\phi -  n^{a} \nabla_{a}n_{d} \nabla_{e}\phi\Big)\ \nonumber \\ 
&& + x_{24}\ \Big(n^{a} n_{e} \nabla_{a}\phi \nabla_{d}\phi -  n^{a} n_{d} \nabla_{a}\phi \nabla_{e}\phi\Big)\ \nonumber \\ 
&& + x_{25}\ \Big(n^{a} n^{b} n_{e} \nabla_{b}n_{a} \nabla_{d}\phi -  n^{a} n^{b} n_{d} \nabla_{b}n_{a} \nabla_{e}\phi\Big)\ \nonumber \\ 
&& + x_{26}\ \Big(n^{a} \nabla_{d}\phi \nabla_{e}n_{a} -  n^{a} \nabla_{d}n_{a} \nabla_{e}\phi\Big)\ + x_{27}\ \Big(n_{e} \nabla_{d}\nabla_{a}n^{a} -  n_{d} \nabla_{e}\nabla_{a}n^{a}\Big)\ \nonumber \\ 
&& + x_{28}\ \Big(n^{a} \nabla_{d}\nabla_{a}n_{e} -  n^{a} \nabla_{e}\nabla_{a}n_{d}\Big)\ + x_{29}\ \Big(n^{a} n_{e} \nabla_{d}\nabla_{a}\phi -  n^{a} n_{d} \nabla_{e}\nabla_{a}\phi\Big)\ \nonumber \\ 
&& + x_{30}\ \Big(n^{a} n^{b} n_{e} \nabla_{d}\nabla_{b}n_{a} -  n^{a} n^{b} n_{d} \nabla_{e}\nabla_{b}n_{a}\Big)\ \nonumber \\ 
&& + x_{31}\ \Big(n_{e} \nabla_{d}\nabla_{i}n^{i} -  n_{d} \nabla_{e}\nabla_{i}n^{i}\Big)\ + x_{32}\ \Big(n_{e} \nabla_{d}\nabla_{i}n^{i} -  n_{d} \nabla_{e}\nabla_{i}n^{i}\Big)\ \nonumber \\ 
&& + x_{33}\ \Big(n^{a} n_{e} \nabla_{a}n_{d} \nabla_{i}n^{i} -  n^{a} n_{d} \nabla_{a}n_{e} \nabla_{i}n^{i}\Big)\  + x_{34}\ \Big(n^{a} n_{e} \nabla_{d}n_{a} \nabla_{i}n^{i} -  n^{a} n_{d} \nabla_{e}n_{a} \nabla_{i}n^{i}\Big)\ \nonumber \\ 
&& + x_{35}\ \Big(\nabla_{d}n_{e} \nabla_{i}n^{i} -  \nabla_{e}n_{d} \nabla_{i}n^{i}\Big)\ + x_{36}\ \Big(n_{e} \nabla_{d}\phi \nabla_{i}n^{i} -  n_{d} \nabla_{e}\phi \nabla_{i}n^{i}\Big)
\eeqa
where $x_2,x_2,\cdots$ are arbitrary parameters. In the above relations $n^\mu$ is the Einstein frame unite vector.  After using the Einstein frame equations of motion \reef{eomE}, imposing orientifold projection, various Bianchi identities, and identities corresponding to the unit vector $n^\mu$,  we impose the condition that there must no odd number of dilation when B-field is zero. This fixes the parameter in the bulk total derivative \reef{PTD2} to be 
\beqa
\alpha=-4
\eeqa
The S-duality constraint also produces the following  4  multiplets in the string frame  which are   T-dual and S-dual invariant:
\beqa
\prt \cL_p&\!\!\!\!\!\!\!=\!\!\!\!\!\!\!& a_{28} \Big[-6 H_{a}{}^{ci} H_{bci} K^{ab} - 32 K_{a}{}^{c} K^{ab} K_{bc} + \frac{40}{3} K^{a}{}_{a} K_{bc} K^{bc} -  \frac{136}{27} K^{a}{}_{a} K^{b}{}_{b} K^{c}{}_{c}\nn\\
&& \qquad  + \frac{52}{9} K^{a}{}_{a} K^{b}{}_{b} K^{i}{}_{i} + 9 H_{abj} H^{ab}{}_{i} K^{ij} - 3 H_{i}{}^{kl} H_{jkl} K^{ij} - 16 K_{i}{}^{k} K^{ij} K_{jk} + \frac{32}{9} K^{a}{}_{a} K^{i}{}_{i} K^{j}{}_{j} \nn\\
&& \qquad  + 12 K^{i}{}_{i} K_{jk} K^{jk} -  \frac{116}{27} K^{i}{}_{i} K^{j}{}_{j} K^{k}{}_{k} - 6 H_{a}{}^{di} H_{bdi} K^{c}{}_{c} n^{a} n^{b} + 12 H_{ac}{}^{i} H_{bdi} K^{cd} n^{a} n^{b}  \nn\\
&& \qquad + 6 H_{a}{}^{cj} H_{bcj} K^{i}{}_{i} n^{a} n^{b} - 12 H_{a}{}^{c}{}_{i} H_{bcj} K^{ij} n^{a} n^{b} - 48 K^{cd} n^{a} n^{b} R_{acbd} + \frac{56}{3} K^{ab} \mathcal{R}_{ab} \nn\\
&& \qquad  + \frac{40}{3} K^{c}{}_{c} n^{a} n^{b} \mathcal{R}_{ab} - 8 K^{i}{}_{i} n^{a} n^{b} \mathcal{R}_{ab} - 24 K^{ij} \mathcal{R}_{ij} + \frac{16}{3} K^{i}{}_{i} n^{a} \nabla_{a}K^{b}{}_{b}+ \frac{16}{3} K^{b}{}_{b} n^{a} \nabla_{a}K^{i}{}_{i}   \nn\\
&& \qquad -  \frac{16}{3} K^{i}{}_{i} n^{a} \nabla_{a}K^{j}{}_{j} -  \frac{8}{3} n^{a} \nabla_{a}\mathcal{R}^{b}{}_{b} + \frac{28}{3} K^{b}{}_{b} K^{c}{}_{c} n^{a} \nabla_{a}\phi -  \frac{56}{3} K^{b}{}_{b} K^{i}{}_{i} n^{a} \nabla_{a}\phi  \nn\\
&& \qquad  + \frac{28}{3} K^{i}{}_{i} K^{j}{}_{j} n^{a} \nabla_{a}\phi - 16 \nabla_{a}\nabla^{a}K^{i}{}_{i} + \frac{8}{3} n^{a} \nabla_{a}\nabla_{b}\nabla^{b}\phi - 16 \nabla_{a}K^{b}{}_{b} \nabla^{a}\phi + 16 \nabla_{a}K^{i}{}_{i} \nabla^{a}\phi  \nn\\
&& \qquad  - 4 K^{i}{}_{i} n^{a} \nabla_{b}K_{a}{}^{b} + \frac{16}{3} \nabla_{b}\nabla_{a}K^{ab} -  \frac{32}{3} n^{a} n^{b} \nabla_{b}\nabla_{a}K^{c}{}_{c} + 16 n^{a} n^{b} \nabla_{b}\nabla_{a}K^{i}{}_{i} \nn\\
&& \qquad  -  \frac{16}{3} K^{c}{}_{c} n^{a} n^{b} \nabla_{b}\nabla_{a}\phi + \frac{32}{3} \nabla_{b}\nabla^{b}K^{a}{}_{a} -  \frac{16}{3} n^{a} n^{b} \nabla_{b}\nabla_{c}K_{a}{}^{c} + 12 H_{a}{}^{di} n^{a} n^{b} n^{c} \nabla_{c}H_{bdi}  \nn\\
&& \qquad  + 12 n^{a} n^{b} \nabla_{a}\phi \nabla_{c}K_{b}{}^{c} + \frac{16}{3} n^{a} n^{b} n^{c} \nabla_{c}\mathcal{R}_{ab} -  \frac{16}{3} n^{a} n^{b} n^{c} \nabla_{c}\nabla_{b}\nabla_{a}\phi -  \frac{16}{3} n^{a} n^{b} \nabla_{b}K_{ac} \nabla^{c}\phi   \nn\\
&& \qquad + 24 K^{b}{}_{b} n^{a} \nabla_{i}K_{a}{}^{i} - 12 n^{a} n^{b} \nabla_{a}\phi \nabla_{i}K_{b}{}^{i} + \frac{16}{3} K^{ab} \nabla_{b}\nabla_{a}\phi \Big] \nonumber \\ 
&& + b_{53} \Big[- \frac{1}{4} H_{a}{}^{ci} H_{bci} K^{ab} -  \frac{1}{3} K^{a}{}_{a} K_{bc} K^{bc} -  \frac{13}{81} K^{a}{}_{a} K^{b}{}_{b} K^{c}{}_{c} + \frac{5}{54} K^{a}{}_{a} K^{b}{}_{b} K^{i}{}_{i}   \nn\\
&& \qquad + \frac{8}{27} K^{a}{}_{a} K^{i}{}_{i} K^{j}{}_{j} -  \frac{1}{6} K^{i}{}_{i} K_{jk} K^{jk} + \frac{17}{162} K^{i}{}_{i} K^{j}{}_{j} K^{k}{}_{k} + \frac{1}{12} H_{a}{}^{di} H_{bdi} K^{c}{}_{c} n^{a} n^{b}   \nn\\
&& \qquad -  \frac{1}{12} H_{a}{}^{cj} H_{bcj} K^{i}{}_{i} n^{a} n^{b} + \frac{7}{18} K^{b}{}_{b} K^{c}{}_{c} n^{a} \nabla_{a}\phi -  \frac{7}{9} K^{b}{}_{b} K^{i}{}_{i} n^{a} \nabla_{a}\phi -  \frac{11}{18} K^{i}{}_{i} K^{j}{}_{j} n^{a} \nabla_{a}\phi  \nn\\
&& \qquad  + \frac{1}{6} K^{i}{}_{i} n^{a} \nabla_{b}K_{a}{}^{b} + K^{i}{}_{i} n^{a} n^{b} \nabla_{a}\phi \nabla_{b}\phi + \frac{1}{4} H_{a}{}^{ci} H_{bci} n^{a} \nabla^{b}\phi + \frac{1}{4} H^{bci} n^{a} \nabla_{c}H_{abi}   \nn\\
&& \qquad + \frac{1}{4} H_{a}{}^{bi} n^{a} \nabla_{c}H_{b}{}^{c}{}_{i} + \frac{1}{2} H_{a}{}^{di} n^{a} n^{b} n^{c} \nabla_{c}H_{bdi} -  \frac{1}{2} n^{a} n^{b} \nabla_{a}\phi \nabla_{c}K_{b}{}^{c}  \nn\\
&& \qquad -  \frac{1}{3} n^{a} n^{b} n^{c} \nabla_{a}\phi \nabla_{b}\phi \nabla_{c}\phi  + \frac{1}{3} K^{b}{}_{b} n^{a} \nabla_{i}K_{a}{}^{i} -  \frac{1}{2} n^{a} n^{b} \nabla_{a}\phi \nabla_{i}K_{b}{}^{i}\Big] \nonumber \\ 
&& + b_{52} \Big[\frac{1}{8} H_{a}{}^{ci} H_{bci} K^{ab} + \frac{1}{6} K^{a}{}_{a} K_{bc} K^{bc} + \frac{37}{162} K^{a}{}_{a} K^{b}{}_{b} K^{c}{}_{c} + \frac{19}{108} K^{a}{}_{a} K^{b}{}_{b} K^{i}{}_{i}   \nn\\
&& \qquad -  \frac{1}{27} K^{a}{}_{a} K^{i}{}_{i} K^{j}{}_{j} + \frac{1}{12} K^{i}{}_{i} K_{jk} K^{jk} -  \frac{11}{324} K^{i}{}_{i} K^{j}{}_{j} K^{k}{}_{k} -  \frac{1}{24} H_{a}{}^{di} H_{bdi} K^{c}{}_{c} n^{a} n^{b}  \nn\\
&& \qquad  + \frac{1}{24} H_{a}{}^{cj} H_{bcj} K^{i}{}_{i} n^{a} n^{b} -  \frac{31}{36} K^{b}{}_{b} K^{c}{}_{c} n^{a} \nabla_{a}\phi -  \frac{5}{18} K^{b}{}_{b} K^{i}{}_{i} n^{a} \nabla_{a}\phi + \frac{5}{36} K^{i}{}_{i} K^{j}{}_{j} n^{a} \nabla_{a}\phi   \nn\\
&& \qquad -  \frac{1}{12} K^{i}{}_{i} n^{a} \nabla_{b}K_{a}{}^{b} + K^{c}{}_{c} n^{a} n^{b} \nabla_{a}\phi \nabla_{b}\phi -  \frac{1}{8} H_{a}{}^{ci} H_{bci} n^{a} \nabla^{b}\phi -  \frac{1}{8} H^{bci} n^{a} \nabla_{c}H_{abi}   \nn\\
&& \qquad -  \frac{1}{8} H_{a}{}^{bi} n^{a} \nabla_{c}H_{b}{}^{c}{}_{i} -  \frac{1}{4} H_{a}{}^{di} n^{a} n^{b} n^{c} \nabla_{c}H_{bdi} + \frac{1}{4} n^{a} n^{b} \nabla_{a}\phi \nabla_{c}K_{b}{}^{c}  \nn\\
&& \qquad  -  \frac{1}{3} n^{a} n^{b} n^{c} \nabla_{a}\phi \nabla_{b}\phi \nabla_{c}\phi -  \frac{1}{6} K^{b}{}_{b} n^{a} \nabla_{i}K_{a}{}^{i} + \frac{1}{4} n^{a} n^{b} \nabla_{a}\phi \nabla_{i}K_{b}{}^{i}\Big] \nonumber \\ 
&& + b_{67} \Big[ -  \frac{11}{18} K^{a}{}_{a} K_{bc} K^{bc} -  \frac{1}{54} K^{a}{}_{a} K^{b}{}_{b} K^{c}{}_{c} + \frac{1}{36} K^{a}{}_{a} K^{b}{}_{b} K^{i}{}_{i} + \frac{1}{12} K^{i}{}_{i} K_{jk} K^{jk}  \nn\\
&& \qquad  -  \frac{1}{108} K^{i}{}_{i} K^{j}{}_{j} K^{k}{}_{k} -  \frac{1}{24} H_{a}{}^{di} H_{bdi} K^{c}{}_{c} n^{a} n^{b} + \frac{1}{24} H_{a}{}^{cj} H_{bcj} K^{i}{}_{i} n^{a} n^{b} -  \frac{8}{9} K^{ab} \mathcal{R}_{ab}   \nn\\
&& \qquad + \frac{2}{9} K^{c}{}_{c} n^{a} n^{b} \mathcal{R}_{ab} -  \frac{1}{3} K^{i}{}_{i} n^{a} n^{b} \mathcal{R}_{ab} -  \frac{1}{9} K^{i}{}_{i} n^{a} \nabla_{a}K^{b}{}_{b} -  \frac{1}{9} K^{b}{}_{b} n^{a} \nabla_{a}K^{i}{}_{i}  \nn\\
&& \qquad  + \frac{1}{9} K^{i}{}_{i} n^{a} \nabla_{a}K^{j}{}_{j} -  \frac{4}{9} n^{a} \nabla_{a}\mathcal{R}^{b}{}_{b} + \frac{1}{36} K^{b}{}_{b} K^{c}{}_{c} n^{a} \nabla_{a}\phi -  \frac{1}{18} K^{b}{}_{b} K^{i}{}_{i} n^{a} \nabla_{a}\phi   \nn\\
&& \qquad + \frac{1}{36} K^{i}{}_{i} K^{j}{}_{j} n^{a} \nabla_{a}\phi + \frac{1}{3} \nabla_{a}\nabla^{a}K^{i}{}_{i} + \frac{4}{9} n^{a} \nabla_{a}\nabla_{b}\nabla^{b}\phi + \frac{1}{3} \nabla_{a}K^{b}{}_{b} \nabla^{a}\phi   \nn\\
&& \qquad -  \frac{1}{3} \nabla_{a}K^{i}{}_{i} \nabla^{a}\phi + \frac{1}{4} K^{i}{}_{i} n^{a} \nabla_{b}K_{a}{}^{b} -  \frac{1}{9} \nabla_{b}\nabla_{a}K^{ab} + \frac{2}{9} n^{a} n^{b} \nabla_{b}\nabla_{a}K^{c}{}_{c}   \nn\\
&& \qquad + \frac{8}{9} K^{ab} \nabla_{b}\nabla_{a}\phi -  \frac{8}{9} K^{c}{}_{c} n^{a} n^{b} \nabla_{b}\nabla_{a}\phi -  \frac{2}{9} \nabla_{b}\nabla^{b}K^{a}{}_{a} + \frac{1}{9} n^{a} n^{b} \nabla_{b}\nabla_{c}K_{a}{}^{c}  \nn\\
&& \qquad  -  \frac{1}{8} H_{a}{}^{ci} H_{bci} n^{a} \nabla^{b}\phi + n^{a} \mathcal{R}_{ab} \nabla^{b}\phi -  n^{a} \nabla_{b}\nabla_{a}\phi \nabla^{b}\phi -  \frac{1}{8} H^{bci} n^{a} \nabla_{c}H_{abi}   \nn\\
&& \qquad -  \frac{1}{8} H_{a}{}^{bi} n^{a} \nabla_{c}H_{b}{}^{c}{}_{i} -  \frac{1}{4} H_{a}{}^{di} n^{a} n^{b} n^{c} \nabla_{c}H_{bdi} -  \frac{3}{4} n^{a} n^{b} \nabla_{a}\phi \nabla_{c}K_{b}{}^{c} + \frac{8}{9} n^{a} n^{b} n^{c} \nabla_{c}\mathcal{R}_{ab}  \nn\\
&& \qquad  + n^{a} n^{b} n^{c} \nabla_{a}\phi \nabla_{c}\nabla_{b}\phi -  \frac{8}{9} n^{a} n^{b} n^{c} \nabla_{c}\nabla_{b}\nabla_{a}\phi + \frac{1}{9} n^{a} n^{b} \nabla_{b}K_{ac} \nabla^{c}\phi -  \frac{1}{6} K^{b}{}_{b} n^{a} \nabla_{i}K_{a}{}^{i}   \nn\\
&& \qquad + \frac{1}{4} n^{a} n^{b} \nabla_{a}\phi \nabla_{i}K_{b}{}^{i}+\frac{1}{8} H_{a}{}^{ci} H_{bci} K^{ab}-  \frac{1}{3} n^{a} n^{b} \nabla_{b}\nabla_{a}K^{i}{}_{i} \Big]
\labell{t12}
\eeqa
The form of the couplings are not unique. If one chooses another scheme for the independent couplings in \reef{b73}, then the form of the above four multiplets would be changed. In other words, by using the total derivative terms and various Banchi identities and the identities corresponding to the unite vector $n^\mu$, one can rewrite the above couplings in various other forms. However, there is always four multiplets. 

\section{Discussion}

In this paper we have shown that imposing the gauge symmetry on the world-volume couplings of O$_p$-plane in type II superstring theories, one finds at least 48 independent NS-NS couplings with arbitrary coefficients. We then reduce the theory on a circle to impose the T-duality on these couplings. We find that the T-duality constraint  fixes all 48 parameters in terms of one overall factor. The T-duality, however,  is not fully satisfied because one finds some total derivative terms in the base space if the O$_p$-plane is extended to the boundary. We have also shown that the bulk couplings that the gauge symmetry and the T-duality fix are consistent with the S-duality, again, up to some total derivative terms.  Using the Stokes's theorem, one realizes that the presence of the residual total derivative terms  dictates that there must be some couplings on the boundary of O$_p$-plane as well. 

We have shown that imposing the gauge symmetry on the couplings in the boundary of O$_p$-plane, one finds at least 78 independent  NS-NS couplings with arbitrary coefficients. We then impose the T-duality on these couplings and add the residual total derivative terms from the T-duality of bulk couplings. The T-duality then fixes the 78 parameters in terms of  the overall factor of the bulk couplings and 17 other  parameters. We  then impose the S-duality constraint on the remaining couplings and add the residual total derivative terms from the S-duality of the bulk couplings. The S-duality finally fixes  the boundary couplings up to 3 parameters and up to the overall factor of the bulk couplings. The final result for the bulk and boundary couplings in the string frame are
\beqa
\bS_p+\prt\!\! \bS_p&=&-\frac{T_p\pi^2\alpha'^2}{48}\Big[\int_{M^{(p+1)}} d^{p+1}\sigma\, e^{-\phi}\sqrt{-\tg}\,{\cal L}_p+\int_{\prt M^{(p+1)}} d^{p}\tau\, e^{-\phi}\sqrt{|\hg|}\,\prt{\cal L}_p\Big]\labell{Genf}
\eeqa
where $\cL_p$ is
\beqa
{\cal L}_p&\!\!\!\!\!\!\!\!\!\!\!\!\!=\!\!\!\!\!\!\!\!\!\!\!\!\!\!\!&a_{28}\Big[- \frac{3}{4} H_{a}{}^{cj} H^{abi} H_{b}{}^{d}{}_{j} H_{cdi} -  \frac{3}{2} H_{ab}{}^{j} H^{abi} H_{cdj} H^{cd}{}_{i} + H_{a}{}^{cj} H^{abi} H_{bc}{}^{k} H_{ijk} \nonumber \\ 
&& + \frac{3}{2} H_{ab}{}^{j} H^{abi} H_{i}{}^{kl} H_{jkl} -  \frac{1}{4} H_{i}{}^{lm} H^{ijk} H_{jl}{}^{n} H_{kmn}  + 6 H^{abi} H^{cd}{}_{i} R_{abcd} \nonumber \\ 
&&- 6 H^{abi} H_{i}{}^{jk} R_{abjk} - 6 R_{abcd} R^{abcd} + 6 R_{abij} R^{abij} - 6 H_{a}{}^{ci} H_{bci} \mathcal{R}^{ab} \nonumber \\ 
&& + 12 \mathcal{R}_{ab} \mathcal{R}^{ab} + 9 H_{abj} H^{ab}{}_{i} \mathcal{R}^{ij} - 3 H_{i}{}^{kl} H_{jkl} \mathcal{R}^{ij} - 12 \mathcal{R}_{ij} \mathcal{R}^{ij} \nonumber \\ 
&& + \nabla_{a}H_{ijk} \nabla^{a}H^{ijk} - 3 \nabla_{c}H_{abi} \nabla^{c}H^{abi} + 2 \nabla_{i}H_{abc} \nabla^{i}H^{abc} \Big]\labell{bulkf1}
\eeqa
 The boundary Lagrangian 
 $\prt \cL_p$ is given in \reef{t12}. The bulk Lagrangian  $\cL_p$ is consistent  with linear T-duality without using total derivative terms, however, the boundary Lagrangian is not. Since one is free to add total derivative terms to the boundary, one can write  \reef{t12} in other forms as well. The form of the boundary Lagrangian which is consistent with the linear T-duality is  
\beqa
\prt{\cal L}_p&\!\!\!\!\!\!\!\!\!\!\!\!\!=\!\!\!\!\!\!\!\!\!\!\!\!\!\!\!& a_{28} \Big[16 \mathcal{D}_{a}\nabla^{a}\bK -  \frac{136}{27} \bK^3  -  \frac{28}{3} \bK^2\cK - 6 H_{a}{}^{ci} H_{bci} K^{ab} + 8 \bK K_{ab} K^{ab} + 12 \cK K_{ab} K^{ab} \nn\\
&&\qquad - 32 K_{a}{}^{c} K^{ab} K_{bc} + 9 H_{abj} H^{ab}{}_{i} K^{ij} - 3 H_{i}{}^{kl} H_{jkl} K^{ij} - 24 \bK K_{ij} K^{ij} - 12 \cK K_{ij} K^{ij}  \nn\\
&&\qquad- 16 K_{i}{}^{k} K^{ij} K_{jk}  - 6 H_{a}{}^{ci} H_{bci} \bK n^{a} n^{b} + 12 H_{ac}{}^{i} H_{bdi} K^{cd} n^{a} n^{b} - 12 H_{a}{}^{c}{}_{i} H_{bcj} K^{ij} n^{a} n^{b}  \nn\\
&&\qquad - 48 K^{cd} n^{a} n^{b} R_{acbd} + 24 K^{ab} \mathcal{R}_{ab} + 8 \bK n^{a} n^{b} \mathcal{R}_{ab} - 24 K^{ij} \mathcal{R}_{ij}   -  \frac{16}{3} \bK n^{a} \nabla_{a}\bK  \nn\\
&&\qquad- 16 n^{a} n^{b} \nabla_{b}\nabla_{a}\bK + 12 H_{a}{}^{di} n^{a} n^{b} n^{c} \nabla_{c}H_{bdi}\Big]\nn\\
&&+ b_{52} \Big[\frac{37}{162} \bK^3 + \frac{31}{36} \bK^2\cK + \bK\cK^2 + \frac{1}{3} \cK^3  + \frac{1}{8} H_{a}{}^{ci} H_{bci} K^{ab} + \frac{1}{6} \bK K_{ab} K^{ab} + \frac{1}{4} \cK K_{ab} K^{ab}  \nn\\
&&\qquad+ \frac{1}{6} \bK K_{ij} K^{ij}   + \frac{1}{4} \cK K_{ij} K^{ij} -  \frac{1}{24} H_{a}{}^{ci} H_{bci} \bK n^{a} n^{b} -  \frac{1}{8} H^{bci} n^{a} \nabla_{c}H_{abi}  \nn\\
&&\qquad -  \frac{1}{8} H_{a}{}^{bi} n^{a} \mathcal{D}_{c}H_{b}{}^{c}{}_{i} -  \frac{1}{4} H_{a}{}^{di} n^{a} n^{b} n^{c} \nabla_{c}H_{bdi}\Big] \\ 
&& + b_{53} \Big[- \frac{13}{81} \bK^3 -  \frac{7}{18} \bK^2\cK  + \frac{1}{3} \cK^3 -  \frac{1}{4} H_{a}{}^{ci} H_{bci} K^{ab} -  \frac{1}{3} \bK K_{ab} K^{ab}  -  \frac{1}{2} \cK K_{ab} K^{ab} \nn\\
&&\qquad -  \frac{1}{3}\bK K_{ij} K^{ij} -  \frac{1}{2} \cK K_{ij} K^{ij}   + \frac{1}{12} H_{a}{}^{ci} H_{bci} \bK n^{a} n^{b}  + \frac{1}{4} H^{bci} n^{a} \nabla_{c}H_{abi}  \nn\\
&&\qquad+ \frac{1}{4} H_{a}{}^{bi} n^{a} \mathcal{D}_{c}H_{b}{}^{c}{}_{i}   + \frac{1}{2} H_{a}{}^{di} n^{a} n^{b} n^{c} \nabla_{c}H_{bdi}\Big]\nonumber \\ 
&&+ b_{67} \Big[-\frac{1}{54}\bK^3 -  \frac{1}{36} \bK^2\cK -  \frac{1}{2} \bK K_{ab} K^{ab} -  \frac{3}{4} \cK K_{ab} K^{ab} + \frac{1}{6} \bK K_{ij} K^{ij}+ \frac{1}{4} \cK K_{ij} K^{ij} \nn\\
&&\qquad -  \frac{2}{3} \bK n^{a} n^{b} \mathcal{R}_{ab} -  \cK n^{a} n^{b} \mathcal{R}_{ab}  -  \frac{2}{9} \bK n^{a} \nabla_{a}\bK -  \frac{1}{3} \cK n^{a} \nabla_{a}\bK
 +\frac{1}{8} H_{a}{}^{ci} H_{bci} K^{ab}\nn\\&&\qquad -  \frac{1}{24} H_{a}{}^{ci} H_{bci} \bar{K} n^{a} n^{b} -  \frac{1}{8} H^{bci} n^{a} \nabla_{c}H_{abi} -  \frac{1}{8} H_{a}{}^{bi} n^{a} \mathcal{D}_{c}H_{b}{}^{c}{}_{i} -  \frac{1}{4} H_{a}{}^{di} n^{a} n^{b} n^{c} \nabla_{c}H_{bdi} \Big]\nn
\eeqa
where ${\cal D}_a\equiv \nabla_a-\nabla_a\Phi$ is dilaton-derivative and  $ \mathcal{R}_{\mu\nu}\equiv\tG^{\rho\sigma}{R}_{\rho\mu\sigma\nu}+\nabla_{\mu}\nabla_{\nu} \phi $ is dilaton-Riemann which are consistent with the linear T-duality \cite{Mashhadi:2020mzf}. We have also defined $\bK\equiv K^a{}_a-K^i{}_i$ and $\cK\equiv K^i{}_i-n^a\nabla_a\phi$ which are also invariant under the linear T-duality. Note that while there is derivative of Riemann curvature in \reef{t12}, this term has been cancelled in the above form of the boundary couplings by using the boundary total derivative terms.  The  gauge invariant action \reef{Genf} is fully invariant under T-duality and is consistent with S-duality up to some  terms in the boundary of boundary which are zero. There are four parameters in the above action, \ie $a_{28}$, $b_{52}$, $b_{53}$, $b_{67}$. For $a_{28}=-\frac{1}{6}$, the bulk couplings are consistent with the $PR_2$-level or disk-level S-matrix element of two NS-NS vertex operator \cite{Garousi:2006zh,Bachas:1999um}. 

At the leading order of $\alpha'$, the bulk action is given by the DBI action and there is no boundary couplings. However, the  action \reef{Genf} indicates that at order $\alpha'^2$, there are more couplings in the boundary than in the bulk. One may expect that this is a general feature  of boundary couplings at the higher orders of derivative. Moreover, a general feature of higher derivative couplings is that they are depend on the scheme \cite{Metsaev:1987zx}. The metric couplings in the bosonic string theory in a particular  scheme  at order $\alpha'$ is given by the Gauss-Bonnet couplings.  The corresponding boundary couplings have been found in \cite{Myers:1987yn}. It is known how  to include the B-field and dilaton to the bulk couplings \cite{Metsaev:1987zx,Garousi:2019wgz}, however, it is not known how to include these fields in the boundary. It would be interesting to use the gauge symmetry and T-duality constraint to find the boundary action in the bosonic string theory at order $\alpha'$ which includes the metric, B-field and dilaton,  as in \reef{Genf}.

{\bf Acknowledgments}: We would like to thank A. Ghodsi for useful discussions. This work is supported by Ferdowsi University of Mashhad under grant  3/41774(1395/07/13).

\vskip 0.8 cm

{\Large \bf Appendix: Stokes's theorem}

\vskip 0.5 cm

In this appendix we use  the Stokes's theorem to find the formulas \reef{Stokes} and \reef{bstokes} that we have used in this paper (see Appendix E in \cite{Carroll:2004st} for more details). For an $D$-dimensional spacetime  manifold $M$ with boundary $\prt M$, the Stokes's theorem is the following:
\beqa
\int_Md\omega^{(D-1)}&=&\int_{\prt M}\omega^{(D-1)}\labell{Apen1}
\eeqa
where $\omega^{(D-1)}$ is an arbitrary $(D-1)$-form. 
If one chooses it as $\omega^{(D-1)}=*A^{(1)}$ where $A^{(1)}$ is a one-form,  and uses $x^0,\cdots, x^{D-1}$ as the spacetime coordinates, then in terms of $x$-components, one  has
\beqa
\omega_{\mu_1\cdots\mu_{D-1}}&=&A^{\nu}\epsilon_{\mu_1\cdots\mu_{D-1}\nu}\nn\\
(d\omega)_{\lambda\mu_1\cdots\mu_{D-1}}&=&\nabla_{[\lambda}A^{\nu}\epsilon_{\mu_1\cdots\mu_{D-1}]\nu}
\eeqa
where $\epsilon^{(D)}$ is the volume-form of the spacetime manifold $M$. On the other hand, since $d\omega^{(D-1)}$ is an $D$-form, it can be written as $d\omega^{(D-1)}=h\epsilon^{(D)}=*h$ where $h$ is a $0$-form. Using the fact that $**=1$, one finds the function $h$ is $h=**h=*d\omega^{(D-1)}=\epsilon^{\lambda\mu_1\cdots\mu_{D-1}}\nabla_{[\lambda}A^{\nu}\epsilon_{\mu_1\cdots\mu_{D-1}]\nu}$. Using the contraction of two volume-forms, one finds $h=\nabla_\nu A^\nu$. Using the fact that the volume-form in terms of $x$-coordinates  is $\epsilon^{(D)}=\sqrt{|G|}d^Dx$, one can write the integrand on the left-hand side of \reef{Apen1} as
\beqa
d\omega^{(D-1)}=\nabla_\nu A^\nu\sqrt{|G|}d^Dx\labell{do}
\eeqa
On the right-hand side of \reef{Apen1}, one can write $\omega^{(D-1)}$ in terms of the volume-form of the boundary space, \ie $\omega^{(D-1)}=g\hat{\epsilon}^{(D-1)}=\hat{*}g$. Using the fact that $\hat{*}\hat{*}=1$, one can write $g=\hat{*}\hat{*} g=\hat{*}\omega^{(D-1)}$. Then using the relation between $x$-components of volume-forms $\epsilon^{(D)}$ and $\hat{\epsilon}^{(D-1)}$, \ie $\hat{\epsilon}^{\mu_1\cdots\mu_{D-1}}=n_{\lambda}\epsilon^{\lambda\mu_1\cdots\mu_{D-1}}$ where $n^\mu$ is unite vector orthogonal to the boundary,  one finds the $0$-form $g$ to be $g=n_\lambda A^\lambda$. On the other hand, using the fact that the boundary volume-form is $\hat{\epsilon}^{(D-1)}=\sqrt{|\gamma|}d^{D-1}y$ where $\gamma$ is determinate of induced metric on the boundary and the boundary with coordinates $y^0,\cdots, y^{D-2}$ is specified by the functions $x^\mu=x^\mu(y)$, one can write the integrand  on the right-hand side of \reef{Apen1} as 
\beqa
\omega^{(D-1)}&=&n_\nu A^\nu \sqrt{|\gamma|}d^{D-1}y\labell{o}
\eeqa
Replacing \reef{do} and \reef{o} in \reef{Apen1}, one finds the Stokes's theorem in terms of $x$-components is 
\beqa
\int_M\nabla_\nu A^\nu\sqrt{|G|}d^Dx&=&\int_{\prt M}n_\nu A^\nu \sqrt{|\gamma|}d^{D-1}y\labell{Apen2}
\eeqa
This is the formula that we have used in \reef{Stokes}.

For the boundary $\prt M$, the Stokes's theorem is the following:
\beqa
\int_{\prt M}d\Omega^{(D-2)}&=&\int_{\prt\prt M}\Omega^{(D-2)}\labell{Apen3}
\eeqa
where $\Omega^{(D-2)}$ is an arbitrary $(D-2)$-form. Since boundary of boundary is zero, \ie $\prt\prt M=0$, the right-hand side this time is zero. 

If one chooses  $\Omega^{(D-2)}=*F^{(2)}$ where $F^{(2)}$ is a two-form, then in terms of $x$-components, one  has
\beqa
\Omega_{\mu_1\cdots\mu_{D-2}}&=&F^{\alpha\beta}\epsilon_{\mu_1\cdots\mu_{D-2}\alpha\beta}\nn\\
(d\Omega)_{\lambda\mu_1\cdots\mu_{D-2}}&=&\nabla_{[\lambda}F^{\alpha\beta}\epsilon_{\mu_1\cdots\mu_{D-2}]\alpha\beta}
\eeqa
Since $d\Omega^{(D-2)}$ is an $(D-1)$-form, it can be written as $d\Omega^{(D-2)}=k\hat{\epsilon}^{(D-1)}=\hat{*}k$ where $k$ is a $0$-form. Using the fact that $\hat{*}\hat{*}=1$, one finds the function $k$ is $k=\hat{*}\hat{*}k=\hat{*}d\Omega^{(D-2)}=\hat{\epsilon}^{\lambda\mu_1\cdots\mu_{D-2}}\nabla_{[\lambda}F^{\alpha\beta}\epsilon_{\mu_1\cdots\mu_{D-1}]\alpha\beta}$. Then using the relation between $x$-components of volume-forms $\epsilon^{(D)}$ and $\hat{\epsilon}^{(D-1)}$, \ie $\hat{\epsilon}^{\mu_1\cdots\mu_{D-1}}=n_{\lambda}\epsilon^{\lambda\mu_1\cdots\mu_{D-1}}$, and using the contraction of two volume-forms, one finds $k=n_{\alpha}\nabla_\beta F^{\alpha\beta}$. On the other hand, using the relation for the  boundary volume-form  $\hat{\epsilon}^{(D-1)}=\sqrt{|\gamma|}d^{D-1}y$, one can write the Stokes's theorem in the boundary as 
\beqa
\int_{\prt M}n_\alpha\nabla_\beta F^{\alpha\beta} \sqrt{|\gamma|}d^{D-1}y&=&0\labell{Apen4}
\eeqa
where $F^{\alpha\beta}$ is an arbitrary antisymmetric tensor. This is the formula that we have used in  \reef{bstokes}.

\end{document}